\documentclass[a4paper, 12pt]{article}
\usepackage[english]{babel}
\usepackage{amsmath, amsfonts, amsthm, mathrsfs, latexsym}
\usepackage[dvips]{graphicx}

\newcommand{\abs}[1]{\lvert#1\rvert}
\newcommand{\norm}[1]{\lVert#1\rVert}
\newcommand{\field}[2]{\mathbb{#1}^{#2}}
\newcommand{\deriv}[2]{\partial_{#1}^{#2}}
\newcommand{\hilbert}[1]{\mathscr{#1}}
\newcommand{\fock}{\hilbert{F}}
\newcommand{\coinf}{C_{0}^{\infty}}

\renewcommand{\vec}[1]{\boldsymbol{#1}}

\newcommand{\N}{\mathrm{N}}
\newcommand{\E}{{\mathrm{e}}}
\newcommand{\I}{\mathrm{i}}
\newcommand{\Or}{{\mathcal{O}}}
\newcommand{\qn}{Q_{\mathrm{N}}}
\newcommand{\qm}{Q_{M}}

\newcommand{\pepsim}{P_{M}^{\epsi}}
\newcommand{\qless}[1]{Q_{\leq\mathrm{#1}}}
\newcommand{\dgamma}{\mathrm{d\Gamma}}
\newcommand{\hfree}{H_{0}^{\epsi}}
\newcommand{\hdress}{H_{\rm eff}^\epsi}
\newcommand{\heff}{H_{\mathrm{dres}}^{\epsi}}
\newcommand{\heffsigma}{H_{\mathrm{dres}}^{\epsi, \sigma}}
\newcommand{\hefftwochi}{H_{\mathrm{dres},\chi}^{(2)}}
\newcommand{\hdiag}{H_{\mathrm{D}}^{(2)}}
\newcommand{\hdiagp}{H_{\mathrm{D}, \mathrm{p}}^{(2)}}
\newcommand{\hfield}{H_{\mathrm{f}}}
\newcommand{\hp}{h_{\mathrm{p}}}
\newcommand{\Hp}{H_{\mathrm{p}}}
\newcommand{\Hepsip}{\Hp^{\epsi}}
\newcommand{\id}{\mathbf{1}}
\newcommand{\rfield}{R_{\mathrm{f}}^{\perp}}
\newcommand{\hfib}{H_{\mathrm{fib}}}
\newcommand{\hfibsigma}{H_{\mathrm{fib}, \sigma}}
\newcommand{\rfib}{R_{\mathrm{fib}}^{\perp}}
\newcommand{\pop}{\hat{p}}
\newcommand{\popb}{\hat{\vec{p}}}
\newcommand{\epsi}{\varepsilon}
\newcommand{\hepsi}{H^{\epsi}}
\newcommand{\hepsisigma}{H^{\epsi, \sigma}}
\newcommand{\hepsisigmaepsi}{H^{\epsi, \sigma(\epsi)}}

\newcommand{\proj}[2]{\hat{\pi}^{#1}_{#2}}
\newcommand{\pizn}{\hat{\pi}_{0}^{M}}
\newcommand{\uone}{\hilbert{U}}
\newcommand{\uepsi}{\hilbert{U}_{\epsi}}
\newcommand{\uepsisigma}{\hilbert{U}_{\epsi, \sigma}}
\newcommand{\uonemchi}{U^{(1)}_{\mathrm{J}, \chi}}
\newcommand{\uonemchistar}{U^{(1)\,*}_{\mathrm{J}, \chi}}

\newcommand{\phim}{\Phi^{\mathrm{J}}}

\newcommand{\phisigma}{\Phi_{\sigma}}
\newcommand{\phisigmaj}{\Phi_{\sigma}^{\mathrm{J}}}
\newcommand{\phisigmal}{\Phi_{\sigma, l}}
\newcommand{\phisigmajl}{\Phi_{\sigma, l}^{\mathrm{J}}}

\newcommand{\phizerojl}{\Phi^{\mathrm{J}}_{0, l}}
\newcommand{\eqex}{\overset{!}{=}}
\newcommand{\opw}{\mathrm{Op}^{W}_{\epsi}}
\newcommand{\omegap}{\omega_{\mathrm{p}}}
\newcommand{\lh}{\mathcal{L}(\hilbert{H})}

\newcommand{\tr}{\mathrm{tr}}
\newcommand{\vsigma}{V_{\sigma}}
\newcommand{\fockvac}{\Omega_{\mathrm{F}}}

\newcommand{\fockfin}{\fock_{\mathrm{fin}}}
\newcommand{\Proj}[2]{\Pi^{#1}_{#2}}
\newcommand{\hdiagtilde}{\tilde{H}_{\mathrm{D}}^{(2)}}
\newcommand{\hiepsi}[1]{h_{#1}}

\newtheorem{theorem}{Theorem}
\newtheorem*{theorem*}{Theorem}
\newtheorem{lemma}{Lemma}
\newtheorem*{prop}{Proposition}
\newtheorem{proposition}{Proposition}
\newtheorem{corollary}{Corollary}

\theoremstyle{remark}
\newtheorem{remark}{Remark}

\title{Effective dynamics for particles coupled to a quantized scalar field}
\author{L. Tenuta \and S. Teufel}
\date{}

\begin{document}

\maketitle

\begin{center}
Mathematisches Institut, Eberhard-Karls-Universit\"at, Auf der Morgenstelle 10, 72076, T\"ubingen, Germany.

e-mail: lucattilio.tenuta@uni-tuebingen.de, stefan.teufel@uni-tuebingen.de
\end{center}

\begin{abstract}
We consider a system of $N$ non-relativistic spinless quantum particles (``electrons'') interacting with a quantized scalar Bose field (whose excitations we call ``photons''). We examine the case when the velocity $v$ of the electrons is small with respect to the one of the photons, denoted by $c$ ($v/c=\varepsilon\ll 1$). We show that dressed particle states exist (particles surrounded by ``virtual photons''), which, up to terms of order $(v/c)^3$, follow Hamiltonian dynamics. The effective $N$-particle Hamiltonian  contains the kinetic  energies of the particles and Coulomb-like  pair potentials at order $(v/c)^0$ and  the velocity dependent Darwin interaction and a mass renormalization  at  order $(v/c)^{2}$. Beyond that order the effective dynamics are expected to be dissipative. 

 The main mathematical tool we use is adiabatic perturbation theory. However, in the present case there is no    eigenvalue which is separated by a gap from the rest of the spectrum, but its role is taken by the bottom of the absolutely continuous spectrum, which is not an eigenvalue.
 Nevertheless we construct    approximate  dressed electrons subspaces, which are adiabatically invariant for the dynamics up to   order $(v/c)\sqrt{\ln [(v/c)^{-1}]}$. We also give an explicit expression for the non adiabatic transitions corresponding to emission  of free photons. For the radiated energy we obtain the quantum analogue of the Larmor formula of classical electrodynamics.
\end{abstract}

\section{Introduction}
In a system of classical charges the interactions are mediated through the electromagnetic field. In the case when the velocities of the particles are small with respect to the speed of light, it is possible, loosely speaking, to expand their dynamics in powers of $v/c$. 

The qualitative picture which emerges has three main features. Up to terms of order $(v/c)^{3}$, the dynamics of the particles are still of Hamiltonian form. Moreover,    
\begin{itemize}
\item at leading order, $(v/c)^{0}$, the retardation effects can be neglected and the particles interact through an instantaneous pair potential;
\item at order $(v/c)^{2}$, the particles acquire an effective mass, due to the contribution of the electromagnetic mass and, to take into account the retardation effects, one has to add to the potential a velocity-dependent term (the so-called Darwin term).
\end{itemize} 

Including the terms of order $(v/c)^{3}$, the dynamics are not Hamiltonian anymore, instead
\begin{itemize}
\item there is dissipation of energy through radiation. In the dipole approximation, the rate of emitted energy is proportional to the acceleration of a particle squared.
\end{itemize}

A formal derivation of this picture, which does not consider the problem of mass renormalization, can be found in \cite{LaLi}. A mathematical analysis in the framework of the Abraham model, i.\ e., for charges which have a rigid charge distribution, is given in \cite{KuSp1} \cite{KuSp2}.

The above description is expected to remain true also for nonrelativistic quantum electrodynamics, where, neglecting the possibility of pair creation, one considers a system of $N$ nonrelativistic particles interacting with the quantized Maxwell field. In physical terms, the particles carry now a cloud of virtual photons, which makes them heavier, and interact exchanging them or dissipate energy through photons travelling freely to infinity. However, in quantum mechanics one describes the interaction of charged particles in good approximation  by introducing instantaneous pair potentials and without treating the field as dynamical variable. If the particles move sufficiently slowly this is known to be a very good approximation. One goal of our paper is  the mathematical derivation of quantum mechanics from a model of particles that are  coupled to a quantized field, but do not interact directly.

Instead of nonrelativistic QED we consider the massless Nelson model, which describes $N$ spinless particles (which will be called ``electrons'') coupled to a scalar Bose field of zero mass (whose excitations will be called ``photons''). In spite of the simplifications introduced, this model is expected to retain the main physical features of the original one. Therefore, since its introduction by Nelson \cite{Ne}, who analyzed its ultraviolet behavior, it has been extensively studied to get information about the spectral and scattering features of QED, mostly concerning its infrared behavior (not pretending to be exhaustive, some papers related to this aspect are \cite{Ar} \cite{Fr} \cite{LMS} \cite{Pi} and references therein). The recent monograph by Spohn \cite{Sp} contains detailed descriptions of the classical and the quantum mechanical models and results.

In this paper we define and analyze  the dynamics of dressed electron states in the Nelson model in the limit of small particle velocities. Loosely speaking a  dressed electron  is a bare electron dragging with it a cloud of ``virtual'' photons. We show that the dynamics of dressed electrons have the features discussed above: an instantaneous  pair interaction at leading order $(v/c)^0$ and  a renormalized mass together with the velocity-dependent Darwin interaction at order $(v/c)^2$. We also provide an analogue to the classical Larmor formula   for the radiated energy, i.e.\ for the energy carried away by ``real'' photons travelling freely to infinity (the heuristic concept of ``real'' and ``virtual'' photons will be made precise below).  It is important to note that we consider  the massless Nelson model with an ultraviolet cutoff but \emph{no infrared cutoff}. Indeed, the leading order effective dynamics were already analyzed in \cite{Te2} assuming an infrared regularization.
One expects, and we will show it in this paper, that the dynamics of the dressed electrons even at higher orders are essentially independent of an infrared regularization, but the radiation is instead very sensitive to it, which makes the mathematical analysis much more delicate.

To explain in more detail the kind of scaling we are interested in it is convenient to look first at the classical case.
The classical equations of motion for $N$ particles with positions $\vec{q}_{j}$, mass $m_{j}$ and a rigid ``charge'' distribution $\varrho_{j}$ coupled to a scalar field $\phi(\vec{x}, t)$ \footnote[1]{We use bold italic font, $\vec{x}$, to denote vectors in $\field{R}{3}$. The only exception to this, since there is no possibility of misunderstanding, is the three-dimensional momentum of the photons, denoted by $k$. The lightface font, $x$, will be used to denote vectors in $\field{R}{3N}$.} with propagation speed $c$ are
\begin{eqnarray}
\frac{1}{c^{2}}\deriv{t}{2}\phi(\vec{x}, t) & = & \Delta_{\vec{x}}\phi(\vec{x}, t) - \sum_{j=1}^{N}\varrho_{j}(\vec{x}-\vec{q}_{j}(t)), \label{classicaleqfield}\\
m_{j}\ddot{\vec{q}}_{j}(t) & = & -\int dx\, (\nabla_{\vec{x}}\phi)(\vec{x}, t)\varrho_{j}(\vec{x}-\vec{q}_{j}(t)),\label{classicaleqpart} \qquad 1\leq j \leq N \, . 
\end{eqnarray} 

We assume for simplicity that
\begin{equation*}
\varrho_{j}(\vec{x}) = e_{j}\varphi(\vec{x}),
\end{equation*}
where the form factor $\varphi$ gives rise to a sharp ultraviolet cutoff, 
\begin{equation}\label{formfactor}
\hat{\varphi}(k)=\begin{cases} (2\pi)^{-3/2} & \abs{k}<\Lambda\, ,\\
0 & \text{otherwise}\, .
                \end{cases}
\end{equation}

Taking formally the limit $c\to\infty$ in \eqref{classicaleqfield}, we get the Poisson equation for the field, and so, eliminating the field from \eqref{classicaleqpart}, we obtain equations of motion describing $N$ particles interacting through smeared Coulomb potentials. Mass renormalization does not appear at the leading order.

Instead of taking $c\to\infty$, we regard as more natural to explore the regime of particle properties which gives rise to effective equations. Indeed, if we look at \emph{heavy particles} for \emph{long times}, i.\ e., if we substitute $t'=\epsi t$ and $m_{j}'=\epsi^{2}m_{j}$ in \eqref{classicaleqfield} and \eqref{classicaleqpart}, we find that, up to rescaling, the limit $\epsi\to 0$ is equivalent to the limit $c\to\infty$. After quantization however, the two limiting procedures are not equivalent anymore. 

The case $c\to\infty$ was analyzed by Davies \cite{Da} and by Hiroshima \cite{Hi}, who at the same time removed the ultraviolet cutoff, applying methods from the theory of the weak coupling limit. In this paper we analyze the limit $\epsi\to 0$.

We define now briefly the massless Nelson model (whose presentation will be completed in section \ref{nelsonmodel}), we state our main results and the principal ideas of the proof and then compare them to the above mentioned weak coupling limit.

The model is obtained through canonical quantization of the classical system described by \eqref{classicaleqfield} and \eqref{classicaleqpart}. The state space for $N$ spinless particles is $L^{2}(\field{R}{3N})$ and the Hamiltonian, assuming for simplicity that all the particles have equal mass, is given by (we switch to natural units, fixing $\hbar=1$ and $c=1$)
\begin{equation*}
\Hp:= -\frac{1}{2m}\sum_{j=1}^{N}\Delta_{\vec{x}_{j}} . 
\end{equation*}

As explained above, we consider the case of heavy particles, i.\ e.,
\begin{equation}
m = \epsi^{-2}, \qquad 0<\epsi\ll 1,
\end{equation}
therefore the Hamiltonian becomes
\begin{equation}
 \Hepsip = -\frac{\epsi^{2}}{2}\Delta_{x} =: \frac{1}{2}\,\pop^{2}\,,
\qquad  x\in\field{R}{3N}\,,\quad  \pop := -\I\epsi\nabla_{x}\,.
\end{equation}

The particles are coupled to a scalar field, whose states are elements of the bosonic Fock space over $L^{2}(\field{R}{3})$, defined by
\begin{equation}\label{fockspace}
\fock := \oplus_{M=0}^{\infty}\otimes_{(s)}^{M}L^{2}(\field{R}{3}),
\end{equation}
where $\otimes_{(s)}^{M}$ denotes the $M$-times symmetric tensor product and $\otimes_{(s)}^{0}L^{2}(\field{R}{3}):=\field{C}{}$. We denote by $\qm$ the projector on the $M$-particles subspace of $\fock$ and by $\fockvac$ the vector $(1, 0, \ldots)\in\fock$, called the Fock vacuum.

The Hamiltonian for the free bosonic field is
\begin{equation}\label{hfield}
\hfield := \dgamma(\abs{k}),
\end{equation}
where $k$ is the momentum of the photons (the reader who is not familiar with the notation can find more details in section \ref{nelsonmodel}).

The particle $j$ is linearly coupled to the field through the interaction Hamiltonian
\begin{equation}\label{hintj}
H_{\mathrm{I}, j}:= \int_{\field{R}{3}}d\vec{y}\, \phi(\vec{y})\varrho_{j}(\vec{y}-\vec{x}_{j}),
\end{equation}
where $\phi$ is the field operator in position representation.

The state space of the combined system particles $+$ field is 
\begin{equation}
\hilbert{H}:= L^{2}(\field{R}{3N})\otimes\fock \simeq L^{2}(\field{R}{3N}, \fock)
\end{equation}
and its time evolution is generated by the Hamiltonian
\begin{equation}\label{hepsi}\begin{split}
& \hepsi := \hfree + \sum_{j=1}^{N}H_{\mathrm{I}, j},\quad \hfree := \Hepsip\otimes\id + \id\otimes\dgamma(\abs{k})\, ,
\end{split}
\end{equation}
with domain
\begin{equation}\label{hilbertzero}
\hilbert{H}_{0}:= H^2(\field{R}{3N}, \fock) \cap L^2(\field{R}{3N}, D(\hfield)) \, , \quad  H^2(\field{R}{3N}, \fock) := D(\pop^{2}\otimes\id)\, ,
\end{equation}
which is a Hilbert space with the graph norm associated to $\hfree$.

Note that there are no direct forces acting between the particles, all the interactions are mediated through the field.

Our goal is to understand  the dynamics of  the particles  for times of order $\epsi^{-1}$. It is necessary to look at  long times in order to see nontrivial dynamics of the particles, because, since their mass is of order $\Or(\epsi^{-2})$ and we consider states of finite kinetic energy, their velocity is of order $\Or(\epsi)$. However, since the coupling between the electrons and the field is {\em not} small, standard perturbation theory is of no use initially. Indeed, since the charge of the particles is of order one, the local deformation of the field, i.\ e.\ the ``cloud of virtual photons'', is of order one. However, the influence of real photons with finite energy and momentum on the dynamics of the heavy  electrons, whose mass is of order $\epsi^{-2}$, is small. 
Hence one expects    that the coupling between properly defined dressed electron states and real photons is small. 
To make this precise, we construct  a dressing transformation $\uepsi:\hilbert{H}\to\hilbert{H}$, which allows us to define the dressed particle states as follows. In the new representation the vacuum sector $L^2(\field{R}{3N})\otimes \fockvac$ of $\hilbert{H}$ corresponds to states of dressed electrons without real photons, while in the original Hilbert space a state of  dressed electrons with $M$ real photons would be a linear combination of states of the form
\[
\uepsi^{-1}\, ( \psi \otimes a(f_1)^*\cdots a(f_M)^* \fockvac ) \quad\mbox{with}\quad \psi\in L^2(\field{R}{3N})\,,\,\,f_1,\ldots,\,f_M \in L^2(\field{R}{3})\,.
\]
Recall that   $\qm$ denotes the projector on the $M$-particles subspace of Fock space, then the projector on the subspace corresponding to dressed electrons with $M$ real photons is
\[
\pepsim :=\uepsi^*(\id\otimes\qm)\uepsi\,.
\]
In a nutshell, the main results we prove are the following: the subspaces $\pepsim \hilbert{H}$ are approximately invariant under the dynamics generated by $\hepsi$ for times of order $\epsi^{-1}$. For states inside such a subspace the dynamics of the particles can be described by an effective Hamiltonian for the particles alone on the above time scale and with errors of order $\Or\big(\epsi^2\log(\epsi^{-1})\big)$. Finally we can compute the leading order part of the state which leaves the subspace $P^\epsi_0 \hilbert{H}$ under the time evolution, which corresponds to emission of real photons. The formula for the energy of the real photons traveling to infinity, i.e.\ the radiated energy, yields a quantum mechanical analogue of the classical Larmor formula for the radiation of accelerated charges.

%

Before we can state our results in detail, we need to explain the adiabatic structure of the problem in some detail. The Hamiltonian $\hepsi$ is  the perturbation of a fibered Hamiltonian, because, since $H_{\mathrm{I}, j}$ depends only on the configuration $\vec{x}_{j}$ of the $j$th particle, the operator
\begin{equation}
\hfib(x) := \dgamma(\abs{k}) + \sum_{j=1}^{N}H_{\mathrm{I}, j}(\vec{x}_{j})
\end{equation}
acts on $\fock$ for every fixed $x\in\field{R}{3N}$. This means that
\begin{equation*}
\hepsi = -\frac{\epsi^{2}}{2}\Delta_{x}\otimes\id + \int_{\field{R}{3N}}^{\oplus}\hfib(x)\, .
\end{equation*}

\begin{figure}\begin{center}
\includegraphics[width=0.7\textwidth]{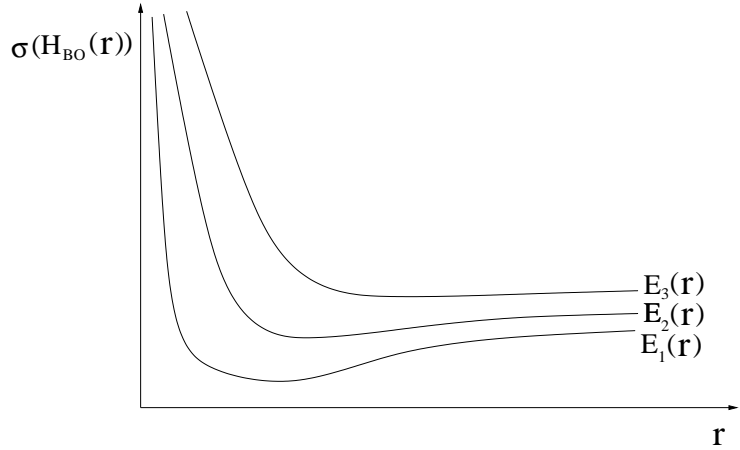}\caption{\footnotesize Schematic spectrum of the fibered Hamiltonian in the case of a diatomic molecule for energies below the dissociation threshold ($r=\abs{\vec{x}_{1}-\vec{x}_{2}}$). The different eigenvalues are pointwise separated by a gap.}\label{diatomic}
\end{center}
\end{figure}

Note the structural similarity with the Born-Oppenheimer approximation. There the Hamiltonian describes the interaction between nuclei and electrons in a molecule and the former have a mass of order $\Or(\epsi^{-2})$ with respect to the latter (the typical spectrum of $\hfib(x)$ for a diatomic molecule is shown in figure \ref{diatomic}). In the present case, the particles take the role of the nuclei, and the bosons the one of the electrons. 

\begin{figure}\begin{center}
\includegraphics[width=0.7\textwidth]{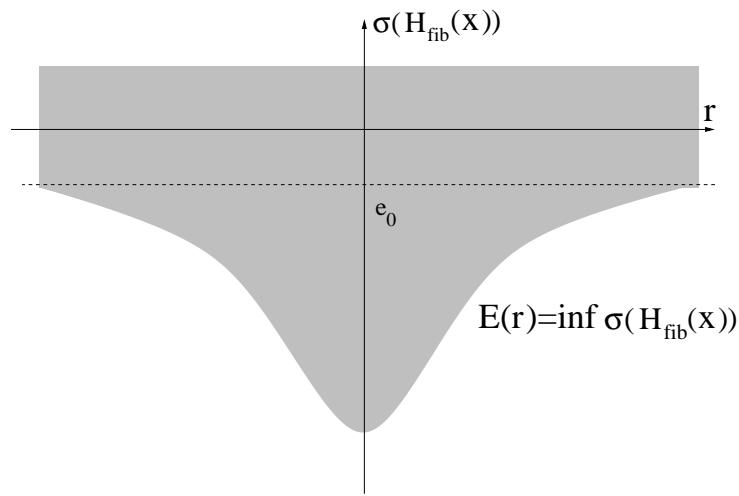}\caption{\footnotesize Spectrum of $\hfib(x)$ for $N=2$ ($r=\abs{\vec{x}_{1}-\vec{x}_{2}}$). The spectrum is absolutely continuous, there is \emph{no eigenvalue at the bottom}. The oscillations in $r$ caused by the sharp ultraviolet cutoff are irrelevant, and therefore we do not show them.}\label{nelson} 
\end{center}\end{figure}

In contrast to the molecular case however, in the Nelson model $\hfib(x)$ has typically (see lemma \ref{fiberedspectrum} and corollaries \ref{corollaryfiberedspectrum} and \ref{corollaryfiberedspectrumtwo})  purely absolutely continuous spectrum, which does not display a structure with pointwise separated bands (\emph{absence of both eigenvalues and spectral gap}) (see figure \ref{nelson}).

The bottom of the spectrum, $E(x)$, can be explicitly calculated,
\begin{equation}\label{ex}
E(x) = \frac{1}{2}\underset{i\neq j}{\sum_{i, j=1}^{N}}V_{ij}(\vec{x}_{i}-\vec{x}_{j}) + e_{0},
\end{equation}
where 
\begin{equation}
V_{ij}(\vec{z}):= -\int_{\field{R}{3}\times\field{R}{3}}d\vec{v}d\vec{w}\, \frac{\varrho_{i}(\vec{v}-\vec{z})\varrho_{j}(\vec{w})}{4\pi\abs{\vec{v}-\vec{w}}},
\end{equation}
and 
\begin{equation}
e_{0}:= -\frac{1}{2}\sum_{j=1}^{N}\int_{\field{R}{3}\times\field{R}{3}}d\vec{v}d\vec{w}\, \frac{\varrho_{j}(\vec{v})\varrho_{j}(\vec{w})}{4\pi\abs{\vec{v}-\vec{w}}}\, .
\end{equation}
The effective pair-potential $V_{ij}(\vec{z})$ coincides, up to the sign, with the electrostatic interaction energy of the charge distributions $\varrho_{i}$ and $\varrho_{j}$ at distance $\vec{z}$, while $e_{0}$ is the sum of all the self-energies.

$E(x)$ becomes an eigenvalue of $\hfib(x)$ if the total charge of the $N$ particles system is equal to zero, as it happens for example in presence of an infrared cutoff. In this case, it is possible to build for every $x$ a unitary operator, $V(x)$, called the \emph{dressing operator}, which diagonalizes $\hfib(x)$ in the sense that
\begin{equation}
\hfib(x) = V(x)\hfield V(x)^{*} + E(x)\,.
\end{equation} 

Exploiting this remark, we approximate the time evolution generated by $\hepsi$ in two steps. 
First we define an infrared regular Hamiltonian, $\hepsisigma$, where the form factor $\hat{\varphi}$ (equation \eqref{formfactor}) is replaced by
\begin{equation}\label{formfactorsigma}
\hat{\varphi}_{\sigma}(k):=\begin{cases} (2\pi)^{-3/2} & \sigma<\abs{k}<\Lambda\, ,\\
0 & \text{otherwise}\, .
                \end{cases}
\end{equation}

\begin{prop}(see proposition \ref{approxsigma}) Let $\mathcal{L}(\hilbert{H}_{0}^{1/2}, \hilbert{H})$ be the space of bounded linear operators from $\hilbert{H}_{0}^{1/2}$ to $\hilbert{H}$ equipped with the operator norm. It holds then that
\begin{equation}
\big\lVert\E^{-\I t\hepsi/\epsi} - \E^{-\I t \hepsisigma/\epsi}\big\rVert_{\mathcal{L}(\hilbert{H}_{0}^{1/2}, \hilbert{H})}\leq C\abs{t}\frac{\sigma^{1/2}}{\epsi},
\end{equation}
where $\hilbert{H}_{0}^{1/2}:= D\big((\hfree)^{1/2}\big)$ with the corresponding graph norm.
\end{prop}

Choosing $\sigma$ as a power of $\epsi$ we can then replace the original dynamics with infrared regular ones. The latter contain however two parameters, $\epsi$ and $\sigma$, therefore it is necessary to control the behavior of all the quantities that appear with respect to both.

The advantage is that for $\hepsisigma$ we can build an approximate dressing operator $\uepsisigma$, acting on the whole Hilbert space $\hilbert{H}$, which is defined for every positive $\sigma$, but whose limit when $\sigma\to 0^{+}$ does not exist if the system has a total charge different from zero. $\uepsisigma$ is unitary and can be expanded in a series of powers of $\epsi$, with coefficients which are \emph{at most logarithmically divergent} in $\sigma$. Moreover, the zero order coefficient is given by $\vsigma$, the dressing operator associated to the infrared regular fibered Hamiltonian, which has therefore the property that
\begin{equation*}
\hepsisigma = \frac{1}{2}\,\pop^{2}\otimes\id + \vsigma(x)\hfield \vsigma(x)^{*} + E_{\sigma}(x).
\end{equation*}

Using $\uepsisigma$, we define the transformed Hamiltonian
\begin{equation}
\heffsigma := \uepsisigma\hepsisigma\uepsisigma^{*}
\end{equation}
which can be expanded in a series of powers of $\epsi$ in $\mathcal{L}(\hilbert{H}_{0}, \hilbert{H})$, with coefficients which are also at most logarithmically divergent in $\sigma$.
Thus the gain in switching to the representation defined by $\uepsisigma$ is twofold. First we can easily separate dressed electrons from real photons and second we can expand the Hamiltonian in powers of the small parameter $\epsi$ and thereby separate different physical effects according to their order of magnitude. 

The first result we find is that, even though the fibered Hamiltonian has no eigenvalues and no spectral gap, there are \emph{approximate $M$-photons dressed subspaces} which are almost invariant for the dynamics.

\begin{theorem*}[Adiabatic invariance of $M$-photons dressed particles subspaces] 
(see theorem \ref{zeroordertime} and remark \ref{adiabaticinvm})\\[2mm]
Given any $\chi\in\coinf(\field{R}{})$ and any function $\sigma(\epsi)$ such that
\begin{equation}\label{conditionsigma}
\sigma(\epsi)^{1/2}\epsi^{-3}\to 0, \qquad \epsi\sqrt{\log(\sigma(\epsi)^{-1})}\to 0, \qquad \epsi\to 0^{+}\, , 
\end{equation} 
then 
\begin{equation}\label{almostinvsub}
\big\lVert[e^{-\I t\hepsi/\epsi}, \pepsim]\chi(\hepsi)\big\rVert_{\lh}\leq C\sqrt{M+1}\abs{t}\epsi\sqrt{\log(\sigma(\epsi)^{-1})},
\end{equation}
where 
\begin{equation}
\pepsim :=\hilbert{U}^*_{\epsi, \sigma(\epsi)} (\id\otimes\qm)\hilbert{U}_{\epsi, \sigma(\epsi)} \,.
\end{equation}
\end{theorem*}

The physical mechanism which leads to the almost invariance of the subspaces is adiabatic decoupling, i.\ e., the separation of scales for the motion of the different parts of the system, which lets the fast degrees of freedom, in our case the photons, instantaneously adjust to the motion of the slow degrees of freedom, the electrons. 
It is a well-known fact however, that the decoupling becomes poorer and poorer when the kinetic energies and thus the velocities of the heavy particles grow. The quadratic dispersion relation for the electrons allows them to become arbitrarily fast, therefore the decoupling holds uniformly just on states of bounded kinetic energy. This is the reason why we introduce a uniform bound on the total energy of the system through the function $\chi$, which obviously implies a bound on the kinetic energy of the electrons.

For the following we fix the function $\sigma(\epsi)$ in some way compatible with \eqref{conditionsigma}, say $\sigma(\epsi) = \epsi^8$. Then we can approximate the dynamics of the particles for states in the range of $\pepsim$ in the following sense.

\begin{theorem*}[Effective dynamics of the particles](see theorem \ref{densitymatrix})

Let $\omega$  be a (mixed) dressed electrons state with finite energy and a fixed number of real photons, i.\ e.\
$\omega\in \hilbert{I}_{1}(P_{M}^{\epsi}\chi(\hepsi)\hilbert{H})$ and 
\begin{equation}\label{omegaddef}
\omega_{\rm d} :=  \tr_\fock \left( \uepsi\,\omega\, \uepsi^{*}\right)
\end{equation}
the partial trace over the real photons, i.\ e.\  the reduced dressed electron density matrix. 
Let the time evolution of $\omega$ be the full time evolution
\begin{equation}
\omega(t):= e^{-\I t\hepsi/\epsi}\,\omega \,e^{\I t\hepsi/\epsi}\,,
\end{equation}
and define the effective time evolution of $\omega_{\rm d}$   by
\[
\omega_{\rm d}(t) := e^{-\I t\hdress/\epsi}\,\omega_{\rm d}\, e^{\I t\hdress/\epsi}\,,
\]
with effective dressed electrons Hamiltonian $\hdress$ given below.

Let $a\in C^\infty_{\rm b} (\field{R}{3N}\times \field{R}{3N})$ be a   ``macroscopic'' observable on the classical phase space of the electrons and $\opw(a)$ its Weyl quantization acting on $L^2(\field{R}{3N})$. Then 
\begin{eqnarray}\label{approx}
 \tr_{\hilbert{H}}\bigg(\big(\opw(a)\otimes\id_{\fock}\big)\,\omega(t)\bigg)&=&\tr_{L^{2}(\field{R}{3N})}\bigg(\opw(a) \,\omega_{\rm d}(t)\bigg) +\hspace{-0.3mm} \Or(\epsi^{3/2}\abs{t})(1 - \delta_{M0}) \nonumber \\[2mm] &&  +\, \Or\big(\epsi^{2}\log( \epsi^{-1}) (\abs{t}+\abs{t}^{2})\big),
\end{eqnarray}
where $\delta_{M0}=1$, when $M=0$, $0$ otherwise.

The effective dressed electrons Hamiltonian is infrared regular and given by  
\begin{eqnarray}
\hdress &:=& \sum_{j=1}^{N}\frac{1}{2m_{j}^{\epsi}}\vec{\pop}_{j}^{2}+E(x)\nonumber \\
&&-\,\frac{\epsi^{2}}{4}\underset{(l \neq j )}{\sum_{l , j =1}^{N}}\int_{\field{R}{3}}dk\, \frac{\hat{\varrho}_{l }(k)^{*}\hat{\varrho}_{j }(k)}{\abs{k}^{2}}\bigg[\E^{\I k\cdot(\vec{x}_{j }-\vec{x}_{l })}(\kappa\cdot\vec{\pop}_{l })(\kappa\cdot\vec{\pop}_{j })\\ &&\,\qquad\qquad\hspace{3cm}+(\kappa\cdot\vec{\pop}_{l })(\kappa\cdot\vec{\pop}_{j })\E^{\I k\cdot(\vec{x}_{j }-\vec{x}_{l })}\bigg],\nonumber
\end{eqnarray}
with $m_{j}^{\epsi}:=1/(1+\frac{\epsi^{2}}{2}\tilde{e}_{j})$ and 
\begin{equation}
\tilde{e}_{j}:= \frac{1}{4\pi}\int_{\field{R}{3}\times\field{R}{3}}d\vec{x}\ d\vec{y}\ \frac{\varrho_{j}(\vec{x})\varrho_{j}(\vec{y})}{\abs{\vec{x}-\vec{y}}} \quad \,.
\end{equation}
\end{theorem*}
\begin{remark}

The Hamiltonian $\hdress$ is equal to the Weyl quantization of the Darwin Hamiltonian, which, as we mentioned above, appears in classical electrodynamics when one expands the dynamics for the particles including term of order $\epsi^2\cong (v/c)^{2}$.   At leading order the dressed electrons interact through instantaneous pair potentials $E(x)$ given in \eqref{ex}. At order $\epsi^{2}$ the mass of the electrons is modified (renormalized) and a velocity dependent interaction, the so called Darwin interaction, appears.
\end{remark}

\begin{remark}
The statement \eqref{approx} remains true if one replaces in \eqref{omegaddef} the dressing operator $\uepsi$ by its leading order approximation $V_\epsi$.
\end{remark}

The dressed electrons subspaces $\pepsim\hilbert{H}$ are only approximately invariant and transitions between them correspond to emission or absorption of real photons. The following theorem describes at leading order the radiated part of the wave function for an initial state in the dressed vacuum.

\begin{theorem*}[Radiation](see corollary \ref{radiatedpiece} and the subsequent remark)

For a system starting in the approximate dressed vacuum $(M=0)$ we have,
\begin{eqnarray} \lefteqn{\hspace{-0cm}\label{transwave}
\Psi_{\rm rad}(t) :=   (\id - P_{0}^{\epsi})e^{-\I \frac{t}{\epsi}\hepsi}P_{0}^{\epsi}\chi(\hepsi)\Psi=}\\
&&\hspace{-3pt} -\frac{\epsi}{\sqrt{2}}P_{1}^{\epsi}\, V_{\epsi}\, \E^{-\I \frac{t}{\epsi}(\pop^{2}/2+E(x))}\,\E^{-\I \frac{t}{\epsi} |k|}  \sum_{j=1}^{N}\frac{e_{j}\hat\varphi_\epsi(k)}{\abs{k}^{3/2}}\kappa\cdot\hspace{-1mm}\int_{0}^{t}ds\, \E^{\I \frac{s}{\epsi}\abs{k}}\opw(\ddot{\vec{x}}_{j}^{c}(s; x, p))\,\psi_{\rm d} \,\nonumber \\
&&\hspace{-3pt} + R(t, \epsi), \nonumber
\end{eqnarray}
where 
\begin{equation*}
\norm{R(t, \epsi)}_{\hilbert{H}}\leq C\epsi^{2}\log(\epsi^{-1})(\abs{t}+\abs{t}^{2})(\norm{\psi_{\rm d}}_{\hilbert{H}} + \norm{\abs{x}\psi_{\rm d}}_{\hilbert{H}} + \norm{\abs{\pop}\psi_{\rm d}}_{\hilbert{H}})\, ,
\end{equation*}
$\vec{x}_{j}^{c}$ is the solution to the classical equations of motion
\begin{equation}\begin{split}
& \ddot{\vec{x}}_{j}^{c}(s; x, p) = -\nabla_{\vec{x}_{j}}E(x^{c}(s; x, p)),\\
& \vec{x}_{j}^{c}(0; x, p)=\vec{x}_{j},\qquad \dot{\vec{x}}_{j}^{c}(0; x, p)=\vec{p}_{j},\quad j=1, \ldots, N\, ,
\end{split}
\end{equation}
$\kappa:= k/\abs{k}$ and 
\begin{equation}\label{psidressed}
\psi_{\rm d} := \langle V_{ \epsi }\Omega_{\mathrm{F}}, \chi(\hepsi)\Psi\rangle_{\fock} 
\end{equation}
is the projection of the initial state on the dressed vacuum in the field component. 
\end{theorem*} 

\begin{remark}
Generically (for the precise meaning of this we refer to corollary \ref{radiatedpiece}) the norm of the radiated piece is bounded below by $\Or\big(\epsi\log(\epsi^{-1})\big)$, which means that the subspace $P_{0}^{\epsi}$ is near \emph{optimal}, in the sense that the transitions are at least of order $\Or\big(\epsi\log(\epsi^{-1})\big)$. 
\end{remark}

\begin{remark} The radiated wave function 
$\Psi_{\rm rad}(t)$ as given in \eqref{transwave} has no limit as $\epsi\to 0$ because $\lim_{\epsi\to 0}\hat\varphi_\epsi/|k|^{3/2}= \hat\varphi_0/|k|^{3/2}\notin L^2(\field{R}{3})$. However, the radiated energy, i.\ e.\ the energy of the real photons in \eqref{transwave} has a  limit. Indeed, if we compute the time derivative of the radiated energy in \eqref{transwave} we obtain the quantum analogue of the Larmor formula from classical electrodynamics. Let 
\begin{equation}\label{radiatedenergy}
E_{\rm rad}(t) := \langle \Psi_{\rm rad}(t) , V_\epsi\, \hfield\,V_\epsi^*\,\Psi_{\rm rad}(t)\rangle\,,
\end{equation}
then 
\begin{eqnarray}\label{radiatedpower}
P_{\rm rad}(t) &:=& \frac{d}{dt}E_{\rm rad}(t) \cong \frac{\epsi^3}{12\pi} \sum_{i,j=1}^N e_i e_j \langle \psi_{\rm d} , \opw \big(\ddot{\vec{x}}^c_i(t;x,p)\cdot  \ddot{\vec{x}}^c_j(t;x,p)\big)\,\psi_{\rm d}\rangle\nonumber\\ &=&
\frac{\epsi^3}{12\pi} \langle \psi_{\rm d} , \opw \big( |\vec{\ddot{d}}(t;x,p)|^2 \big)\,\psi_{\rm d}\rangle
\,,
\end{eqnarray}
where $\vec{\ddot{d}}(t;x,p)$ is the second time derivative of the dipole moment
\[
\vec{d}(t;x,p) := \sum_{i=1}^N e_i \vec{x}_i^{c}(t;x,p)\,,
\]
and the symbol $\cong$ means that we keep just the leading order in the expansion in powers of $\epsi$.
A formal derivation of equation \eqref{radiatedpower} is given in remark \ref{remradiatedpower}.
\end{remark}

The technique used for proving these theorems is based on space-adiabatic perturbation theory, a general scheme designed to expand the dynamics generated by a pseudodifferential operator with a semiclassical symbol \cite{NeSo} \cite{PST1} \cite{Te3}. However, this method exploits the assumption that the fibered Hamiltonian has a spectral gap, of the kind showed in figure \ref{diatomic}, therefore is not directly applicable and needs some modifications. In particular, the infrared cutoff $\sigma$ plays in our context the role of an effective gap.

In the more usual framework of time-dependent Hamiltonians, adiabatic theorems without gap condition were first proven by Bornemann \cite{Bo} and Avron and Elgart \cite{AvEl1} (their proof was simplified and the result somewhat strengthened in \cite{Te1}). 

Using similar techniques, a space-adiabatic theorem without gap for the Nelson model was proven in \cite{Te2}. In the context of quantum statistical mechanics, an adiabatic theorem without gap for the Liouvillian (the generator of the dynamics of the states in a suitable representation of the von Neumann algebra associated to the system) is discussed in \cite{AbFr1}, while  in \cite{AbFr2} a general time-adiabatic theorem for resonances is proved. 

All these results are of the first order type, i.\ e., they describe the leading order adiabatic evolution and give upper bounds for the transitions. 

In our case instead, we build adiabatic dynamics including terms of the second order, and give an explicit expression for the non-adiabatic transitions, which allows us to give also a lower bound for them. In the time-dependent case, lower bounds for the transitions were calculated by Avron and Elgart \cite{AvEl2} for the Friedrichs model, which describes a ``small system'', whose Hilbert space is one-dimensional, interacting with a ``reservoir'', whose Hilbert space is $L^{2}(\field{R}{}_{+}, d\mu(k))$, with a suitable spectral density $\mu$.

Finally let us  compare our results with those obtained by Davies \cite{Da}. He considers the limit $c\to\infty$ for the Hamiltonian
\begin{equation*}
H^{c} := H_{\mathrm{p}}\otimes\id + \id\otimes\dgamma(c\abs{k}) + \sqrt{c}H_{I}
\end{equation*}
which is equal to $H^{\epsi=1}$ not setting $c=1$ as we did before. Davies proves then that for all $t\in\field{R}{}$,
\begin{equation*}
\lim_{c\to\infty}\E^{-\I H^{c}t}(\psi\otimes\Omega_{\mathrm{F}})= (\E^{-\I (-\Delta/2 + E(x))t}\psi)\otimes\Omega_{\mathrm{F}}\, .
\end{equation*}

This shows that, while at the classical level the limit $\epsi\to 0$ and $c\to\infty$ are equivalent, they differ at the quantum one. Indeed, the limit $\epsi\to 0$ is singular, because no limiting dynamics for $\epsi=0$ exist. Moreover, the effective dynamics we get refer to dressed states, which contain a non zero number of photons, while the $c\to\infty$ limit is taken on states which contain no photons. 

We summarize the structure of our paper. After some basic facts about the model recalled in section \ref{nelsonmodel}, we explain how to construct the approximate dressing operator $\hilbert{U}$ in section \ref{constructionunitary}. In section \ref{effectivehamiltonian}, we analyze the expansion of the transformed Hamiltonian and then apply these results to the study of the effective dynamics and the radiation in section~\ref{effectivedynamics}.

\section{Description of the model and preliminary facts}\label{nelsonmodel}

In this section we complete the presentation of the Nelson model, discussing also the spectrum of the fibered Hamiltonian, and collect some basic facts we will use in the following. 

\subsection{Fock space and field operator}

(The proofs of the statements we claim can be found in (\cite{ReSi2}, section X.7)).

We denote by $\fockfin$ the subspace of the Fock space, defined in \eqref{fockspace}, for which $\Psi^{(M)}=0$ for all but finitely many $M$. Given $f\in L^{2}(\field{R}{3})$, one defines on $\fockfin$ the annihilation operator by
\begin{equation}
(a(f)\Psi)^{(M)}(k_{1}, \ldots, k_{M}):= \sqrt{M+1}\int_{\field{R}{3}}dk\, \bar{f}(k)\Psi^{(M+1)}(k, k_{1}, \ldots, k_{M}) \, .
\end{equation}
The adjoint of $a(f)$ is called the creation operator, and its domain contains $\fockfin$. On this subspace they satisfy the canonical commutation relations
\begin{equation}\begin{split}
& [a(f), a(g)^{*}]=\langle f, g\rangle_{L^{2}(\field{R}{3})},\\
& [a(f), a(g)]=0, \quad [a(f)^{*}, a(g)^{*}]=0 \, .
\end{split}
\end{equation}
Since the commutator between $a(f)$ and $a(f)^{*}$ is bounded, it follows that $a(f)$ can be extended to a closed operator on the same domain of $a(f)^{*}$.

On this domain one defines the Segal field operator
\begin{equation}
\Phi(f) := \frac{1}{\sqrt{2}}(a(f) + a(f)^{*})
\end{equation}
which is essentially self-adjoint on $\fockfin$. Moreover, $\fockfin$ is a set of analytic vectors for $\Phi(f)$. From the canonical commutation relations it follows that
\begin{equation}\label{commphi}
[\Phi(f), \Phi(g)] = \I\Im\langle f, g\rangle_{L^{2}(\field{R}{3})}\, .
\end{equation}

Given a self-adjoint multiplication operator by the function $\omega$ on the domain $D(\omega)\subset L^{2}(\field{R}{3})$, we define
\begin{equation}
\fock_{\omega, \mathrm{fin}} := \mathcal{L}\{\fockvac, a(f_{1})^{*}\cdots a(f_{M})^{*}\fockvac: M\in\field{N}{}, f_{j}\in D(\omega), j=1, \dots, M\},
\end{equation}
where $\mathcal{L}$ means ``finite linear combinations of''.

On $\fock_{\omega, \mathrm{fin}}$ we define the second quantization of $\omega$, $\dgamma(\omega)$, by
\begin{equation}
(\dgamma(\omega)\Psi)^{(M)}(k_{1}, \ldots, k_{M}) := \sum_{j=1}^{M}\omega(k_{j})\Psi^{(M)}(k_{1}, \ldots, k_{M}),
\end{equation}
which is essentially self-adjoint. In particular, the free field Hamiltonian $\hfield$, equation \eqref{hfield}, acts as
\begin{equation*}
(\hfield\Psi)^{(M)}(k_{1}, \ldots, k_{M}) = \sum_{j=1}^{M}\abs{k_{j}}\Psi^{(M)}(k_{1}, \ldots, k_{M})
\end{equation*}
and is self-adjoint on its maximal domain.

From the previous definitions, given $f\in D(\omega)$, one gets the commutation properties
\begin{equation}\label{commutators}\begin{split}
& [\dgamma(\omega), a(f)^{*}] = a(\omega f)^{*}, \quad [\dgamma(\omega), a(f)] = -a(\omega f),\\
& [\dgamma(\omega), \I\Phi(f)] = \Phi(\I\omega f) \, .
                \end{split}
\end{equation}

\subsection{The Nelson model} Using the Segal field operator (which involves taking a Fourier transform) the interaction Hamiltonian, equation \eqref{hintj} and \eqref{hepsi}, can be written as
\begin{equation}
\sum_{j=1}^{N}H_{\mathrm{I}, j}(\vec{x}_{j}) = \Phi(\abs{k}v(x, k)),
\end{equation}  
where
\begin{equation}
v(x, k) := \sum_{j=1}^{N}\E^{\I k\cdot \vec{x}_{j}}\frac{\hat{\varrho}_{j}(k)}{\abs{k}^{3/2}} = \sum_{j=1}^{N}\E^{\I k\cdot \vec{x}_{j}}e_{j}\frac{\hat{\varphi}(k)}{\abs{k}^{3/2}} \, .
\end{equation}

This form is useful to prove some standard properties of $\hepsi$ and $\hfib(x)$.

\begin{lemma}
\begin{enumerate}
\item $\hepsi$ is self-adjoint on $\hilbert{H}_{0}$ (see eq. \eqref{hilbertzero}).
\item For every $x\in\field{R}{3N}$, $\hfib(x)$ is self-adjoint on $D(\hfield)$.
\end{enumerate}
\end{lemma}

\begin{proof}
The claims are based on the standard estimates (see, e.\ g., \cite{Be}, proposition 1.3.8)
\begin{equation}
\norm{\Phi(f)\psi}^{2}_{\fock}\leq 2\norm{f/\sqrt{\abs{k}}}_{L^{2}(\field{R}{3})}^{2}\langle\psi, \hfield\psi\rangle_{\fock}+2\norm{f}^{2}_{L^{2}(\field{R}{3})}\norm{\psi}_{\fock}^{2},
\end{equation}
for $\psi\in D(\hfield)$, and
\begin{equation}\label{inequalityphi}
\norm{\Phi(f)\Psi}^{2}_{\hilbert{H}}\leq 2\norm{f/\sqrt{\abs{k}}}_{L^{2}(\field{R}{3})}^{2}\langle\Psi, (\mathbf{1}\otimes\hfield)\Psi\rangle_{\hilbert{H}}+2\norm{f}^{2}_{L^{2}(\field{R}{3})}\norm{\Psi}_{\hilbert{H}}^{2},
\end{equation}
for $\Psi\in L^{2}(\field{R}{3N}, D(\hfield))$, which imply respectively that, for fixed $x$, $\Phi(\abs{k}v(x, k))$ is infinitesimally small with respect to $\hfield$ and, as an operator on $\hilbert{H}$, is infinitesimally small with respect to $\hfree$, uniformly in $\epsi$.\qed
\end{proof}

The fibered Hamiltonian has the form of a quadratic part, which corresponds to the free field, plus a term linear in the annihilation and creation operators. Hamiltonians of this form are usually called in the literature
``van Hove Hamiltonians'' (a review on the subject is given in \cite{De}).

Their simple form allows to analyze in detail their spectral and dynamical features. 

In the finite dimensional case, if one considers an harmonic oscillator with a linear perturbation, like an external electric field for example, the natural way to recover the spectrum of the Hamiltonian is to translate the $x$ variable, transforming the initial Hamiltonian into a purely quadratic one. 

In quantum field theory, the analogous strategy would be to find a unitary operator which translates the annihilation and creation operators. Such operator would implement what is called a Bogoliubov transformation. While for a finite number of degrees of freedom the Bogoliubov transformation is always implementable, this may not be the case if the phase space is infinite dimensional. In particular this is not possible if the van Hove Hamiltonian is not sufficiently regular in the infrared region.

After this preliminary remarks, we can state
\begin{lemma}\label{fiberedspectrum} Given the fibered Hamiltonian
\begin{equation*}
K_{\mathrm{fib}}(x) := \hfield + \Phi(z(x, k))
\end{equation*}
where $z(x, k)$ satisfies
\begin{equation*}
\alpha(x):=-\frac{1}{2}\int_{\field{R}{3}}dk\, \frac{\abs{z(x, k)}^{2}}{\abs{k}} < \infty\qquad \forall x\in\field{R}{3N}
\end{equation*}
one has:
\begin{enumerate}
\item The spectrum of $K_{\mathrm{fib}}(x)$ is given by $[\alpha(x), +\infty)$, and 
\begin{equation}
\sigma_{\mathrm{ac}}(K_{\mathrm{fib}}(x))= (\alpha(x), +\infty) \, .
\end{equation}
\item The infimum of the spectrum, $\alpha(x)$, is an eigenvalue if and only if
\begin{equation*}
\int_{\field{R}{3}}dk\, \frac{\abs{z(x, k)}^{2}}{\abs{k}^{2}} <\infty
\end{equation*} 

In this case, moreover, the unitary operator 
\begin{equation}
V(x) := \E^{\I\Phi(\I z(x, k)\abs{k}^{-1})}
\end{equation}
is well-defined on the Fock space and 
\begin{equation}
K_{\mathrm{fib}}(x) = V(x)\hfield V(x)^{*} + \alpha(x).
\end{equation}
If $z(x, k)\abs{k}^{-1}\notin L^{2}(\field{R}{3}, dk)$, then the spectrum is absolutely continuous.
\end{enumerate}
\end{lemma}

\begin{proof}
Point $1$ is proposition $3.10$ of \cite{De} and point $2$ is proposition $3.13$ of the same paper.\qed
\end{proof}

\begin{corollary}\label{corollaryfiberedspectrum}
$\sigma(\hfib(x)) = \sigma_{\mathrm{ac}}(\hfib(x)) = [E(x), +\infty)$, where
\begin{equation}
E(x) = -\frac{1}{2}\int_{\field{R}{3}}dk\, \abs{k}\abs{v(x, k)}^{2}\, .
\end{equation}
A more explicit expression for $E(x)$ is given in \eqref{ex}. We note also for later use that $E$ is a smooth function, bounded with all its derivatives.
\end{corollary}

\begin{corollary}\label{corollaryfiberedspectrumtwo}
The infrared regularized fibered Hamiltonian $H_{\mathrm{fib}, \sigma}(x)$ (see equation \eqref{formfactorsigma}) can be written as
\begin{equation*}
H_{\mathrm{fib}, \sigma}(x) = \vsigma(x)\hfield \vsigma(x)^{*} + E_{\sigma}(x),
\end{equation*}
where
\begin{equation}\label{vsigma}
\vsigma(x) := \E^{\I\Phi(\I v_{\sigma}(x, k))} \, .
\end{equation}
$E_{\sigma}(x)$ is the only eigenvalue, with eigenvector $\Omega_{\sigma}(x):= \vsigma(x)\fockvac$.

Moreover, for every $\alpha\in\field{N}{3N}$,
\begin{equation}\label{estimateex}
\deriv{x}{\alpha}E_{\sigma} = \deriv{x}{\alpha}E + \Or(\sigma^{\abs{\alpha}+1})_{\mathcal{L}(\hilbert{H})}\, .  
\end{equation}
\end{corollary}

\begin{proof}
The only thing left to prove is equation \eqref{estimateex}, which follows immediately from the fact that
\begin{equation*}
\deriv{x}{\alpha}E(x) - \deriv{x}{\alpha}E_{\sigma}(x) = -\frac{1}{2}\int_{\abs{k}<\sigma}dk\, \abs{k}\,\deriv{x}{\alpha}\abs{v(x, k)}^{2}
\end{equation*}
and 
\begin{equation*}
\abs{k}\,\deriv{x}{\alpha}\abs{v(x, k)}^{2}\leq C\abs{k}^{\abs{\alpha}-2} \, . 
\end{equation*}\qed
\end{proof}

As last preliminaries, we show that we can approximate the true dynamics with infrared regular ones and that the same holds for the cutoffs in the energy.

\begin{proposition}\label{approxsigma}\begin{equation}
\big\lVert\E^{-\I t\hepsi/\epsi} - \E^{-\I t \hepsisigma/\epsi}\big\rVert_{\mathcal{L}(\hilbert{H}_{0}^{1/2}, \hilbert{H})}\leq C\abs{t}\frac{\sigma^{1/2}}{\epsi},
\end{equation}
where $\hilbert{H}_{0}^{1/2} = D\big((\hfree)^{1/2}\big)$ with the corresponding graph norm.
\end{proposition}

\begin{proof}
Both Hamiltonians are self-adjoint on $\hilbert{H}_{0}$, therefore, given $\Psi\in\hilbert{H}_{0}$ we can apply the Duhamel formula to get
\begin{equation*}\begin{split}
& (\E^{-\I t\hepsi/\epsi} - \E^{-\I t\hepsisigma/\epsi})\Psi = \frac{\I}{\epsi}\int_{0}^{t}ds\, \E^{\I (s-t)\hepsisigma/\epsi}(\hepsi-\hepsisigma)\E^{-\I s\hepsi/\epsi}\Psi = \\
&= \frac{\I}{\epsi}\int_{0}^{t}ds\, \E^{\I (s-t)\hepsisigma/\epsi}\Phi\big(\abs{k}v(x, k)\id_{(0, \sigma)}(k)\big)\E^{-\I s\hepsi/\epsi}\Psi\\
&\Rightarrow \norm{(\E^{-\I t\hepsi/\epsi} - \E^{-\I t\hepsisigma/\epsi})\Psi}_{\hilbert{H}}\leq \frac{1}{\epsi}\int_{0}^{t}ds\, \norm{\Phi\big(\abs{k}v(x, k)\id_{(0, \sigma)}(k)\big)\E^{-\I s\hepsi/\epsi}\Psi}_{\hilbert{H}}
\end{split}
\end{equation*}
where $\id_{(0, \sigma)}(k)$ is the characteristic function of the interval indicated.
Using equation \eqref{inequalityphi} we find that
\begin{eqnarray*}
\lefteqn{ \hspace{-1cm} \norm{\Phi\big(\abs{k}v(x, k)\id_{(0, \sigma)}(k)\big)\E^{-\I s\hepsi/\epsi}\Psi}_{\hilbert{H}} \leq}\\ & \leq& \sqrt{2}\norm{\abs{k}^{1/2}v(x, k)\id_{(0, \sigma)}(k)}_{L^{2}(\field{R}{3})}\norm{(\hfree)^{1/2}\E^{-\I s\hepsi/\epsi}\Psi}_{\hilbert{H}} \\
&&+\, \sqrt{2}\norm{\abs{k}v(x, k)\id_{(0, \sigma)}(k)}_{L^{2}(\field{R}{3})}\norm{\Psi}_{\hilbert{H}}\,.
\end{eqnarray*}

Moreover, for $\beta>0$,
\begin{equation}\label{normsigma}\begin{split}
&\norm{\abs{k}^{\beta}v(x, k)\id_{(0, \sigma)}(k)}_{L^{2}(\field{R}{3})}\leq \sum_{j=1}^{N}\abs{e_{j}}\cdot\norm{\abs{k}^{\beta-3/2}\hat{\varphi}(k)\id_{(0, \sigma)}(k)}_{L^{2}(\field{R}{3})},\\
& \norm{\abs{k}^{\beta-3/2}\hat{\varphi}(k)\id_{(0, \sigma)}(k)}_{L^{2}(\field{R}{3})}^{2}=\frac{4\pi}{(2\pi)^{3}}\frac{\sigma^{2\beta}}{2\beta}.
\end{split}
\end{equation}

Since $\Phi(\abs{k}v(x, k))$ is infinitesimally small with respect both to $\id\otimes\hfield$ and $\hfree$ (as it follows again from \eqref{inequalityphi}), the graph norm defined by $(\hfree)^{1/2}$ and the one defined by $(\hepsi)^{1/2}$ are equivalent, uniformly in $\epsi$ (theorem X.18, \cite{ReSi2}). This implies that, for $\sigma$ sufficiently small,
\begin{equation*}
\norm{\Phi\big(\abs{k}v(x, k)\id_{(0, \sigma)}(k)\big)\E^{-\I s\hepsi/\epsi}\Psi}_{\hilbert{H}} \leq\tilde{C}\sigma^{1/2}\big(\norm{(\hfree)^{1/2}\Psi}_{\hilbert{H}}^{2} + \norm{\Psi}_{\hilbert{H}}^{2}\big)^{1/2},
\end{equation*}
which proves the statement.\qed
\end{proof}

\begin{lemma}\label{cutoffenergysigma} Given a function $\chi\in\coinf(\field{R}{})$ then
\begin{equation} 
\norm{\chi(\hepsi) - \chi(\hepsisigma)}_{\mathcal{L}(\hilbert{H})}\leq C\sigma^{1/2}\, .
\end{equation}
\end{lemma}

\begin{proof}
Using the Hellfer-Sj\"ostrand formula (see, e. g., \cite{DiSj} chapter $8$), given a self-adjoint operator $A$, we can write 
\begin{equation}\label{hellfersjostrand}
\chi(A) = \frac{1}{\pi}\int_{\field{R}{2}}dx dy\ \bar{\partial}\chi^{a}(z)(A-z)^{-1}, \quad z:=x+\I y,
\end{equation}
where $\chi^{a}\in \coinf(\field{C}{})$ is an almost analytic extension of $\chi$, which satisfies the properties
\begin{eqnarray*}
\forall \bar{N}\in \field{N}{}\quad \exists\,D_{\bar{N}}&:& \abs{\bar{\partial}\chi^{a}}\leq D_{\bar{N}}\abs{\Im z}^{\bar{N}},\\
\chi^{a}_{|_{\field{R}{}}}=\chi \quad .
\end{eqnarray*}
(For the explicit construction of such a $\chi^{a}$ see \cite{DiSj}).

Applied to our case \eqref{hellfersjostrand} yields
\begin{equation*}
\chi(\hepsi) - \chi(\hepsisigma) = \frac{1}{\pi}\int_{\field{R}{2}}dxdy\, \bar{\partial}\chi^{a}(z)\big[(\hepsi-z)^{-1} - (\hepsisigma-z)^{-1}\big].
\end{equation*}
Since both Hamiltonians are self-adjoint on $\hilbert{H}_{0}$ we have 
\begin{eqnarray*}
 (\hepsi-z)^{-1} - (\hepsisigma-z)^{-1} &=& (\hepsisigma-z)^{-1}(\hepsisigma - \hepsi)(\hepsi-z)^{-1} \\ &=&-(\hepsisigma-z)^{-1}\Phi(\abs{k}v\id_{(0, \sigma)})(\hepsi-z)^{-1}\,,
 \end{eqnarray*}
 and hence
 \begin{eqnarray*}\lefteqn{
 \norm{\chi(\hepsi) - \chi(\hepsisigma)}_{\mathcal{L}(\hilbert{H})} \leq}\\
& \leq&\frac{1}{\pi} \int_{\field{R}{2}} dxdy\, \abs{\bar{\partial}\chi^{a}(z)}\norm{(\hepsisigma-z)^{-1}}_{\mathcal{L}(\hilbert{H})}\norm{\Phi(\abs{k}v\id_{(0, \sigma)})(\hepsi-z)^{-1}}_{\mathcal{L}(\hilbert{H})}.
\end{eqnarray*}
In addition we have that
\begin{equation}\label{resolvbounded}
\norm{(\hepsisigma-z)^{-1}}_{\mathcal{L}(\hilbert{H})} \leq \frac{C}{\abs{\Im z}}.
\end{equation}
This follows because $\hilbert{H}_{0}$ is dense in the domain of $H^{\epsi=0, \sigma=0}=L^{2}(\field{R}{3N}, D(\hfield))$, and for every $\Psi\in\hilbert{H}_{0}$
\begin{equation*}
\hepsisigma\Psi \to H^{\epsi=0, \sigma=0}\Psi \quad\mbox{as}\quad  (\epsi, \sigma)\to (0, 0) \, .
\end{equation*}
According to theorem VIII.25 \cite{ReSi1}, this implies that 
\begin{equation*}
(\hepsisigma-z)^{-1}\Psi \to (H^{\epsi=0, \sigma=0}-z)^{-1}\Psi,
\end{equation*}
therefore $\abs{\Im z}\norm{(\hepsisigma-z)^{-1}\Psi}$ is bounded for every $\Psi$ and the uniform boundedness principle gives \eqref{resolvbounded}.

For the second norm we find that for $z\in \mathrm{supp}\,\chi^{a}$
\begin{eqnarray*}\lefteqn{\hspace{-1cm}
 \norm{\Phi(\abs{k}v\id_{(0, \sigma)})(\hepsi-z)^{-1}}_{\mathcal{L}(\hilbert{H})}\leq}\\&\leq &\norm{\Phi(\abs{k}v\id_{(0, \sigma)})}_{\mathcal{L}(\hilbert{H}_{0}, \hilbert{H})}\cdot\norm{(\hepsi-z)^{-1}}_{\mathcal{L}(\hilbert{H}, \hilbert{H}_{0})}\\
& \leq& \frac{C}{\abs{\Im z}}\norm{\Phi(\abs{k}v\id_{(0, \sigma)})}_{\mathcal{L}(\hilbert{H}_{0}, \hilbert{H})},  
\end{eqnarray*}
because $\hepsi$ and $\hfree$ define the same graph norm uniformly in $\epsi$, as it follows from \eqref{inequalityphi}. From the same equation and from \eqref{normsigma} it follows also that
\begin{equation*}
\norm{\Phi(\abs{k}v\id_{(0, \sigma)})}_{\mathcal{L}(\hilbert{H}_{0}, \hilbert{H})}\leq \norm{\abs{k}^{1/2}v\id_{(0, \sigma)}}_{L^{2}(\field{R}{3})} + \sqrt{2}\norm{\abs{k}v\id_{(0, \sigma)}}_{L^{2}(\field{R}{3})}\leq D\sigma^{1/2},
\end{equation*}
which proves the statement.\qed
\end{proof}

\section{The approximate dressing operator}\label{constructionunitary}

In this section we construct the approximate dressing operator $\uepsisigma$ for the infrared regularized Hamiltonian $\hepsisigma$. 

We recall at the beginning the formal procedure to build the so called \emph{superadiabatic projectors}, introduced first   by Berry \cite{Ber} for time-dependent Hamiltonians and generalized by Nenciu and Sordoni \cite{NeSo} to the space-adiabatic setting. 

The basic idea is to construct a sequence of projectors $P^{\epsi}_{j}, j\in\field{N}{}$, whose commutator with the Hamiltonian is of order $\Or(\epsi^{j+1})$. Then the range of $P^{\epsi}_{j}$ is almost invariant for the corresponding time evolution in the sense that 
\begin{equation*}
[\E^{-\I t\hepsi/\epsi}, P^{\epsi}_{j}]=\Or(\epsi^{j}\abs{t})\, .
\end{equation*}

As we have briefly mentioned in the introduction, this procedure, which allows to build adiabatic dynamics beyond the leading order, has been implemented till now only for fibered Hamiltonians whose spectrum is made up of eigenvalues separated by a gap. Exploiting the fact that the infrared cutoff $\sigma$ gives rise effectively to a gap, we can however carry out the calculations also in our case.

We proceed first in a formal way, using the superadiabatic projectors just as a formal tool to deduce the expression of the unitary intertwiner $\hilbert{U}\footnote[1]{To simplify the notation, we drop from now on the dependence of the unitary on $\sigma$ and $\epsi$, denoting it simply by $\uone$ instead that by $\uepsisigma$, as we did in the introduction.}$. We will then show that the expression we find for $\uone$ actually defines a unitary operator on $\hilbert{H}$ and we will prove some of its useful properties, which allow us to fully characterize the domain of the effective Hamiltonian, and to expand it in a series of powers of $\epsi$ which is convergent in $\mathcal{L}(\hilbert{H}_{0}, \hilbert{H})$.

The projectors $P_{M}$ on the almost invariant subspaces are then recovered from $\uone$ via $P_{M} = \uone^{*}Q_{M}\uone$.

It is important to stress that, to characterize the transitions between the different almost invariant subspaces, we need a unitary which maps all of them {\em simultaneously} to the corresponding reference subspaces. For this reason, we cannot use the procedure based on Nagy formula proposed in \cite{MaSo} \cite{NeSo} and we employ instead a simpler one, based on the construction of the projectors.

In what follows, $\sigma$ is considered a parameter independent of $\epsi$. Only at the end, we will choose it as a suitable function of $\epsi$.

\subsection{Formal definition}\label{formaldefinition}

Using the dressing operator $V_{\sigma}(x)$, whose expression is given in \eqref{vsigma}, we can define dressed $M$-particles projections:
\begin{equation}\label{pizn}
\pizn(x):= \vsigma(x)\qm \vsigma(x)^{*}.
\end{equation}
Note that, for $M=0$, we have $\Proj{(0)}{0}(x)=|\Omega_{\sigma}(x)\rangle\langle\Omega_{\sigma}(x)|$, the projection onto the ground state of $\hfibsigma(x)$, but, for $M\neq 0$, $\pizn(x)$ is not a spectral projection of the fibered Hamiltonian.

The standard construction we briefly mentioned above, which is used in the case with gap, is based on an iterative procedure. Starting from a zero order projection $\Proj{(0)}{} = \proj{}{0}$, {\em which corresponds to a spectral subspace of the fibered Hamiltonian}, it allows to build an approximate orthogonal projection $\Proj{(n)}{} := \sum_{j=1}^{n}\epsi^{j}\proj{}{j}$ which satisfies
\begin{equation}\label{pin}\begin{split}
& (\Proj{(n)}{})^{*}=\Proj{(n)}{}, \qquad (\Proj{(n)}{})^{2} - \Proj{(n)}{} = \Or(\epsi^{n+1}),\\
& [\hepsi, \Proj{(n)}{}] = \Or(\epsi^{n+1}).
\end{split}
\end{equation}

As we already stressed, we avoid on purpose to be more precise on the sense in which equation \eqref{pin} holds and on which norm one should use on the right-hand side, because we will use it just as a formal tool to deduce the expression of $\uone$.

To clarify the method we employ, we recall briefly the derivation of the first order almost projection $\Proj{(1)}{}$ in the case with gap, i.\ e., we assume for the moment that $\hfibsigma(x)$ has a nondegenerate eigenvalue $E_{\sigma}(x)$ which is isolated from the rest of the spectrum. Our discussion is taken from \cite{PST2}.

Therefore, we start from a projection $\proj{}{0}(x)$ which projects onto the eigenspace associated to $E_{\sigma}(x)$, and we proceed formally, without worrying about specifying any regularity assumption. It holds then that
\begin{equation}\label{pizeroproj}
\proj{*}{0}=\proj{}{0},\qquad (\proj{}{0})^{2} = \proj{}{0},
\end{equation}
\begin{equation}
 [\proj{}{0}, \hepsisigma ] = [\proj{}{0}, \frac{1}{2}\,\pop^{2}] = \I\epsi(\nabla\proj{}{0})\cdot\pop + \frac{\epsi^{2}}{2}\Delta\proj{}{0},
\end{equation}
where the right-hand side of the commutator is of order $\Or(\epsi)$ when applied to functions of bounded kinetic energy. Applying the inductive scheme, we determine now the coefficient $\proj{}{1}$ in $\Proj{(1)}{} = \proj{}{0} + \epsi\proj{}{1}$ by requiring that
\begin{equation}\label{propertiespione}\begin{split}
& (\Proj{(1)}{})^{*}=\Proj{(1)}{},\qquad (\Proj{(1)}{})^{2} - \Proj{(1)}{} = \Or(\epsi^{2}),\\
& [\Proj{(1)}{}, \hepsi] = \Or(\epsi^{2}).
                \end{split}\end{equation}
The first condition gives
\begin{equation*}
(\proj{}{0} + \epsi\proj{}{1})^{2} - (\proj{}{0} + \epsi\proj{}{1}) = \epsi(\proj{}{0}\proj{}{1} + \proj{}{1}\proj{}{0} - \proj{}{1}) + \Or(\epsi^{2}) \eqex \Or(\epsi^{2}),  
\end{equation*}		
so that we must have
\begin{equation}\label{pioneoffdiag}
\proj{}{1} = \proj{}{0}\proj{}{1} + \proj{}{1}\proj{}{0} + \Or(\epsi).
\end{equation}		

Concerning the commutator, we have
\begin{equation*}
[\Proj{(1)}{}, \hepsisigma] = \I\epsi(\nabla\proj{}{0})\cdot\pop + \epsi[\proj{}{1}, \hfibsigma] + \Or(\epsi^{2})\eqex  \Or(\epsi^{2}),
\end{equation*}
so $\proj{}{1}$ has to satisfy
\begin{equation}\label{commeq}
[\proj{}{1}, \hfibsigma] = -\I(\nabla\proj{}{0})\cdot\pop + \Or(\epsi).
\end{equation}
It can be shown (\cite{Te3}, lemma $3.8$) that the unique solution, up to terms of order $\Or(\epsi)$, of \eqref{commeq} is given by
\begin{equation}\label{solcommeq}
\proj{}{1} = -\I\proj{}{0}(\nabla\proj{}{0})(\hfibsigma - E)^{-1}\proj{\perp}{0}\cdot\pop\ +\ \textrm{adj.}\, ,
\end{equation}
where $\proj{\perp}{0}:= \mathbf{1} - \proj{}{0}$, and $+$ adj. means that the adjoint of everything which lies to the left is added. This ensures also that $\proj{}{1}$ is formally self-adjoint.

If there is a gap, the reduced resolvent is certainly bounded in norm:
\begin{equation}
\norm{(\hfibsigma - E)^{-1}\proj{\perp}{0}}_{\mathcal{L}(\hilbert{H})} = \bigg[\inf_{x\in\field{R}{3N}}\textrm{dist}\bigg(E_{\sigma}(x), \sigma(\hfibsigma(x))\backslash\{E_{\sigma}(x)\}\bigg)\bigg]^{-1},
\end{equation}
but in general the construction breaks down in the case without gap, because the reduced resolvent can be unbounded. 

A possible way out of this difficulty was proposed in \cite{Te1} and \cite{Te2} to prove the adiabatic decoupling for the zero order subspaces. It consists in introducing a second parameter $\delta>0$, by which the reduced resolvent is shifted into the complex plane: instead of $(\hfib-E)^{-1}\proj{\perp}{0}$, one considers $(\hfib-E-\I\delta)^{-1}\proj{\perp}{0}$.

In our case, the infrared cutoff $\sigma$ plays an analogous role. Actually, at the end of our formal calculations, it will result that the reduced resolvent of $\hfibsigma$ is bounded.

We carry out  another generalization of the standard construction, extending it to the situation when $M\neq 0$. To understand the way how we proceed it is useful to analyze more explicitly the case $M=0$. 
According to the above discussion we define
\begin{eqnarray*}
 \Proj{(1)}{0}(\epsi, \sigma)&:= &\Proj{(0)}{0} + \epsi\proj{0}{1}(\sigma),\\
 \proj{0}{1}(\sigma)&:= &-\I(\nabla\Proj{(0)}{0})\rfib(E_{\sigma})\cdot\pop\ +\ \textrm{adj.}\,,
\end{eqnarray*}
where $\rfib(E_{\sigma}):= (\hfibsigma-E_{\sigma})^{-1}\Proj{(0)\,\perp}{0}$.
We have omitted the $\Proj{(0)}{0}$ because of the well-known fact that $\nabla\Proj{(0)}{0}$ is off-diagonal with respect to the block decomposition induced by $\Proj{(0)}{0}$. 

Using the fact that $\nabla\Proj{(0)}{0}=\I V_{\sigma}(x)[\Phi(\I\nabla_{x}v_{\sigma}), Q_{0}]V_{\sigma}(x)^{*}$, and introducing the notation $\rfield(0):= \hfield^{-1}Q_{0}^{\perp}$, we get then
\begin{eqnarray*} 
\proj{0}{1} &=& \vsigma(x)[\Phi(\I\nabla_{x}v_{\sigma}), Q_{0}]\rfield(0) \vsigma(x)^{*}\hspace{-2pt}\cdot\pop
 -\pop\cdot\hspace{-1pt} \vsigma(x)\rfield(0)[\Phi(\I\nabla_{x}v_{\sigma}), Q_{0}]\vsigma(x)^{*}\\
&=& -\vsigma(x)\big[\dgamma(\abs{k}^{-1}), [\Phi(\I\nabla_{x}v_{\sigma}), Q_{0}]\big]\cdot\pop\vsigma(x)^{*} + \Or(\epsi),
\end{eqnarray*}
where we have used  that the commutator between $\pop$ and a function of $x$ is of order $\Or(\epsi)$ (we stress again that we don't worry here about smoothness assumptions, the calculations should be considered as formal) and 
\begin{equation*}
\rfield(0)Q_{\leq 1} = \dgamma(\abs{k}^{-1})Q_{\leq 1},
\end{equation*}
together with
\begin{equation*}
Q_{0}\Phi(\I\nabla_{x}v_{\sigma})=Q_{0}\Phi(\I\nabla_{x}v_{\sigma})Q_{\leq 1}, \quad \Phi(\I\nabla_{x}v_{\sigma})Q_{0}=Q_{\leq 1}\Phi(\I\nabla_{x}v_{\sigma}) Q_{0}.
\end{equation*}
Calculating the commutator using equation \eqref{commutators}, we get
\begin{equation*}
\proj{0}{1} = \I\vsigma[Q_{0}, \phisigma]\cdot\pop\vsigma^{*},
\end{equation*}
where we define
\begin{equation}
\phisigma(x) := \Phi\bigg(\frac{\nabla_{x}v_{\sigma}(x, k)}{\abs{k}}\bigg).
\end{equation}

Since $\nabla_{x}v_{\sigma}(x, k)\abs{k}^{-1}\in L^{2}(\field{R}{3}, dk)$, $\proj{0}{1}$ is well-defined, and one can symmetrize it to get a symmetric operator (the fact that $\proj{0}{1}$ is not bounded due to the presence of $\pop$ will be dealt with below).

It is now fairly clear how to proceed. For any $M$, we define
\begin{equation}\begin{split}
& \Proj{(1)}{M}:= \Proj{(0)}{M} + \epsi\proj{M}{1},\\
& \proj{M}{1}:= \frac{\I}{2}\vsigma\big\{[\qm, \phisigma]\cdot\pop+\pop\cdot[\qm, \phisigma]\big\} \vsigma^{*} \, .
\end{split}
\end{equation}
Formally $\Proj{(1)}{M}$ satisfies \eqref{propertiespione} for any $M$. 
In fact, $\phisigma$ is self-adjoint, therefore $\I[\qm, \phisigma]$ is self-adjoint, and $(\Proj{(1)}{M})^{*}=\Proj{(1)}{M}$. Moreover, since $\proj{M}{1}$ is off-diagonal with respect to $\proj{M}{0}$,
\begin{equation*}
\proj{M}{0}\proj{M}{1}\proj{M}{0} = (\mathbf{1} - \proj{M}{0})\proj{M}{1}(\mathbf{1}-\proj{M}{0})=0,
\end{equation*} 
equation \eqref{pioneoffdiag} holds exactly, without $\Or(\epsi)$ corrections. 

Concerning the commutator equation \eqref{commeq}, we get
\begin{equation*}\begin{split}
& [\proj{M}{1}, \hfibsigma] = \vsigma\big[[\qm, \I\phisigma]\cdot\pop, \hfield + E(x)\big]\vsigma^{*}+\Or(\epsi)=\\ &=\vsigma\big[[\qm, \I\phisigma], \hfield]\big]\cdot\pop \vsigma^{*} +\Or(\epsi)=\vsigma\big[[\hfield, \I\phisigma], \qm\big]\cdot\pop \vsigma^{*}+ \Or(\epsi).
\end{split}
\end{equation*}
Applying again equation \eqref{commutators} we get, on $\fock_{\abs{k}, \mathrm{fin}}$,

\begin{equation}\label{commhfa}
[\hfield, \phisigma] = \Phi(\I\nabla_{x}v_{\sigma}),
\end{equation}
so that
\begin{equation}\label{commeqsatisfied}
[\proj{M}{1}, \hfibsigma] = -\I\nabla\proj{M}{0}\cdot\pop + \Or(\epsi).
\end{equation}

The next step is to find a unitary operator which intertwines the almost projections $\Proj{(1)}{M}$ with the reference projections given by the $\qm$ up to terms of order $\Or(\epsi^{2})$.

Employing a procedure analogous to the one we used for the projections, we assume that we can write an expansion
\begin{equation}
U^{(n)}:= \sum_{j=1}^{n}\epsi^{j}U_{j},
\end{equation}
starting from a known $U_{0}$, and imposing that
\begin{equation}\label{requnitary}\begin{split}
& U^{(n)}U^{(n)*} - \mathbf{1}=\Or(\epsi^{n+1}),\qquad U^{(n)*}U^{(n)} - \mathbf{1}=\Or(\epsi^{n+1})\, ,\\
& U^{(n)}\Proj{(1)}{M}U^{(n)*}= \qm +\Or(\epsi^{n+1})\, , 
\end{split}
\end{equation}
to deduce the coefficients $U_{n}$ iteratively.

An obvious choice for the zero order unitary is given by 
\begin{equation}
U_{0}(x):= \vsigma(x)^{*},
\end{equation}
which satisfies $U_{0}(x)\Proj{(0)}{M}(x)U_{0}(x)^{*}=\qm$.

Without loss of generality, we assume that $U_{1}=TU_{0}$, for some operator $T$ on $\hilbert{H}$. The requirements \eqref{requnitary} then lead to
\begin{eqnarray*}\lefteqn{\hspace{-1cm}
 (U_{0}+\epsi U_{1})(U_{0}^{*}+\epsi U_{1}^{*}) = \mathbf{1} + \epsi(U_{0}U_{1}^{*} + U_{1}U_{0}^{*}) + \Or(\epsi^{2})=}\\ &=& \mathbf{1} + \epsi(T^{*} + T) +
 \Or(\epsi^{2}) \eqex \Or(\epsi^{2})\quad \Rightarrow \quad T^{*}+T=\Or(\epsi)\,,
\end{eqnarray*}
and
\begin{equation*}\begin{split}
& (U_{0}+\epsi U_{1})(\Proj{(0)}{M}+\epsi\proj{M}{1})(U_{0}^{*}+\epsi U^{*}_{1}) = (\mathbf{1}+\epsi T)(\qm + \epsi[\qm, \I\phisigma]\cdot\pop)(\mathbf{1}-\epsi T) \\
&\hspace{2cm}+ \,\Or(\epsi^{2}) = \qm +\epsi([\qm, \I\phisigma]\cdot\pop - [\qm, T]) + \Or(\epsi^{2})\,.
\end{split}
\end{equation*}
Therefore, by  choosing 
\begin{equation*}
T=\I\phisigma\cdot\pop\,,
\end{equation*}
we satisfy both requirements  for every $M$.
The first order almost unitary is then given by
\begin{equation}
U^{(1)}=(\mathbf{1}+\I\epsi \phisigma\cdot\pop)\vsigma^{*}.
\end{equation}

\subsection{Rigorous definition and properties}

To give a meaning to the till now formal expression for $U^{(1)}$ we have to deal with two problems. 

The first is due to the fact that the operator $\phisigma$ is unbounded. We introduce therefore in its definition a cutoff in the number of particles, replacing it by
\begin{equation}\label{defamdelta}
\phisigmaj:=\qless{J}\phisigma\qless{J}.
\end{equation}
The operator 
\begin{equation*}
U^{(1)}_{\mathrm{J}}:=(\mathbf{1}+\I\epsi\phisigmaj\cdot\pop)\vsigma^{*}
\end{equation*}
is again formally almost unitary up to order $\Or(\epsi^{2})$, and intertwines the superadiabatic almost projectors for $M<\mathrm{J}$,
\begin{equation}
U^{(1)}_{\mathrm{J}}\Proj{(1)}{M}U^{(1)*}_{\mathrm{J}}= \qm +\Or(\epsi^{2}), \qquad M<\mathrm{J}. 
\end{equation}
This means that we can use $U^{(1)}_{\mathrm{J}}$ to study the transitions among superadiabatic subspaces up to an arbitrary, but fixed $\mathrm{J}$.

The second problem, already mentioned in the introduction, is related to the presence of the unbounded momentum operator $\pop$. 

To make the whole expression bounded we introduce a cutoff in the total energy. More precisely, given a function $\chi\in\coinf(\field{R}{})$, we define
\begin{eqnarray}\label{uonechi} 
\lefteqn{ \hspace{-0.5cm}U^{(1)}_{\mathrm{J}, \chi}:= \vsigma^{*}[\mathbf{1}+\I\epsi \vsigma\phisigmaj\cdot\pop \vsigma^{*} -\I\epsi(\mathbf{1}-\chi(\hepsisigma))\vsigma\phisigmaj\cdot\pop \vsigma^{*}(\mathbf{1}-\chi(\hepsisigma))]=}\nonumber\\
&=&\vsigma^{*}[\mathbf{1} + \I\epsi\chi(\hepsisigma)\vsigma\phisigmaj\cdot\pop \vsigma^{*} + \I\epsi(\mathbf{1}-\chi(\hepsisigma))\vsigma\phisigmaj\cdot\pop \vsigma^{*}\chi(\hepsisigma)]\, .
\end{eqnarray}
Note that $U^{(1)}_{\mathrm{J}, \chi}\tilde{\chi}(\hepsisigma)=U_{\mathrm{J}}^{(1)}\tilde{\chi}(\hepsisigma)$, for every $\tilde{\chi}\in\coinf(\field{R}{})$ such that $\chi\tilde{\chi}=\tilde{\chi}$. In the context of Born-Oppenheimer approximation, Sordoni \cite{So} applied a similar strategy in order to define bounded superadiabatic projections.

Our aim in this section is first to prove that $U^{(1)}_{\mathrm{J}, \chi}\in\mathcal{L}(\hilbert{H})\cap\mathcal{L}(\hilbert{H}_{0})$. Once we have shown this, we can define a true unitary $\hilbert{U}$ through the formula
\begin{equation*}
\hilbert{U}:= U^{(1)}_{\mathrm{J}, \chi}[U^{(1)\,*}_{\mathrm{J}, \chi}U^{(1)}_{\mathrm{J}, \chi}]^{-1/2}.
\end{equation*}
We will then prove that both $\hilbert{U}$ and $\hilbert{U}^{-1}=\hilbert{U}^{*}$ belong to $\mathcal{L}(\hilbert{H})\cap\mathcal{L}(\hilbert{H}_{0})$, i.\ e., that $\hilbert{U}$ is a bijection on $\hilbert{H}_{0}$. We will in addition show that it can be expanded in a convergent power series both in $\mathcal{L}(\hilbert{H})$ and in $\mathcal{L}(\hilbert{H}_{0})$.

\begin{lemma}\label{lemmav}
$\vsigma$ and $\vsigma^{*}$ belong to $\mathcal{L}(\hilbert{H}_{0})$. Moreover
\begin{equation}
\norm{\vsigma}_{\mathcal{L}(\hilbert{H}_{0})}\leq C\,,
\end{equation}
with a constant $C<\infty$ independent of $\sigma$.
An analogous estimate holds for $\vsigma^{*}$.
\end{lemma}

\begin{proof}({\em We give the proof for $\vsigma$. The one for $\vsigma^{*}$ is the same up to some changes in the signs}).

We can calculate the norm on a dense subset. It is known (\cite{ReSi1}, theorem VIII.33) that $H_{0}$ is essentially self-adjoint on a set of the form 
\begin{equation*}
\mathcal{D}_{\mathrm{p}}\otimes\mathcal{D}_{\mathrm{f}}:=\mathcal{L}\{\psi\otimes\varphi: \psi\in\mathcal{D}_{\mathrm{p}}, \varphi\in\mathcal{D}_{\mathrm{f}}\},
\end{equation*}
where $\mathcal{L}$ means ``finite linear combinations of'', $\mathcal{D}_{\mathrm{p}}$ is a core for $\pop^{2}$ and $\mathcal{D}_{\mathrm{f}}$ is a core for $\hfield$. We can choose then $\mathcal{D}_{\mathrm{p}}=\coinf(\field{R}{3N})$ and (\cite{ReSi2}, section X.7)
\begin{equation}\label{df}
\mathcal{D}_{\mathrm{f}}=\{\varphi\in\fock_{\mathrm{fin}}: \varphi^{(\N)}\in\otimes_{j=1}^{\N}\coinf(\field{R}{3})\cap L^{2}_{\mathrm{s}}(\field{R}{3\N})\},
\end{equation}
where $\otimes_{j=1}^{\N}\coinf(\field{R}{3}):= \mathcal{L}\{\varphi_{1}\otimes\ldots\otimes\varphi_{\N}: \varphi_{j}\in\coinf(\field{R}{3}), j=1, \ldots, \N\}$.

The vectors in $\fock_{\mathrm{fin}}$ are analytic vectors for $\vsigma(x)$ (\cite{ReSi2}, theorem X.41), so, if $\psi\in\mathcal{D}_{\mathrm{p}}$, $\varphi\in\mathcal{D}_{\mathrm{f}}$, we have
\begin{equation*}
\vsigma(\psi\otimes\varphi)(x) = \psi(x)\vsigma(x)\varphi = \sum_{j=0}^{\infty}\I^{j}\frac{\Phi(\I v_{\sigma}(x, \cdot))^{j}}{j!}\psi(x)\varphi. 
\end{equation*}
Moreover
\begin{eqnarray*} 
\lefteqn{\hspace{-1cm} (\pop^{2}\otimes\mathbf{1})\Phi(\I v_{\sigma}(x, \cdot))^{j}\psi(x)\varphi = -\epsi^{2}j\Phi(\I v_{\sigma})^{j-1}\Phi(\I\Delta v_{\sigma})\psi(x)\varphi}\\
&& -\,\epsi^{2}j(j-1)\Phi(\I v_{\sigma})^{j-2}\Phi(\I\nabla_{x}v_{\sigma})\cdot\Phi(\I\nabla_{x}v_{\sigma})\psi(x)\varphi\\
&& -\,2\I\epsi j\Phi(\I v_{\sigma})^{j-1}\Phi(\I\nabla_{x}v_{\sigma})\cdot\pop\psi(x)\varphi + \Phi(\I v_{\sigma})^{j}\pop^{2}\psi(x)\varphi\,,
\end{eqnarray*}
where we have used \eqref{commphi} and the fact that $\Im\langle v_{\sigma}(x, \cdot), \nabla_{x}v_{\sigma}(x, \cdot)\rangle_{L^{2}(\field{R}{3})}=0$. 
From the previous equations it follows that
\begin{equation*}
\sum_{j=0}^{J}(\pop^{2}\otimes\mathbf{1})\I^{j}\frac{\Phi(\I v_{\sigma}(x, \cdot))^{j}}{j!}\psi(x)\varphi
\end{equation*}
is convergent, so that $\vsigma(\psi\otimes\varphi)\subset D(\pop^{2}\otimes\mathbf{1})$ and
\begin{eqnarray}\label{commpv} 
\lefteqn{ \hspace{-0.5cm}(\pop^{2}\otimes\mathbf{1})\vsigma(\psi\otimes\varphi)=-\I\epsi^{2}\vsigma\Phi(\I\Delta v_{\sigma})(\psi\otimes\varphi)+\epsi^{2}\vsigma\Phi(\I\nabla_{x}v_{\sigma})\cdot}\\
&&\cdot\Phi(\I\nabla_{x}v_{\sigma})(\psi\otimes\varphi)+2\epsi \vsigma\Phi(\I\nabla_{x}v_{\sigma})\cdot\pop(\psi\otimes\varphi)+\vsigma(\pop^{2}\otimes\mathbf{1})(\psi\otimes\varphi)\,.\nonumber
\end{eqnarray}
This implies
\begin{eqnarray}\label{estimatepsquare} 
\lefteqn{  \hspace{-0.5cm}\norm{(\pop^{2}\otimes\mathbf{1})\vsigma(\psi\otimes\varphi)}\leq \epsi^{2}\norm{\Phi(\I\Delta v_{\sigma})(\psi\otimes\varphi)}+}\nonumber \\
&&+\, \epsi^{2} \norm{\Phi(\I\nabla_{x}v_{\sigma})\cdot\Phi(\I\nabla_{x}v_{\sigma})(\psi\otimes\varphi)}+2\epsi\norm{\Phi(\I\nabla_{x}v_{\sigma})\cdot\pop(\psi\otimes\varphi)}\\
&&+\,\norm{(\pop^{2}\otimes\mathbf{1})(\psi\otimes\varphi)}\,.\nonumber
\end{eqnarray}
Using equation \eqref{inequalityphi} we can bound the first term with some constant independent of $\sigma$ and $\epsi$ times $\norm{\psi\otimes\varphi}_{\hilbert{H}_{0}}$. 

This happens because the constants involved in \eqref{inequalityphi} contain terms of the form $\norm{f/\sqrt{\abs{k}}}_{L^{2}(\field{R}{3})}$, but when one differentiates $v_{\sigma}$ with respect to $x$, one gets an additional $\abs{k}$, therefore all the terms of the form $\norm{\nabla_{x}v_{\sigma}/\sqrt{\abs{k}}}$ and so on are uniformly bounded in $\sigma$. 
The same reasoning applies also to all the estimates which follow.

For the second term in \eqref{commpv}, using again \eqref{inequalityphi}, and the notation $\Psi=\psi\otimes\varphi$ and $f=\I\deriv{j}{}v_{\sigma}$, we get
\begin{eqnarray*} \lefteqn{\hspace{-5mm}
\norm{\Phi(f)^{2}\Psi}^{2}_{\hilbert{H}} \leq  2\norm{f/\sqrt{\abs{\,\cdot\,}}}_{L^{2}(\field{R}{3})}^{2}\langle\Phi(f)\Psi, (\mathbf{1}\otimes\hfield)\Phi(f)\Psi\rangle_{\hilbert{H}} +}\\ &&+\,2\norm{f}^{2}_{L^{2}(\field{R}{3})}\norm{\Phi(f)\Psi}_{\hilbert{H}}^{2}\\
&\leq& \sqrt{2}\norm{f/\sqrt{\abs{\,\cdot\,}}}_{L^{2}(\field{R}{3})}^{2}\langle\Phi(f)\Psi, \big[a(\abs{\,\cdot\,}f)^{*}-a(\abs{\,\cdot\,}f)\big]\Psi\rangle_{\hilbert{H}} \\
&&+ \,2\norm{f/\sqrt{\abs{\,\cdot\,}}}_{L^{2}(\field{R}{3})}^{2}\langle\Phi(f)^{2}\Psi, \hfield\Psi\rangle_{\hilbert{H}} +2\norm{f}^{2}_{L^{2}(\field{R}{3})}\norm{\Phi(f)\Psi}_{\hilbert{H}}^{2}\\
&\leq &\sqrt{2}\norm{f/\sqrt{\abs{\,\cdot\,}}}_{L^{2}(\field{R}{3})}^{2}\cdot\norm{\Phi(f)\Psi}_{\hilbert{H}}\cdot\norm{\big[a(\abs{\,\cdot\,}f)^{*}-a(\abs{\,\cdot\,}f)\big]\Psi}_{\hilbert{H}}\\
&&+\,2\norm{f/\sqrt{\abs{\,\cdot\,}}}_{L^{2}(\field{R}{3})}^{2}\cdot\norm{\Phi(f)^{2}\Psi}_{\hilbert{H}}\norm{\hfield\Psi}_{\hilbert{H}}+2\norm{f}^{2}_{L^{2}(\field{R}{3})}\norm{\Phi(f)\Psi}_{\hilbert{H}}^{2}\\
&\leq& \frac{a^{2}}{\sqrt{2}}\norm{f/\sqrt{\abs{\,\cdot\,}}}_{L^{2}(\field{R}{3})}^{2}\norm{\Phi(f)\Psi}^{2}_{\hilbert{H}}+\frac{1}{a^{2}\sqrt{2}}\norm{\big[a(\abs{\,\cdot\,}f)^{*}-a(\abs{\,\cdot\,}f)\big]\Psi}_{\hilbert{H}}^{2}\\
&&+\,b^{2}\norm{f/\sqrt{\abs{\,\cdot\,}}}_{L^{2}(\field{R}{3})}^{2}\norm{\Phi(f)^{2}\Psi}^{2}_{\hilbert{H}}+\frac{1}{b^{2}}\norm{f/\sqrt{\abs{\,\cdot\,}}}_{L^{2}(\field{R}{3})}^{2}\norm{\hfield\Psi}^{2}_{\hilbert{H}}\\
&&+\,2\norm{f}^{2}_{L^{2}(\field{R}{3})}\norm{\Phi(f)\Psi}_{\hilbert{H}}^{2},\qquad\forall a, b>0\,.
\end{eqnarray*}
Hence
\begin{eqnarray*}\lefteqn{\hspace{-0.5cm}
 \norm{\Phi(f)^{2}\Psi}^{2}_{\hilbert{H}}\big(1-b^{2}\norm{f/\sqrt{\abs{\,\cdot\,}}}^{2})\leq}\\
&\leq& \frac{a^{2}}{\sqrt{2}}\norm{f/\sqrt{\abs{\,\cdot\,}}}_{L^{2}(\field{R}{3})}^{2}\norm{\Phi(f)\Psi}^{2}_{\hilbert{H}}+\frac{1}{a^{2}\sqrt{2}}\norm{\big[a(\abs{\,\cdot\,}f)^{*}-a(\abs{\,\cdot\,}f)\big]\Psi}_{\hilbert{H}}^{2}\\
&&+\,\frac{1}{b^{2}}\norm{f/\sqrt{\abs{\,\cdot\,}}}_{L^{2}(\field{R}{3})}^{2}\norm{\hfield\Psi}^{2}_{\hilbert{H}}+2\norm{f}^{2}_{L^{2}(\field{R}{3})}\norm{\Phi(f)\Psi}_{\hilbert{H}}^{2}\,.
\end{eqnarray*}
Each term on the right-hand side can be bounded by a constant independent of $\sigma$ times $\norm{\Psi}_{\hilbert{H}_{0}}$.

For the third term in \eqref{estimatepsquare} we get
\begin{eqnarray*}
\lefteqn{\hspace{-9mm} \norm{\Phi(f)\pop_{j}\Psi}_{\hilbert{H}}^{2}\leq 2\norm{f/\sqrt{\abs{\,\cdot\,}}}_{L^{2}(\field{R}{3})}^{2}\langle\pop_{j}\Psi, \hfield\pop_{j}\Psi\rangle + 2\norm{f}_{L^{2}(\field{R}{3})}^{2}\norm{\pop_{j}\Psi}^{2}_{\hilbert{H}}\leq}\\
&\leq& 2\norm{f/\sqrt{\abs{\,\cdot\,}}}_{L^{2}(\field{R}{3})}^{2}\langle\Psi, \hfield\pop_{j}^{2}\Psi\rangle + 2\norm{f}_{L^{2}(\field{R}{3})}^{2}\norm{\pop_{j}\Psi}^{2}_{\hilbert{H}}\leq\\
&\leq &\norm{f/\sqrt{\abs{\,\cdot\,}}}_{L^{2}(\field{R}{3})}^{2}\norm{(\pop_{j}^{2}+\hfield)\Psi}_{\hilbert{H}}^{2} +2\norm{f}_{L^{2}(\field{R}{3})}^{2}\norm{\pop_{j}\Psi}^{2}_{\hilbert{H}}\,,
\end{eqnarray*}
and we can again bound the right-hand side by a constant times $\norm{\Psi}_{\hilbert{H}_{0}}$.

We have now to prove similar estimates for $\norm{(\mathbf{1}\otimes\hfield) \vsigma(\psi\otimes\varphi)}$, and we are done. For this we need that, on $\mathcal{D}_{\mathrm{p}}\otimes\mathcal{D}_{\mathrm{f}}$,
\begin{equation*}
[\hfield, \Phi(\I v_{\sigma})]=\I\Phi(\abs{k}v_{\sigma}),
\end{equation*}
and that
\begin{equation*}\begin{split}
& [\Phi(\abs{k}v_{\sigma}), \Phi(\I v_{\sigma})]=\I\Im\langle\abs{k}v_{\sigma}, \I v_{\sigma}\rangle=\I\int dk\,\abs{k}\abs{v_{\sigma}(x, k)}^{2}=-2\I E_{\sigma}(x)\\
&\Rightarrow\quad [\Phi(\abs{k}v_{\sigma}), \Phi(\I v_{\sigma})^{j}]=-2\I j E_{\sigma}(x)\Phi(\I v_{\sigma})^{j-1}.
\end{split}
\end{equation*}
We have then
\begin{eqnarray} 
\lefteqn{ \hspace{-1cm}(\mathbf{1}\otimes\hfield)\Phi(\I v_{\sigma}(x, \cdot))^{j}(\psi\otimes\varphi)=\I\sum_{l=0}^{j-1}\Phi(\I v_{\sigma})^{l}\Phi(\abs{k}v_{\sigma})\Phi(\I v_{\sigma})^{j-l-1}(\psi\otimes\varphi)+}\nonumber\\
&&+\,\Phi(\I v_{\sigma})^{j}(\mathbf{1}\otimes\hfield)(\psi\otimes\varphi)\nonumber\\
&=&\I j\Phi(\I v_{\sigma})^{j-1}\Phi(\abs{k}v_{\sigma})(\psi\otimes\varphi) +E_{\sigma}(x)j(j-1)\Phi(\I v_{\sigma})^{j-2}(\psi\otimes\varphi)\nonumber\\
&&+\,\Phi(\I v_{\sigma})^{j}(\mathbf{1}\otimes\hfield)(\psi\otimes\varphi)\,.
\end{eqnarray}
This means, as for $\pop^{2}\otimes\mathbf{1}$, that $\vsigma(\psi\otimes\varphi)\subset D(\mathbf{1}\otimes\hfield)$, and that
\begin{equation}
(\mathbf{1}\otimes\hfield)\vsigma(\psi\otimes\varphi)=-\vsigma\Phi(\abs{k}v_{\sigma})(\psi\otimes\varphi)- E_{\sigma}(x)\vsigma(\psi\otimes\varphi)+\vsigma(\mathbf{1}\otimes\hfield)(\psi\otimes\varphi).
\end{equation}
Using equation \eqref{inequalityphi}, we can again bound each term by a constant independent of $\sigma$ times $\norm{\psi\otimes\varphi}_{\hilbert{H}_{0}}$.\qed
\end{proof}

\begin{lemma}\label{lemmaamdelta}\begin{enumerate}
\item For each $x\in\field{R}{3N}$ and $l=1, \ldots, 3N$, the components $\phisigmajl$ of the operator $\phisigmaj$ satisfy $\phisigmajl(x)\in\mathcal{L}(\fock)$ and $\phisigmajl(x)^{*}=\phisigmajl(x)$. 

Moreover,
\begin{equation}
\phisigmajl: \field{R}{3N}\to\mathcal{L}(\fock), \qquad x\mapsto \phisigmajl(x), \quad\in C_{\mathrm{b}}^{\infty}(\field{R}{3N}, \mathcal{L}(\fock)),
\end{equation}
and, for $\sigma$ small enough,
\begin{equation}
\norm{\phisigmajl}_{\mathcal{L}(\hilbert{H})}\leq C\sqrt{\mathrm{J}+1}\sqrt{\log(\sigma^{-1})}.
\end{equation}

Given $\alpha\in\field{N}{3N}$ with $\abs{\alpha}>0$, it holds instead 
\begin{equation}
\deriv{x}{\alpha}\phisigmajl = \deriv{x}{\alpha}\phizerojl + \Or(\sigma^{\abs{\alpha}}\sqrt{\mathrm{J}+1})_{\mathcal{L}(\hilbert{H})},
\end{equation}
where
\begin{equation}
\deriv{x}{\alpha}\phizerojl := (\deriv{x}{\alpha}\phisigmajl)_{|\sigma=0}
\end{equation}
is a well-defined bounded operator on $\hilbert{H}$.

\item The statements of point $1$ (except for the self-adjointness of $\phisigmajl(x)$) remain true if $\fock$ is replaced by $D(\hfield)$.
\end{enumerate} 
\end{lemma}
\begin{proof} The proof follows applying the standard inequality 
\begin{equation*}
\norm{\qless{J}\Phi(f(x, k))\qless{J}}_{\mathcal{L}(\hilbert{H})}\leq 2^{1/2}\sqrt{\mathrm{J}+1}\cdot\sup_{x\in\field{R}{3N}}\norm{f(x, k)}_{L^{2}(\field{R}{3}, dk)},
\end{equation*}
to the case $f = \deriv{x}{\alpha}\deriv{l}{}v_{\sigma}$ \qed 
\end{proof}
\begin{corollary}
The fibered operators $\deriv{x}{\alpha}\phisigmajl$ belong to $\mathcal{L}(\hilbert{H})\cap\mathcal{L}(\hilbert{H}_{0})$ $\forall\alpha\in\field{N}{3N}$.
\end{corollary}

\begin{lemma}\label{lemmauonechi}
$U^{(1)}_{\mathrm{J}, \chi}$ belongs to $\mathcal{L}(\hilbert{H})\cap\mathcal{L}(\hilbert{H}_{0})$ and 
\begin{equation}\label{normujchi}
\norm{U^{(1)}_{\mathrm{J}, \chi}}_{\mathcal{L}(\hilbert{K})}\leq C(1+\epsi\sqrt{\log(\sigma^{-1})}),
\end{equation}
where $\hilbert{K}=\hilbert{H}$ or $\hilbert{H}_{0}$. 
The same holds for $U^{(1)\,*}_{\mathrm{J}, \chi}$.
\end{lemma}

\begin{proof}
We start considering $U_{\mathrm{J}, \chi}^{(1)}$.
We have already shown that $\vsigma$ and $\vsigma^{*}$ belong to $\mathcal{L}(\hilbert{H})\cap\mathcal{L}(\hilbert{H}_{0})$ with uniformly bounded norms, so we have to examine
\begin{equation*}
\epsi\chi(\hepsisigma)\vsigma\phisigmaj\cdot\pop \vsigma^{*}, \quad \epsi(\mathbf{1}-\chi(\hepsisigma))V\phisigmaj\cdot\pop \vsigma^{*}\chi(\hepsisigma) \, .
\end{equation*}
Given $\Psi\in\mathcal{D}_{\mathrm{p}}\otimes\mathcal{D}_{\mathrm{f}}$ (for the definitions, see the proof of lemma \ref{lemmav}) we have
\begin{equation*}
\chi(\hepsisigma)\vsigma\phisigmaj\cdot\pop \vsigma^{*}\Psi=-\I\epsi\chi(\hepsisigma)\{\nabla_{x}\cdot(\vsigma\phisigmaj)\}\vsigma^{*}\Psi+\chi(\hepsisigma)\pop\cdot \vsigma\phisigmaj \vsigma^{*}\Psi.
\end{equation*}
Since $\nabla_{x}\cdot(\vsigma\phisigmaj)$ and $\chi(\hepsisigma)\pop$ are bounded operators (see lemma \ref{lemmav} and \ref{lemmaamdelta}) the left-hand side belongs to $\mathcal{L}(\hilbert{H})\cap\mathcal{L}(\hilbert{H}_{0})$. Moreover, from the previous lemma it follows that
\begin{equation*}
\norm{\phisigmaj}_{\mathcal{L}(\hilbert{K})}\leq C \sqrt{\log(\sigma^{-1})}\, ,
\end{equation*}
while all the other terms have uniformly bounded norms.

For the second one,
\begin{equation*}
\vsigma\phisigmaj\cdot\pop \vsigma^{*}\chi(\hepsisigma)\Psi=\vsigma\phisigmaj\cdot[\pop, \vsigma^{*}]\chi(\hepsisigma)\Psi + \vsigma\phisigmaj\cdot \vsigma^{*}\pop\chi(\hepsisigma)\Psi.
\end{equation*}
$\pop\chi(\hepsi)\in\mathcal{L}(\hilbert{H})\cap\mathcal{L}(\hilbert{H}_{0})$, and so do all the other terms, as it follows from lemma~\ref{lemmav} and lemma~\ref{lemmaamdelta}.

Concerning $U_{\mathrm{J}, \chi}^{(1)\,*}$, from equation \eqref{uonechi} it follows that
\begin{equation}\label{uonechistar}
U_{\mathrm{J}, \chi}^{(1)\,*}=[\mathbf{1}-\I\epsi\chi(\hepsisigma)\vsigma\pop\cdot\phisigmaj \vsigma^{*}-\I\epsi(\mathbf{1}-\chi(\hepsisigma))\vsigma\pop\cdot\phisigmaj \vsigma^{*}\chi(\hepsisigma)]\vsigma,
\end{equation}
so one can apply the same reasoning as above.\qed
\end{proof}

\begin{theorem}\label{theoremuone}
Assume that $\sigma=\sigma(\epsi)$ and that conditions \eqref{conditionsigma} hold, then the operator 
\begin{equation}\label{uone}
\hilbert{U}:= U^{(1)}_{\mathrm{J}, \chi}[U^{(1)\,*}_{\mathrm{J}, \chi}U^{(1)}_{\mathrm{J}, \chi}]^{-1/2}
\end{equation}
is well-defined and unitary, for $\epsi$ small enough. 

Both $\hilbert{U}$ and $\hilbert{U}^{*}$ belong to $\mathcal{L}(\hilbert{H})\cap\mathcal{L}(\hilbert{H}_{0})$, with the property that
\begin{equation}\label{unormbounded}
\norm{\hilbert{U}}_{\mathcal{L}(\hilbert{H}_{0})}, \quad \norm{\hilbert{U}^{*}}_{\mathcal{L}(\hilbert{H}_{0})} \quad \leq C \, ,
\end{equation} 
where $C$ is independent of $\epsi$ and $\sigma$.

Moreover we can expand them in a power series
which converges both in $\mathcal{L}(\hilbert{H})$ and in $\mathcal{L}(\hilbert{H}_{0})$. 
\end{theorem}

\begin{proof}
Combining equations \eqref{uonechi} and \eqref{uonechistar}, and using the fact that $\pop\cdot\phisigmaj=\phisigmaj\cdot\pop -\I\epsi\nabla\cdot\phisigmaj$, we have that
\begin{equation*}
U_{\mathrm{J}, \chi}^{(1)\,*}U_{\mathrm{J}, \chi}^{(1)}=\mathbf{1}+\epsi^{2}B_{\epsi},
\end{equation*}
where
\begin{eqnarray}\label{bepsi}
 B_{\epsi}&:= &B_{0}+\epsi B_{1},\nonumber\\ \nonumber\\
 B_{0}&:=& -\chi(\hepsisigma)\vsigma\nabla\cdot\phisigmaj\vsigma^{*} -(\mathbf{1}-\chi(\hepsisigma))\vsigma\nabla\cdot\phisigmaj\vsigma^{*}\chi(\hepsisigma)\nonumber\\
&& +\, [\chi(\hepsisigma)\vsigma(\phisigmaj\cdot\pop)\vsigma^{*}+(\mathbf{1}-\chi(\hepsisigma))\vsigma(\phisigmaj\cdot\pop)\vsigma^{*}\chi(\hepsisigma)]^{2},\\ \nonumber\\
 B_{1}&:= &-\I[\chi(\hepsisigma)\vsigma\nabla\cdot\phisigmaj\vsigma^{*}+(\mathbf{1}-\chi(\hepsisigma))\vsigma\nabla\cdot\phisigmaj\vsigma^{*}\chi(\hepsisigma)]\cdot\nonumber\\
&&\cdot[\chi(\hepsisigma)\vsigma(\phisigmaj\cdot\pop)\vsigma^{*}+(\mathbf{1}-\chi(\hepsisigma))\vsigma(\phisigmaj\cdot\pop)\vsigma^{*}\chi(\hepsisigma)]\,.\nonumber
\end{eqnarray}
>From lemma \ref{lemmauonechi} it follows that $B_{\epsi}$ is a bounded operator both on $\hilbert{H}$ and $\hilbert{H}_{0}$, self-adjoint on $\hilbert{H}$. From equation \eqref{normujchi} we have also that
\begin{equation*}
\norm{B_{\epsi}}_{\mathcal{L}(\hilbert{K})}= \Or\big(\log(\sigma^{-1})\big), \qquad \hilbert{K}=\hilbert{H}, \hilbert{H}_{0}\, ,
\end{equation*}
so, under conditions \eqref{conditionsigma}, we can assume that $\norm{\epsi^{2}B_{\epsi}}_{\mathcal{L}(\hilbert{K})}<1$.  $U_{\mathrm{J}, \chi}^{(1)\,*}U_{\mathrm{J}, \chi}^{(1)}$ is therefore strictly positive, and the square root in \eqref{uone} is well-defined. 

Moreover, we can express it through a convergent power series, both in $\mathcal{L}(\hilbert{H})$ and in $\mathcal{L}(\hilbert{H}_{0})$:
\begin{equation}\label{expansionsquareroot}
[\uonemchistar\uonemchi]^{-1/2} = (\mathbf{1}+\epsi^{2}B_{\epsi})^{-1/2}=\sum_{j=0}^{\infty}(-1)^{j}\frac{(2j-1)!!}{(2j)!!}\epsi^{2j}B_{\epsi}^{j}\;\footnote[1]{$(2j-1)!!=1\cdot 3\cdot 5\cdots (2j-1)$; $(2j)!!=2\cdot 4 \cdot 6\cdots 2j = 2^{j}j!$}\, .
\end{equation}
We can also explicitly calculate that
\begin{equation*}\begin{split}
& \hilbert{U}\hilbert{U}^{*}= \uonemchi[\uonemchistar\uonemchi]^{-1}\uonemchistar=\mathbf{1};\\
& \hilbert{U}^{*}\hilbert{U}= [\uonemchistar\uonemchi]^{-1/2}\uonemchistar\uonemchi[\uonemchistar\uonemchi]^{-1/2}=\mathbf{1},
\end{split}
\end{equation*}
where we have used in the first equation the fact that both $\uonemchi$ and $\uonemchistar$ are invertible, since they differ by a term of order $\Or(\epsi\sqrt{\log(\sigma^{-1}}))$ from a unitary operator.

Putting together \eqref{uone} and \eqref{expansionsquareroot} we get in the end the expansion for $\uone$.
\qed
\end{proof}

\section{The dressed Hamiltonian}\label{effectivehamiltonian}
Using the results of the last section, we can define the dressed Hamiltonian just as the unitary transform of $\hepsisigma$,
\begin{equation}\label{heff}
\heff:=\uone\hepsisigma\hilbert{U}^{*}.
\end{equation}
It follows from theorem \ref{theoremuone} that $\uone$ is a bijection on $\hilbert{H}_{0}$, so $\heff$ is self-adjoint on$\hilbert{H}_{0}$. Moreover, since $\uone$ can be expanded in a power series in $\mathcal{L}(\hilbert{H}_{0})$, we can expand $\heff$ in $\mathcal{L}(\hilbert{H}_{0}, \hilbert{H})$. However, putting directly the expansion for $\uone$ in \eqref{heff} we get $\epsi-$dependent coefficients, because of the $\epsi-$dependence in $\hepsisigma$. To get the correct expansion we need then to rearrange some terms, but the remainder we get will in the end be bounded in $\mathcal{L}(\hilbert{H}_{0}, \hilbert{H})$.

\begin{theorem}\label{thmheff}
The expansion up to the second order of the dressed Hamiltonian is given by
\begin{equation}
\heff = h_{0}+\epsi h_{1, \chi} + \epsi^{2}h_{2, \chi} + \Or(\sigma)_{\mathcal{L}(\hilbert{H}_{0}, \hilbert{H})} +\Or(\epsi^{3}\big(\log(\sigma^{-1})\big)^{3/2})_{\mathcal{L}(\hilbert{H}_{0}, \hilbert{H})},
\end{equation}
where $h_{i, \chi}\in\mathcal{L}(\hilbert{H}_{0}, \hilbert{H})$,
\begin{equation}
h_{0}= \frac{1}{2} \pop^{2}+\hfield + E(x),
\end{equation}
$h_{1,\chi}$ is given in equation \eqref{honechi} and $h_{2,\chi}$ is given in equation \eqref{htwochi}.
\end{theorem}

\begin{remark}
The coefficients in the expansion of $\heff$ depend explicitly on the cutoff function $\chi$ (and also on the cutoff $\mathrm{J}$ in the number of particles, even though we have not stressed it in the notation).
 
However, as we have already mentioned, we expect the adiabatic decoupling to be meaningful only on states in the range of $\chi(\hepsisigma)$ (or $\chi(\hepsi)$, according to lemma \eqref{cutoffenergysigma} we can interchange the two), which becomes in the representation space $\chi(\heff)$. 

With this in mind, we will prove later that the effective dynamics on the range of $\qm\chi(\heff)$ ($M<\mathrm{J}+1$) is generated by an Hamiltonian independent of $\chi$ and $\mathrm{J}$.    
\end{remark}

\begin{proof}
To simplify the following reasoning we write
\begin{eqnarray*}
 \uonemchi&=&\vsigma^{*}(\mathbf{1}+\epsi T)\,,\\ 
T&:=& \I\chi(\hepsisigma)\vsigma\phisigmaj\cdot\pop \vsigma^{*} + \I(\mathbf{1}-\chi(\hepsisigma))\vsigma\phisigmaj\cdot\pop\vsigma^{*}\chi(\hepsisigma)\,,\\
 \uonemchistar&=&(\mathbf{1}-\epsi T +\I\epsi^{2}S)\vsigma\,,\\
 S&:= &\I\chi(\hepsisigma)\vsigma\nabla\cdot\phisigmaj\vsigma^{*}+\I(\mathbf{1}-\chi(\hepsisigma))\vsigma\nabla\cdot\phisigmaj\vsigma^{*}\chi(\hepsisigma)\,.
\end{eqnarray*}
We omit the dependence on $\sigma$ and $\epsi$ in $T$ and $S$ to streamline the presentation.
We can then write
\begin{equation*}
B_{\epsi}=\I S -T^{2}+\I\epsi ST .
\end{equation*}
Expanding the square root according to formula \eqref{expansionsquareroot} we get
\begin{eqnarray*}
\lefteqn{ \heff = \uone\hepsisigma\hilbert{U}^{*}=\vsigma^{*}(\mathbf{1}+\epsi T)\bigg[\mathbf{1}-\frac{\epsi^{2}}{2}(\I S -T^{2})\bigg]\hepsisigma\cdot}\\
&&\cdot\bigg[\mathbf{1}-\frac{\epsi^{2}}{2}(\I S -T^{2})\bigg](\mathbf{1}-\epsi T +\I\epsi^{2}S)\vsigma + \Or\big(\epsi^{3}(\log(\sigma^{-1}))^{3/2}\big)_{\mathcal{L}(\hilbert{H}_{0}, \hilbert{H})}\\
&=&\vsigma^{*}\hepsisigma \vsigma +\epsi \vsigma^{*}T\hepsisigma \vsigma -\epsi \vsigma^{*}\hepsisigma T\vsigma +\I\epsi^{2}\vsigma^{*}\hepsisigma S\vsigma  \\
&&-\, \frac{\epsi^{2}}{2}\vsigma^{*}\hepsisigma(\I S -T^{2})\vsigma -\epsi^{2}\vsigma^{*}T\hepsisigma T\vsigma-\frac{\epsi^{2}}{2}\vsigma^{*}(\I S -T^{2})\hepsisigma \vsigma \\ 
&&+\, \Or\big(\epsi^{3}(\log(\sigma^{-1}))^{3/2}\big)_{\mathcal{L}(\hilbert{H}_{0}, \hilbert{H})}= \vsigma^{*}\hepsisigma \vsigma + \epsi \vsigma^{*}[T, \hepsisigma]\vsigma  \\
&&+\, \frac{\I\epsi^{2}}{2}\vsigma^{*}[\hepsisigma, S]\vsigma +\frac{\epsi^{2}}{2}\vsigma^{*}\big[[\hepsisigma, T], T\big]\vsigma +\Or\big(\epsi^{3}(\log(\sigma^{-1}))^{3/2}\big)_{\mathcal{L}(\hilbert{H}_{0}, \hilbert{H})}\,.
\end{eqnarray*}
We examine now separately the terms coming from different powers of $\epsi$.
\begin{eqnarray} \label{vhepsiv}
(0)=\vsigma^{*}\hepsisigma \vsigma&=& \frac{1}{2}\vsigma(x)^{*}\pop^{2}\vsigma(x) + \hfield + E_{\sigma}(x)\\& =& \frac{1}{2} \pop^{2}+\hfield + E_{\sigma}(x) 
+ \frac{1}{2}\vsigma(x)^{*}[\pop^{2}, \vsigma(x)]\nonumber\\&\stackrel{  \eqref{commpv} }{=}&\ \frac{1}{2} \pop^{2}+\hfield + E_{\sigma}(x)+\,\epsi\Phi(\I\nabla_{x}v_{\sigma})\cdot\pop \nonumber\\&&+ \,\frac{\epsi^{2}}{2}\Phi(\I\nabla_{x}v_{\sigma})\cdot\Phi(\I\nabla_{x}v_{\sigma}) -\I\frac{\epsi^{2}}{2}\Phi(\I\Delta v_{\sigma})\nonumber\\
&\stackrel{\eqref{estimateex}}{=}& \frac{1}{2} \pop^{2}+\hfield + E(x) +\frac{\epsi}{2}\Phi(\I\nabla_{x}v_{\sigma})\cdot\pop + \frac{\epsi}{2}\pop\cdot \Phi(\I\nabla_{x}v_{\sigma}) \nonumber\\
&&+ \,\frac{\epsi^{2}}{2}\Phi(\I\nabla_{x}v_{\sigma})\cdot\Phi(\I\nabla_{x}v_{\sigma}) + \Or(\sigma)_{\mathcal{L}(\hilbert{H})},\nonumber
\end{eqnarray}
%
This implies  that
\begin{equation*}
h_{0}=\frac{1}{2} \pop^{2}+\hfield + E(x)\,.
\end{equation*}

Summing the term of order $\Or(\epsi)$ coming from $(0)$ to the ones coming from the expansion of $\uone$ we get
\begin{equation*}
(1)= \frac{\epsi}{2}\Phi(\I\nabla_{x}v_{\sigma})\cdot\pop + \frac{\epsi}{2}\pop\cdot \Phi(\I\nabla_{x}v_{\sigma})+ \epsi \vsigma^{*}[T, \hepsisigma]\vsigma\, .
\end{equation*}
Here
\begin{eqnarray*}
\lefteqn{ [T, \hepsisigma]=}\\ &=&\I\,\chi(\hepsisigma)[\vsigma\phisigmaj\cdot\pop \vsigma^{*}, \hepsisigma] +\I(\mathbf{1}-\chi(\hepsisigma))[\vsigma\phisigmaj\cdot\pop \vsigma^{*}, \hepsisigma] \chi(\hepsisigma)
\\
&=&\frac{\I}{2}\chi(\hepsisigma)[\vsigma\phisigmaj\cdot\pop \vsigma^{*}, \pop^{2}] + \frac{\I}{2}(\mathbf{1}-\chi(\hepsisigma))[\vsigma\phisigmaj\cdot\pop \vsigma^{*} , \pop^{2}]\chi(\hepsisigma)+\\
&&+\,\I\,\chi(\hepsisigma)\vsigma[\phisigmaj\cdot\pop , \hfield + E_{\sigma}(x)]\vsigma^{*} \\&&+ \,\I(\mathbf{1}-\chi(\hepsisigma))\vsigma[\phisigmaj\cdot\pop , \hfield + E_{\sigma}(x)] \vsigma^{*}\chi(\hepsisigma)\,\,.
\end{eqnarray*}
The commutator appearing in the last two lines gives
\begin{equation*}\begin{split}
&\hspace{-1cm} [\phisigmaj\cdot\pop , \hfield + E_{\sigma}(x)]=[\phisigmaj, \hfield]\cdot\pop + \phisigmaj\cdot[\pop, E_{\sigma}(x)]=\\ &=\I\qless{J}\Phi(\I\nabla_{x}v_{\sigma})\qless{J}\cdot\pop -\I\epsi\phisigmaj\cdot\nabla E_{\sigma}(x)\,,
\end{split}
\end{equation*}
where we have used equation \eqref{commhfa} to calculate the first commutator.

Putting together all the terms we have therefore
\begin{eqnarray*} 
(1) &=& \frac{\epsi}{2}\Phi(\I\nabla_{x}v_{\sigma})\cdot\pop + \frac{\epsi}{2}\pop\cdot \Phi(\I\nabla_{x}v_{\sigma}) -\epsi \chi(\vsigma^{*}\hepsisigma \vsigma)[\phim(\I\nabla_{x}v_{\sigma})\cdot\pop \\
&&-\,\epsi\phisigmaj\cdot\nabla E_{\sigma}(x)] - \epsi(\mathbf{1}-\chi(\vsigma^{*}\hepsisigma \vsigma))[\phim(\I\nabla_{x}v_{\sigma})\cdot\pop \\
&&-\,\epsi\phisigmaj\cdot\nabla E_{\sigma}(x)]\chi(\vsigma^{*}\hepsisigma\vsigma) + \Or(\epsi^{2}\sqrt{\log(\sigma^{-1})})_{\mathcal{L}(\hilbert{H}_{0}, \hilbert{H})}\,,
\end{eqnarray*}
where the terms containing the commutator with $\pop^{2}$ give rise to higher order contributions and we abbreviated  $\phim(f):=\qless{J}\Phi(f)\qless{J}$.

Analyzing the terms of order $\Or(\epsi^{2})$ we will see that they yield, as it happens for the zero order one, terms of the form $\pop\cdot\Phi$, which make the previous expression a symmetric operator. We have therefore in the end that
\begin{eqnarray} \label{honechi}
h_{1, \chi}&=& \frac{1}{2}\Phi(\I\nabla_{x}v_{\sigma})\cdot\pop -\frac{1}{2} \chi(\vsigma^{*}\hepsisigma \vsigma)\phim(\I\nabla_{x}v_{\sigma})\cdot\pop \\ 
&&-\,\frac{1}{2}(\mathbf{1}-\chi(\vsigma^{*}\hepsisigma\vsigma))\phim(\I\nabla_{x}v_{\sigma})\cdot\pop \chi(\vsigma^{*}\hepsisigma\vsigma) +\textrm{``}\pop\cdot\Phi\textrm{''}\,,\nonumber
\end{eqnarray}
where the symbol ``$\pop\cdot\Phi$'' means that, associated to each term of the form $\Phi\cdot\pop$, there is another one of the form $\pop\cdot\Phi$ which makes the sum a symmetric operator. 

To calculate $h_{2}$ we follow the same route,
\begin{eqnarray*}
 (2)&=& \frac{\epsi^{2}}{2}\Phi(\I\nabla_{x}v_{\sigma})\cdot\Phi(\I\nabla_{x}v_{\sigma}) +\epsi^{2}\chi(\vsigma^{*}\hepsisigma\vsigma)\phisigmaj\cdot\nabla E_{\sigma}(x) \\&
&+\,\epsi^{2}(\mathbf{1}-\chi(\vsigma^{*}\hepsisigma\vsigma))\phisigmaj\cdot\nabla E_{\sigma}(x)\chi(\vsigma^{*}\hepsisigma\vsigma) +\frac{\I\epsi}{2}\chi(\vsigma^{*}\hepsisigma\vsigma)\\&
&\cdot[\phisigmaj\cdot\pop, \vsigma^{*}\pop^{2}\vsigma] +\frac{\I\epsi}{2}(\mathbf{1}-\chi(\vsigma^{*}\hepsisigma\vsigma))[\phisigmaj\cdot\pop, \vsigma^{*}\pop^{2}\vsigma]\chi(\vsigma^{*}\hepsisigma\vsigma)\\&
&+\,\frac{\I\epsi^{2}}{2}\vsigma^{*}[\hepsisigma, S]\vsigma +\frac{\epsi^{2}}{2}\vsigma^{*}\big[[\hepsisigma, T], T\big]\vsigma \,.
\end{eqnarray*}
We examine separately the different terms.
\begin{eqnarray*} 
\bullet && \frac{\epsi^{2}}{2}\Phi(\I\nabla_{x}v_{\sigma})\cdot\Phi(\I\nabla_{x}v_{\sigma})=\frac{\epsi^{2}}{4}\big[a(\I\deriv{j}{}v_{\sigma}) + a(\I\deriv{j}{}v_{\sigma})^{*}\big]\big[a(\I\deriv{j}{}v_{\sigma})  +a(\I\deriv{j}{}v_{\sigma})^{*}\big]\\
&=&\frac{\epsi^{2}}{4}\sum_{j=1}^{3N}\big[a(\I\deriv{j}{}v_{\sigma})^{2} + a(\I\deriv{j}{}v_{\sigma})^{*\,2} + a(\I\deriv{j}{}v_{\sigma})^{*}a(\I\deriv{j}{}v_{\sigma})\big] + \norm{\nabla v_{\sigma}}_{L^{2}(\field{R}{3}, dk)\otimes\field{C}{3N}}^{2}\,.
\end{eqnarray*}
We remark that the last term, being $x$-independent, is a $c$-number.
\begin{eqnarray}\label{commaphi} 
  \bullet && [\phisigmaj\cdot\pop, \vsigma^{*}\pop^{2}\vsigma]= [\phisigmaj\cdot\pop, \pop^{2}] + [\phisigmaj\cdot\pop, \vsigma^{*}[\pop^{2}, \vsigma]]=\nonumber\\
&=&2\I\epsi\deriv{j}{}\phisigmajl\pop_{l}\pop_{j} +\epsi[\phisigmaj\cdot\pop, \Phi(\I\nabla_{x}v_{\sigma})\cdot\pop + \pop\cdot \Phi(\I\nabla_{x}v_{\sigma})]+\Or(\epsi^{2})_{\mathcal{L}(\hilbert{H}_{0}, \hilbert{H})}\nonumber\\
&=&2\I\epsi\deriv{j}{}\phisigmajl\pop_{l}\pop_{j} +\epsi[\phisigmajl, \Phi(\I\deriv{j}{}v)]\pop_{l}\pop_{j} + \epsi\pop_{l}\pop_{j}[\phisigmajl, \Phi(\I\deriv{j}{}v_{\sigma})] +\Or(\epsi^{2})_{\mathcal{L}(\hilbert{H}_{0}, \hilbert{H})}\nonumber\\
&=&2\I\epsi\deriv{j}{}\phizerojl\pop_{l}\pop_{j} +\epsi[\phisigmajl, \Phi(\I\deriv{j}{}v_{\sigma})]\pop_{l}\pop_{j} + \epsi\pop_{l}\pop_{j}[\phisigmajl, \Phi(\I\deriv{j}{}v_{\sigma})] +\Or(\epsi^{2})_{\mathcal{L}(\hilbert{H}_{0}, \hilbert{H})}\nonumber \\
&&+ \,\Or(\epsi\sigma)_{\mathcal{L}(\hilbert{H}_{0}, \hilbert{H})},
\end{eqnarray}
where we have used lemma~\ref{lemmaamdelta}.

\begin{eqnarray*} 
   \bullet& \ [\phisigmajl,& \Phi(\I\deriv{j}{}v_{\sigma})]= [\qless{J}\phisigmal\qless{J}, \Phi(\I\deriv{j}{}v)] = [\qless{J}, \Phi(\I\deriv{j}{}v_{\sigma})]\phisigmaj\qless{J}\\
&&+\,\qless{J}\phisigmaj[\qless{J}, \Phi(\I\deriv{j}{}v_{\sigma})]+\qless{J}[\phisigmaj, \Phi(\I\deriv{j}{}v_{\sigma})]\qless{J}\\
&=&\frac{1}{\sqrt{2}}Q_{\mathrm{J}}a(\I\deriv{j}{}v_{\sigma})\phisigmaj\qless{J} - \frac{1}{\sqrt{2}}a(\I\deriv{j}{}v_{\sigma})^{*}Q_{\mathrm{J}}\phisigmaj\qless{J}\\
&&+\,\frac{1}{\sqrt{2}}\qless{J}\phisigmaj Q_{\mathrm{J}}a(\I\deriv{j}{}v_{\sigma})- \frac{1}{\sqrt{2}}\qless{J}\phisigmaj a(\I\deriv{j}{}v_{\sigma})^{*}Q_{\mathrm{J}}\\
&&+\,\I\qless{J}\Re\big\langle\deriv{j}{}v_{\sigma}(x, \cdot), \frac{\deriv{l}{}v_{\sigma}(x, \cdot)}{\abs{\cdot}}\big\rangle_{L^{2}(\field{R}{3}, dk)}\\
&=:&\mathcal{\tilde{R}}_{\mathrm{J}} +\I\qless{J}\Re\big\langle\deriv{j}{}v_{\sigma}(x, \cdot), \frac{\deriv{l}{}v_{\sigma}(x, \cdot)}{\abs{\cdot}}\big\rangle_{L^{2}(\field{R}{3}, dk)}\\
&=&\mathcal{\tilde{R}}_{\mathrm{J}} +\I\qless{J}\Re\big\langle\frac{\deriv{j}{}v(x, \cdot)}{\abs{\cdot}^{1/2}}, \frac{\deriv{l}{}v(x, \cdot)}{\abs{\cdot}^{1/2}}\big\rangle_{L^{2}(\field{R}{3}, dk)} + \Or(\sigma^{1/2})_{\mathcal{L}(\hilbert{H})}\, ,
\end{eqnarray*}
where we have bounded the scalar product using the same procedure applied in lemma \ref{lemmaamdelta}.

The remainder term $\mathcal{\tilde{R}}_{\mathrm{J}}$ vanishes when applied to states in the range of $\qm$ with $M<\mathrm{J}-1$. The scalar product can be written in a clearer way using the explicit expression of the function $v$ ($l_{1}, j_{1}=1, \ldots N$, $l_{2}, j_{2}=1, \ldots 3$):
\begin{eqnarray*}
 \bullet\,\, & \Re&\big\langle\frac{\deriv{(l_{1}, l_{2})}{}v(x, \cdot)}{\abs{\cdot}^{1/2}}, \frac{\deriv{(j_{1}, j_{2})}{}v(x, \cdot)}{\abs{\cdot}^{1/2}}\big\rangle_{L^{2}(\field{R}{3}, dk)}\\
&&=\,\,\Re\int_{\field{R}{3}}dk\, \hat{\varrho}_{l_{1}}(k)^{*}\hat{\varrho}_{j_{1}}(k)\frac{\E^{\I k\cdot(\vec{x}_{j_{1}}-\vec{x}_{l_{1}})}}{\abs{k}^{2}}\kappa_{l_{2}}\kappa_{j_{2}}, \quad \kappa:=\frac{k}{\abs{k}}\, .
\end{eqnarray*} 
When $l_{1}=j_{1}$, since the charge densities are spherically symmetric, we get
\begin{equation*}\begin{split}
& \hspace{-5mm}\Re\int_{\field{R}{3}}dk\, \frac{\abs{\hat{\varrho}_{l_{1}}(k)}^{2}}{\abs{k}^{2}}\kappa_{l_{2}}\kappa_{j_{2}}=\frac{1}{3}\int_{\field{R}{3}}dk\, \frac{\abs{\hat{\varrho}_{l_{1}}(k)}^{2}}{\abs{k}^{2}}\delta_{l_{2}, j_{2}}=\\
&=\frac{1}{12\pi}\int_{\field{R}{3}\times\field{R}{3}}d\vec{x}\ d\vec{y}\ \frac{\varrho_{l_{1}}(\vec{x})\varrho_{l_{1}}(\vec{y})}{\abs{\vec{x}-\vec{y}}}\delta_{l_{2}, j_{2}}=:\frac{1}{3}e_{l_{1}}\delta_{l_{2}, j_{2}}\,.
\end{split}
\end{equation*}
Putting all the terms together, we have in the end
\begin{eqnarray*}
 \lefteqn{ \hspace{-5mm}\frac{\epsi}{2}[\phisigmaj\cdot\pop, V^{*}\pop^{2}V]= \I\epsi^{2}\deriv{l}{}\phizerojl\pop_{j}\pop_{l} +\I\epsi^{2}\qless{J}\sum_{l_{1}=1}^{N}e_{l_{1}}\popb_{l_{1}}^{2} }\\
&&+\,\I\frac{\epsi^{2}}{2}\qless{J}\underset{(l_{1}\neq j_{1})}{\sum_{l_{1}, j_{1}=1}^{N}}\int_{\field{R}{3}}dk\, \frac{\hat{\varrho}_{l_{1}}(k)^{*}\hat{\varrho}_{j_{1}}(k)}{\abs{k}^{2}}\big[\E^{\I k\cdot(\vec{x}_{j_{1}}-\vec{x}_{l_{1}})}(\kappa\cdot\popb_{l_{1}})(\kappa\cdot\popb_{j_{1}}) \\
&&+ \,(\kappa\cdot\popb_{l_{1}})(\kappa\cdot\popb_{j_{1}})\E^{\I k\cdot(\vec{x}_{j_{1}}-\vec{x}_{l_{1}})}\big] +\epsi^{2}\mathcal{R}_{\mathrm{M}} + \Or(\epsi^{3})_{\mathcal{L}(\hilbert{H}_{0}, \hilbert{H})} \\& &+\,\Or(\epsi^{2}\sigma)_{\mathcal{L}(\hilbert{H}_{0}, \hilbert{H})},
\end{eqnarray*}
where $\mathcal{R}_{\mathrm{J}}$ is a remainder term that vanishes when applied to states in the range of $\qm$, with $M<\mathrm{J}-1$, and $\popb_{l_{1}}$ denotes the three-dimensional momentum operator associated to each particle.
\begin{eqnarray*}
\lefteqn{ \bullet\quad \frac{\I\epsi^{2}}{2}\vsigma^{*}[\hepsisigma, S]\vsigma= -\frac{\epsi^{2}}{2}\vsigma^{*}\chi(\hepsisigma)[\hepsisigma, \vsigma\nabla\cdot\phisigmaj \vsigma^{*}]\vsigma }\\
&&-\, \frac{\epsi^{2}}{2}\vsigma^{*}(\mathbf{1}-\chi(\hepsisigma))[\hepsisigma, \vsigma\nabla\cdot\phisigmaj \vsigma^{*}]\chi(\hepsisigma)\vsigma\\
&=&-\frac{\epsi^{2}}{2}\chi(\vsigma^{*}\hepsisigma\vsigma)[\vsigma^{*}\hepsisigma\vsigma, \nabla\cdot\phisigmaj] - \frac{\epsi^{2}}{2}(\mathbf{1}-\chi(\vsigma^{*}\hepsisigma\vsigma))\cdot\\
&&\cdot[\vsigma^{*}\hepsisigma\vsigma, \nabla\cdot\phisigmaj]\chi(\vsigma^{*}\hepsisigma\vsigma)\\
&=&-\frac{\epsi^{2}}{2}\chi(\vsigma^{*}\hepsisigma\vsigma)[\hfield, \nabla\cdot\phisigmaj] - \frac{\epsi^{2}}{2}(\mathbf{1}-\chi(\vsigma^{*}\hepsisigma\vsigma))\cdot\\
&&\cdot[\hfield, \nabla\cdot\phisigmaj]\chi(\vsigma^{*}\hepsisigma\vsigma) +\Or(\epsi^{3})_{\mathcal{L}(\hilbert{H}_{0}, \hilbert{H})}\\
&=&\frac{\I\epsi^{2}}{2}\chi(\vsigma^{*}\hepsisigma\vsigma)\Phi^{\mathrm{J}}(\I\Delta v_{\sigma}) + \frac{\I\epsi^{2}}{2}(\mathbf{1}-\chi(\vsigma^{*}\hepsisigma\vsigma))\Phi^{\mathrm{J}}(\I\Delta v_{\sigma})\chi(\vsigma^{*}\hepsisigma\vsigma) \\
&&+\,\Or(\epsi^{3})_{\mathcal{L}(\hilbert{H}_{0}, \hilbert{H})}\, .
\end{eqnarray*}
This gives exactly the terms needed to make $h_{1}$ a symmetric operator.
\begin{equation*}
\bullet\qquad \frac{\epsi^{2}}{2}\vsigma^{*}\big[[\hepsisigma, T], T\big]\vsigma\, .
\end{equation*}
Using the calculations for the first order part $h_{1}$, we get that
\begin{equation*}\begin{split}
&\hspace{-5mm} [\hepsisigma, T] = \chi(\hepsisigma)\vsigma\phim(\I\nabla_{x}v_{\sigma})\cdot\pop \vsigma^{*} +\\ & +(\mathbf{1}-\chi(\hepsisigma))\vsigma\phim(\I\nabla_{x}v_{\sigma})\cdot\pop \vsigma^{*}\chi(\hepsisigma) + \Or(\epsi\sqrt{\log(\sigma^{-1})}))_{\mathcal{L}(\hilbert{H}_{0}, \hilbert{H})},
\end{split}
\end{equation*}
we can keep therefore just the first two terms, omitting the remainder.
\begin{eqnarray*}
\lefteqn{ \bullet\quad \frac{\epsi^{2}}{2}\vsigma^{*}\big[[\hepsisigma, T], T\big]\vsigma=\frac{\epsi^{2}}{2}\vsigma^{*}\big[\chi(\hepsi)\vsigma\phim(\I\nabla_{x}v_{\sigma})\cdot\pop \vsigma^{*} }\\ 
&&+\,(\mathbf{1}-\chi(\hepsisigma))\vsigma\phim(\I\nabla_{x}v_{\sigma})\cdot\pop \vsigma^{*}\chi(\hepsisigma), T\big]\vsigma + \Or(\epsi^{3}\sqrt{\log(\sigma^{-1})})\\
&=&\frac{\I\epsi^{2}}{2}\vsigma^{*}\big[\chi(\hepsisigma)\vsigma\phim(\I\nabla_{x}v_{\sigma})\cdot\pop \vsigma^{*},\ \chi(\hepsisigma)\vsigma\phisigmaj\cdot\pop \vsigma^{*}\big]\vsigma \\
&&+\,\frac{\I\epsi^{2}}{2}\vsigma^{*}\big[(\mathbf{1}-\chi(\hepsisigma))\vsigma\phim(\I\nabla_{x}v_{\sigma})\cdot\pop \vsigma^{*}\chi(\hepsisigma),\ \chi(\hepsisigma)\vsigma\phisigmaj\cdot\pop \vsigma^{*}\big]\vsigma\\
&&+\,\frac{\I\epsi^{2}}{2}\vsigma^{*}\big[\chi(\hepsisigma)\vsigma\phim(\I\nabla_{x}v_{\sigma})\cdot\pop \vsigma^{*},\ (\mathbf{1}-\chi(\hepsisigma))\vsigma\phisigmaj\cdot\pop \vsigma^{*}\chi(\hepsisigma)\big]\vsigma\\
&&+\,\frac{\I\epsi^{2}}{2}\vsigma^{*}\big[(\mathbf{1}-\chi(\hepsisigma))\vsigma\phim(\I\nabla_{x}v_{\sigma})\cdot\pop \vsigma^{*}\chi(\hepsisigma),\\ &&\hspace{1cm}(\mathbf{1}-\chi(\hepsisigma))\vsigma\phisigmaj\cdot\pop \vsigma^{*}\chi(\hepsisigma)\big]\vsigma +\Or(\epsi^{3}\sqrt{\log(\sigma^{-1})})_{\mathcal{L}(\hilbert{H}_{0}, \hilbert{H})}\, .
\end{eqnarray*}
Summing up, we get in the end that
\begin{eqnarray}\label{htwochi}\lefteqn{\hspace{-3mm}
h_{2,\chi}= \frac{1}{4}\sum_{j=1}^{3N}\big[a(\I\deriv{j}{}v_{\sigma})^{2} + a(\I\deriv{j}{}v_{\sigma})^{*\,2} + a(\I\deriv{j}{}v_{\sigma})^{*}a(\I\deriv{j}{}v_{\sigma})\big] }\\
&&+\, \norm{\nabla v_{\sigma}}_{L^{2}(\field{R}{3}, dk)\otimes\field{C}{3n}}^{2} +\chi(\vsigma^{*}\hepsisigma\vsigma)\bigg\{\phisigmaj\cdot\nabla E(x)\nonumber\\
&&-\, \frac{1}{2}\big(\deriv{j}{}\Phi_{0, l}\pop_{l}\pop_{j} + \pop_{l}\pop_{j}\deriv{j}{}\Phi_{0, l}\big) - \qless{J}\sum_{l_{1}=1}^{N}e_{l_{1}}\popb_{l_{1}}^{2}\nonumber\\
&&-\,\frac{1}{2}\qless{J}\underset{(l_{1}\neq j_{1})}{\sum_{l_{1}, j_{1}=1}^{N}}\int_{\field{R}{3}}dk\, \frac{\hat{\varrho}_{l_{1}}(k)^{*}\hat{\varrho}_{j_{1}}(k)}{\abs{k}^{2}}\big[\E^{\I k\cdot(\vec{x}_{j_{1}}-\vec{x}_{l_{1}})}(\kappa\cdot\popb_{l_{1}})(\kappa\cdot\popb_{j_{1}})\nonumber \\
&&+\, (\kappa\cdot\popb_{l_{1}})(\kappa\cdot\popb_{j_{1}})\E^{\I k\cdot(\vec{x}_{j_{1}}-\vec{x}_{l_{1}})}\big] +\mathcal{R}_{\mathrm{J}}\bigg\} \nonumber\\&&+\, (\mathbf{1}-\chi(\vsigma^{*}\hepsisigma\vsigma))\bigg\{\cdots\bigg\}\chi(\vsigma^{*}\hepsisigma \vsigma)\nonumber\\
&&+\,\frac{\I}{2}\vsigma^{*}\big[\chi(\hepsisigma)\vsigma\phim(\I\nabla_{x}v_{\sigma})\cdot\pop \vsigma^{*},\ \chi(\hepsisigma)\vsigma\phisigmaj\cdot\pop \vsigma^{*}\big]\vsigma\nonumber \\
&&+\,\frac{\I}{2}\vsigma^{*}\big[(\mathbf{1}-\chi(\hepsisigma))\vsigma\phim(\I\nabla_{x}v_{\sigma})\cdot\pop \vsigma^{*}\chi(\hepsisigma),\ \chi(\hepsisigma)\vsigma\phisigmaj\cdot\pop \vsigma^{*}\big]\vsigma
\nonumber\\
&&+\,\frac{1}{2}\vsigma^{*}\big[\chi(\hepsisigma)\vsigma\phim(\I\nabla_{x}v_{\sigma})\cdot\pop \vsigma^{*},\ (\mathbf{1}-\chi(\hepsisigma))\vsigma\phisigmaj\cdot\pop \vsigma^{*}\chi(\hepsisigma)\big]\vsigma\nonumber\\
&&+\,\frac{1}{2}\vsigma^{*}\big[(\mathbf{1}-\chi(\hepsisigma))\vsigma\phim(\I\nabla_{x}v_{\sigma})\cdot\pop \vsigma^{*}\chi(\hepsisigma),\nonumber\\ &&\hspace{1.5cm}(\mathbf{1}-\chi(\hepsisigma))\vsigma\phisigmaj\cdot\pop \vsigma^{*}\chi(\hepsisigma)\big]\vsigma \, ,\nonumber
\end{eqnarray}
where we have symmetrized $\deriv{j}{}\Phi_{0, l}\pop_{l}\pop_{j}$, which is possible up to terms of order $\Or(\epsi)$.
The expression is fairly lengthy, but we will show below that many terms vanish when applied to a state in the range of $\qm\chi(\heff)$ ($M<\mathrm{J}-1$).\qed
\end{proof}

\section{The effective dynamics}\label{effectivedynamics}

We start with a number of lemmas we need to analyze the effective time evolution.

\begin{lemma}\label{approxchi}
Assume that $\sigma$ satisfies conditions \eqref{conditionsigma}, then

$1.$\ Given a function $\tilde{\chi}\in\coinf(\field{R}{})$, we have
\begin{equation}
\tilde{\chi}(\vsigma^{*}\hepsisigma \vsigma)-\tilde{\chi}(h_{0})=\epsi\mathcal{R}^{\epsi}_{\chi},
\end{equation}
where $\mathcal{R}^{\epsi}_{\chi}\in\mathcal{L}(\hilbert{H}, \hilbert{H}_{0})$, $\norm{\mathcal{R}^{\epsi}_{\chi}}_{\mathcal{L}(\hilbert{H}, \hilbert{H}_{0})}=\Or(1)$ and
\begin{equation}
\mathcal{R}^{\epsi}_{\chi}\qm=(\mathrm{Q}_{M+1}+\mathrm{Q}_{M-1})\mathcal{R}^{\epsi}_{\chi}\qm + \Or(\epsi^{2})_{\mathcal{L}(\hilbert{H}, \hilbert{H}_{0})}.
\end{equation}

$2.$\ Moreover, we have that
\begin{equation}\label{diffheffhzero}
\tilde{\chi}(\heff)-\tilde{\chi}(h_{0})=\Or(\epsi)_{\mathcal{L}(\hilbert{H}, \hilbert{H}_{0})},
\end{equation}
and that
\begin{equation}\label{diffqnheffhzero}
\qm\tilde{\chi}(\heff)=\qm\tilde{\chi}(h_{0})\tilde{\tilde{\chi}}(\heff) +\Or(\epsi^{2}\sqrt{\log(\sigma^{-1})})_{\mathcal{L}(\hilbert{H}, \hilbert{H}_{0})},
\end{equation}
where $\tilde{\tilde{\chi}}$ is any $\coinf(\field{R}{})$ function such that $\tilde{\chi}\tilde{\tilde{\chi}}=\tilde{\chi}$ and $\tilde{\tilde{\chi}}\chi=\tilde{\tilde{\chi}}$, $M<\mathrm{J}-1$.
\end{lemma}

\begin{proof}
Applying the Hellfer-Sj\"ostrand formula, equation \eqref{hellfersjostrand}, we get
\begin{equation*}
\tilde{\chi}(\vsigma^{*}\hepsisigma\vsigma)-\tilde{\chi}(h_{0})= \frac{1}{\pi}\int_{\field{R}{2}}dx dy\ \bar{\partial}\tilde{\chi}^{a}(z)\big[(\vsigma^{*}\hepsisigma\vsigma-z)^{-1} - (h_{0}-z)^{-1}\big].
\end{equation*} 
Since both Hamiltonians are self-adjoint on $\hilbert{H}_{0}$, we get
\begin{eqnarray*}
\lefteqn{ \hspace{-0.5cm}(\vsigma^{*}\hepsisigma\vsigma-z)^{-1} - (h_{0}-z)^{-1}=}
\\&=&(h_{0}-z)^{-1}(h_{0}-\vsigma^{*}\hepsisigma \vsigma)(\vsigma^{*}\hepsisigma \vsigma-z)^{-1}
\\&=&-(h_{0}-z)^{-1}\bigg[\frac{\epsi}{2}\Phi(\I\nabla_{x}v_{\sigma})\cdot\pop + \frac{\epsi}{2}\pop\cdot \Phi(\I\nabla_{x}v_{\sigma})
\\&&+\,\frac{\epsi^{2}}{2}\Phi(\I\nabla_{x}v_{\sigma})\cdot\Phi(\I\nabla_{x}v_{\sigma})\bigg](\vsigma^{*}\hepsisigma \vsigma-z)^{-1} +\Or(\sigma\abs{\Im z}^{-2})_{\mathcal{L}(\hilbert{H}, \hilbert{H}_{0})}\,,
\end{eqnarray*}
where we have used equation \eqref{vhepsiv} to calculate the difference of the two Hamiltonians and estimated the integrand proceeding in the same way as in lemma~\ref{cutoffenergysigma}. 
Moreover, iterating the formula we get
\begin{eqnarray}\label{expansionresolvent} 
\lefteqn{ \hspace{-5mm}(\vsigma^{*}\hepsisigma\vsigma-z)^{-1} - (h_{0}-z)^{-1}=}\\
&=&
(h_{0}-z)^{-1}(h_{0}-\vsigma^{*}\hepsisigma\vsigma) (\vsigma^{*}\hepsisigma\vsigma-z)^{-1}\nonumber
\\&=& 
-(h_{0}-z)^{-1}\bigg[\frac{\epsi}{2}\Phi(\I\nabla_{x}v)\cdot\pop +\frac{\epsi}{2}\pop\cdot \Phi(\I\nabla_{x}v)\bigg](h_{0}-z)^{-1}\nonumber\\&&+\, \Or(\epsi^{2}\abs{\Im z}^{-3})_{\mathcal{L}(\hilbert{H}, \hilbert{H}_{0})} \,,\nonumber
\end{eqnarray}
so 
\begin{eqnarray*}
\lefteqn{ \hspace{-5mm} \big[(\vsigma^{*}\hepsisigma \vsigma-z)^{-1} - (h_{0}-z)^{-1}\big]\qm =}\\&=& (\mathrm{Q}_{M +1}+\mathrm{Q}_{M-1})\big[(\vsigma^{*}\hepsisigma \vsigma-z)^{-1} - (h_{0}-z)^{-1}\big]\qm\\&& +\,\Or(\epsi^{2}\abs{\Im z}^{-3})_{\mathcal{L}(\hilbert{H}, \hilbert{H}_{0})}.
\end{eqnarray*}
Concerning point $2$, equation \eqref{diffheffhzero} follows immediately from the fact that 
\begin{equation*}\begin{split}
&\hspace{-5mm}\heff-z)^{-1} - (h_{0}-z)^{-1}=(h_{0}-z)^{-1}(h_{0}-\heff)(\heff-z)^{-1}=\\
&=\epsi(h_{0}-z)^{-1}h_{1,\chi}(\heff-z)^{-1} +\Or(\epsi^{2}\sqrt{\log(\sigma^{-1})})_{\mathcal{L}(\hilbert{H}_{0}, \hilbert{H})},
\end{split}
\end{equation*}
while, for equation \eqref{diffqnheffhzero}, we have by definition that
\begin{equation*}\begin{split}
&\hspace{-9mm}\qm\tilde{\chi}(\heff)=\qm\tilde{\chi}(\heff)\tilde{\tilde{\chi}}(\heff)=\\&= \qm\tilde{\chi}(h_{0})\tilde{\tilde{\chi}}(\heff) + \qm[\tilde{\chi}(\heff)-\tilde{\chi}(h_{0})]\tilde{\tilde{\chi}}(\heff)\, .
\end{split}
\end{equation*}
Proceeding as above we find
\begin{equation}\begin{split}\label{estimatechi}
& \hspace{-5mm}\qm[\tilde{\chi}(\heff)-\tilde{\chi}(h_{0})]\tilde{\tilde{\chi}}(\heff) =\\&= \frac{1}{\pi}\int_{\field{R}{2}}dx dy\ \bar{\partial}\tilde{\chi}^{a}(z)\qm\big[(\heff-z)^{-1} 
- (h_{0}-z)^{-1}\big]\tilde{\tilde{\chi}}(\heff)\,,
\end{split}
\end{equation}
so, if we show that
\begin{equation}\label{qnhonechichi}
\qm h_{1,\chi}\tilde{\tilde{\chi}}(\heff)=\Or(\epsi\sqrt{\log(\sigma^{-1})})_{\mathcal{L}(\hilbert{H}_{0}, \hilbert{H})}
\end{equation}
we are done.

From equation \eqref{honechi} (omitting the ``$\pop\cdot\Phi$'' part, which can be treated in the same way) we get
\begin{equation*}\begin{split}
&\qm h_{1, \chi} \tilde{\tilde{\chi}}(\heff) = \frac{1}{2}\qm\Phi(\I\nabla_{x}v_{\sigma})\cdot\pop \tilde{\tilde{\chi}}(\heff) -\frac{1}{2} \qm\chi(\vsigma^{*}\hepsisigma\vsigma)\cdot\\
&\cdot\phim(\I\nabla_{x}v_{\sigma})\cdot\pop\tilde{\tilde{\chi}}(\heff) -\frac{1}{2}\qm(\mathbf{1}-\chi(\vsigma^{*}\hepsisigma\vsigma))\phim(\I\nabla_{x}v_{\sigma})\cdot\pop\chi(\vsigma^{*}\hepsisigma\vsigma)\cdot\\
&\cdot\tilde{\tilde{\chi}}(\heff),
\end{split}
\end{equation*}
but it follows from point $1$, that we can replace $\chi(V^{*}\hepsi V)$ with $\chi(h_{0})$ up to terms of order $\Or(\epsi)$, and from equation \eqref{estimatechi} that we can replace $\tilde{\tilde{\chi}}(\heff)$ with $\tilde{\tilde{\chi}}(h_{0})$ up to terms of order $\Or(\epsi)$, therefore we have
\begin{eqnarray*}
\lefteqn{ \hspace{-5mm}\qm h_{1, \chi} \tilde{\tilde{\chi}}(\heff)=\frac{1}{2}\qm\Phi(\I\nabla_{x}v_{\sigma})\cdot\pop \tilde{\tilde{\chi}}(h_{0}) -\frac{1}{2} \qm\chi(\vsigma^{*}\hepsisigma\vsigma)\phim(\I\nabla_{x}v_{\sigma})\cdot}\\
&&\cdot\pop\tilde{\tilde{\chi}}(h_{0}) -\frac{1}{2}\qm(\mathbf{1}-\chi(\vsigma^{*}\hepsisigma \vsigma))\phim(\I\nabla_{x}v_{\sigma})\cdot\pop\tilde{\tilde{\chi}}(h_{0}) +\Or(\epsi)\\
&=&\frac{1}{2}\qm\Phi(\I\nabla_{x}v_{\sigma})\cdot\pop \tilde{\tilde{\chi}}(h_{0}) -\frac{1}{2} \qm\phim(\I\nabla_{x}v_{\sigma})\cdot\pop\tilde{\tilde{\chi}}(h_{0})+\Or(\epsi)=\Or(\epsi)\,,
\end{eqnarray*}
so point $2$ is proved.\qed
\end{proof}

The following lemma was proved in (\cite{FGS}, Appendix B) and we will use it to characterize the range of $\chi_{(-\infty, c)}(h_{0})$, where $\chi_{(-\infty, c)}$ is the characteristic function of the indicated interval.
\begin{lemma}\label{lemmafgs}
Let $\tilde{H}$ be a Hamiltonian of the form
\begin{equation}
\tilde{H}:= \mathbf{1}\otimes d\Gamma(\abs{k}) + H\otimes\mathbf{1},
\end{equation} 
acting on the Hilbert space $\tilde{\hilbert{H}}=\hilbert{H}\otimes\fock$, where $\fock$ is the bosonic Fock space over $L^{2}(\field{R}{d})$ and $\hilbert{H}$ is a generic Hilbert space. 

Then the set of all linear combinations of vectors of the form
\begin{equation}\label{rangechi}
\varphi\otimes a(g_{1})^{*}\cdots a(g_{\N})^{*}\Omega_{\mathrm{F}}, \quad \lambda + \sum_{j=1}^{\N}M_{j}< c, \quad (c>0),
\end{equation}
where $\varphi=\chi_{(-\infty, \lambda)}(H)\varphi$ for some $\lambda<c$, $\N\in\field{N}{}$ and
\begin{equation*}
M_{j}:=\sup\{\abs{k}: k\in\mathrm{supp}\ g_{j}\}
\end{equation*}
is dense in $\chi_{(-\infty, c)}(\tilde{H})\tilde{\hilbert{H}}$.
\end{lemma}

\begin{lemma}\label{lemmapseudo}
Let $\chi\in\coinf(\field{R}{})$ and $\tilde{h}_{0}:=\mathbf{1}\otimes\hfield + \pop^{2}/2\otimes\mathbf{1}$. Then, $\exists\xi\in\coinf(\field{R}{})$ such that $\chi\xi=\chi$ and 
\begin{equation}
\xi^{c}(\tilde{h}_{0})\chi(h_{0})= \Or(\epsi^{\infty})_{\mathcal{L}(\hilbert{H})},
\end{equation} 
where $\xi^{c}:= 1-\xi$.

Moreover, denoting by 
\begin{equation*}
c_{\chi}:=\sup\{\abs{k}: k\in\mathrm{supp}\ \chi\},
\end{equation*}
and defining
\begin{equation}
c_{\xi}:= 2c_{\chi} + E_{\infty},
\end{equation}
where $E_{\infty}:= \sup_{x\in\field{R}{3n}}\abs{E(x)}$, we can choose $\sup\{\abs{k}: k\in\mathrm{supp}\ \xi\}$ arbitrarily close to $c_{\xi}$.

The statement remains true also if we invert the roles of $h_{0}$ and $\tilde{h}_{0}$.
\end{lemma}

\begin{proof}
It follows immediately from the spectral theorem and the fact that $\hfield$ is a nonnegative operator that, if $\chi^{(s)}_{(-E_{\infty}, c_{\chi})}$ denotes a smoothed version of the characteristic function of the interval indicated, then
\begin{equation*}
\chi^{(s)}_{(-E_{\infty}, c_{\chi})}(h_{\mathrm{p}})\chi(h_{0})=\chi(h_{0}),
\end{equation*}
where 
\begin{equation*}
h_{\mathrm{p}}:= \frac{1}{2}\,\pop^{2} + E(x).
\end{equation*}

Our aim is now to use the functional calculus for pseudodifferential operators developed in \cite{HeRo} (see also \cite{DiSj}, chapter 8) to show that, if $\xi\in\coinf(\field{R}{})$, $\xi=1$ on a neighborhood of $[0, c_{\chi} + E_{\infty}]$, then
\begin{equation}\label{pseudodiff}
\xi^{c}(\tilde{h}_{\mathrm{p}})\chi^{(s)}_{(-E_{\infty}, c_{\chi})}(h_{\mathrm{p}})=\Or(\epsi^{\infty})_{\mathcal{L}(\hilbert{H})},
\end{equation} 
where $\tilde{h}_{\mathrm{p}}:= \pop^{2}/2$.

Once we have shown this, we will have
\begin{equation*}
\chi(h_{0})=\xi(\tilde{h}_{\mathrm{p}})\chi(h_{0}) + \Or(\epsi^{\infty})_{\mathcal{L}(\hilbert{H})},
\end{equation*}
but, applying lemma \ref{lemmafgs}, we can write
\begin{equation*}
\xi(\tilde{h}_{\mathrm{p}})\chi(h_{0})\Psi = \lim_{n\to\infty}\xi(\tilde{h}_{\mathrm{p}})\Psi_{n},
\end{equation*}
where the $\Psi_{n}$ are finite linear combinations of vectors of the form \eqref{rangechi}, with $\varphi=\chi_{(-\infty, \lambda)}(h_{\mathrm{p}})\varphi$, $0<\lambda<c_{\chi}$, so that
\begin{eqnarray*}
\xi(\tilde{h}_{\mathrm{p}})\chi(h_{0})\Psi &=& \lim_{n\to\infty}\xi(\tilde{h}_{\mathrm{p}})\Psi_{n}= \lim_{n\to\infty}\chi^{(s)}_{(0, 2c_{\chi}+E_{\infty})}(\tilde{h}_{0})\xi(\tilde{h}_{\mathrm{p}})\Psi_{n}\\
&=&\chi^{(s)}_{(0, 2c_{\chi}+E_{\infty})}(\tilde{h}_{0})\xi(\tilde{h}_{\mathrm{p}})\chi(h_{0})\Psi\\
\Rightarrow \quad\chi(h_{0})&=&\chi^{(s)}_{(0, 2c_{\chi}+E_{\infty})}(\tilde{h}_{0})\chi(h_{0}) + \Or(\epsi^{\infty})_{\mathcal{L}(\hilbert{H})}\,.
\end{eqnarray*}

We proceed now to prove \eqref{pseudodiff}. We first recall some facts we need from \cite{HeRo} and \cite{DiSj}.
Given a function $\chi\in\coinf(\field{R}{})$ and a pseudodifferential operator $P$ with symbol in a suitable symbol class (for our aims it is enough to say that this holds for both $h_{\mathrm{p}}$ and $\tilde{h}_{\mathrm{p}}$), then also $\chi(P)$ is a pseudodifferential operator, with symbol
\begin{equation*}\begin{split}
&\chi(P)= \opw(a), \quad a \sim \sum_{j=0}^{\infty}\epsi^{j}a_{j},\\
& a_{0}=\chi(p_{0}),\quad a_{j}=\sum_{k=1}^{2j-1}\frac{d_{j, k}}{k!}\chi^{(k)}(p_{0}),
\end{split}
\end{equation*}
where $\opw$ denotes the Weyl quantization, $p_{0}$ is the principal symbol of $P$ and the coefficients $d_{j, k}$ depend on the higher order terms in the expansion of the symbol of $P$ (their precise form is given in \cite{HeRo}).

We remark that the previous expressions are local in $p_{0}$, and that for $h_{p}$ and $\tilde{h}_{p}$ the symbol is just the principal symbol, and is given by
\begin{equation*}
\mathfrak{h}_{0}(x, p):= \frac{1}{2}p^{2} + E(x), \quad \tilde{\mathfrak{h}}_{0}(x, p):=\frac{1}{2}p^{2}. 
\end{equation*}

If we multiply two pseudodifferential operators, the symbol of the product is given by the twisted product of the symbols of the two operators involved:
\begin{equation}\label{productformula}\begin{split}
& \hspace{1cm}\opw(a_{1})\cdot\opw(a_{2}) = \opw(a_{1}\sharp_{\epsi}a_{2}),\\[1mm]
& a_{1}\sharp_{\epsi}a_{2} \sim \sum_{j=0}^{\infty}\frac{1}{j!}\bigg(\bigg(\frac{\I\epsi}{2}(\nabla_{p}\cdot\nabla_{x}-\nabla_{\xi}\cdot\nabla_{q})\bigg)^{j}a_{1}(q, p)a_{2}(x, \xi)\bigg)_{\big|_{x=q, \xi=p}}\, .
\end{split}
\end{equation} 

Applying these formulas to calculate the product $\xi^{c}(\tilde{h}_{\mathrm{p}})\chi^{(s)}_{(-E_{\infty}, c_{\chi})}(h_{\mathrm{p}})$ and using the locality in the principal symbol, we get that all the terms in the expansion of the product vanish, i.e., equation \eqref{pseudodiff}.\qed
\end{proof}

\begin{corollary} Given a function $\chi\in\coinf(\field{R}{})$ and a $\sigma>0$, we have
\begin{equation}\label{energynotincreas}
a\bigg(\frac{\I\deriv{j}{}v_{\sigma}(x, \cdot)}{\abs{k}}\bigg)\qm\chi(h_{0})=\mathrm{Q}_{M-1}\xi(h_{0})a\bigg(\frac{\I\deriv{j}{}v_{\sigma}(x, \cdot)}{\abs{k}}\bigg)\qm\chi(h_{0}) + \Or_{0}(\epsi^{\infty}),
\end{equation}
where $\xi\in\coinf(\field{R}{})$ and
\begin{equation*}
c_{\xi}=2d_{\chi} + E_{\infty},
\end{equation*}
where 
\begin{eqnarray*}
d_{\chi}&:=& 2c_{\chi}+E_{\infty}+\min\{c_{\chi}, \Lambda\}\,,\\
 c_{\chi}&:=&\sup\{\abs{k}: k\in\mathrm{supp}\ \chi\}\,,\\
 E_{\infty}&:=& \sup_{x\in\field{R}{n}}\abs{E(x)}\,,
\end{eqnarray*}
and we can choose $\sup\{\abs{k}: k\in\mathrm{supp}\ \xi\}$ arbitrarily close to $c_{\xi}$.

An analogous statement holds for the creation operator.
\end{corollary}

\begin{proof}
Applying twice lemma \ref{lemmapseudo}, the thesis will follow if we prove \eqref{energynotincreas} replacing $h_{0}$ with $\tilde{h}_{0}$.

Applying lemma \ref{lemmafgs}, we have
\begin{equation*}
\chi_{(-\infty, c_{\chi})}(\tilde{h}_{0})\qm\chi(\tilde{h}_{0})\Psi=\qm\chi(\tilde{h}_{0})\Psi=\lim_{n\to\infty}\Psi_{n},
\end{equation*}
where $\Psi_{n}$ is a linear combination of vectors of the form described in \eqref{rangechi}, with $\varphi=\chi_{(-\infty, \lambda)}(\pop^{2})\varphi$.
We have therefore
\begin{equation*}
a\bigg(\frac{\I\deriv{j}{}v_{\sigma}(x, \cdot)}{\abs{k}}\bigg)\qm\chi(h_{0})\Psi=\lim_{n\to\infty}a\bigg(\frac{\I\deriv{j}{}v_{\sigma}(x, \cdot)}{\abs{k}}\bigg)\Psi_{n}.
\end{equation*}
Applying the operator to a vector of the form \eqref{rangechi} we get
\begin{eqnarray*}
\lefteqn{ \hspace{-2cm}a\bigg(\frac{\I\deriv{j}{}v_{\sigma}(x, \cdot)}{\abs{k}}\bigg)\varphi\otimes a(g_{1})^{*}\cdots a(g_{M})^{*}\Omega_{\mathrm{F}}=\varphi(x)\sum_{l=1}^{\N}\big\langle \frac{\I\deriv{j}{}v_{\sigma}(x,k)}{\abs{k}}, g_{l}(k)\big\rangle_{L^{2}(\field{R}{3}, dk)}\cdot}\\
&&\hspace{3cm} \cdot a(g_{1})^{*}\cdots a(\overset{\vee}{g_{l}})^{*}\cdots a(g_{M})^{*}\Omega_{\mathrm{F}}\,,
\end{eqnarray*}
with
\[
\big\langle \frac{\I\deriv{j}{}v_{\sigma}(x,k)}{\abs{k}}, g_{l}(k)\big\rangle_{L^{2}(\field{R}{3}, dk)}= -\int dk\, \frac{\id_{(\sigma, \infty)}(k)\hat{\varrho}_{j_{1}}(k)^{*}}{\abs{k}^{3/2}}\E^{-\I k\cdot x_{j_{1}}}g_{l}(k)\kappa_{j_{2}}\,.
\]
The functions $g_{l}$ have by hypothesis compact support in $k$, with radius uniformly bounded by $c_{\chi}$.

For the part depending on $x$, calculating the Fourier transform with the convolution theorem, we have
\begin{equation*}\begin{split}
& \mathcal{F}\bigg(\varphi(x)\big\langle \frac{\I\deriv{j}{}v_{\sigma}(x,k)}{\abs{k}}, g_{l}(k)\big\rangle\bigg)(p)=-\int dk\, \frac{\id_{(\sigma, \infty)}(k)\hat{\varrho}_{j_{1}}(k)^{*}}{\abs{k}^{3/2}}g_{l}(k)\kappa_{j_{2}}\hat{\varphi}(p + \tilde{k}_{j_{1}}),
\end{split}
\end{equation*}
where 
\begin{equation*}
\tilde{k}_{j_{1}} = (0, \ldots, k, 0, \ldots 0)\in\field{R}{3N},
\end{equation*}
with the $k$ in the entry $j_{1}$.
Now, since $\hat{\varrho}_{j_{1}}$, $g_{l}$ and $\hat{\varphi}$ have compact support, also the Fourier transform will have compact support in $p$, with radius bounded by $c_{\chi}+\min\{c_{\chi}, K\}=:d_{\xi}$. 
This means that
\begin{equation*}\begin{split}
& \hspace{-1cm}a\bigg(\frac{\I\deriv{j}{}v_{\sigma}(x, \cdot)}{\abs{k}}\bigg)\varphi\otimes a(g_{1})^{*}\cdots a(g_{M})^{*}\Omega_{\mathrm{F}}=\\
&=\chi_{(-\infty, d_{\xi})}(\tilde{h}_{0})a\bigg(\frac{\I\deriv{j}{}v_{\sigma}(x, \cdot)}{\abs{k}}\bigg)\varphi\otimes a(g_{1})^{*}\cdots a(g_{M})^{*}\Omega_{\mathrm{F}},
\end{split}
\end{equation*}
so, remarking that $\tilde{h}_{0}$ is a positive operator and that we can smooth $\chi_{[0, d_{\xi})}$ by a $\coinf$ function with arbitrarily close support, we prove the thesis.\qed
\end{proof}

\begin{theorem}\label{zeroordertime}\emph{(Zero order approximation to the time evolution)} The following two estimates hold:
\begin{equation}
\norm{(\E^{-\I \heff\frac{t}{\epsi}} - \E^{-\I h_{0}\frac{t}{\epsi}})\qm\tilde{\chi}(\heff)}_{\mathcal{L}(\hilbert{H})}=\Or\big(\sqrt{M\hspace{-3pt}+\hspace{-3pt}
1}\abs{t}\epsi\sqrt{\log(\sigma(\epsi)^{-1})}\big),
\end{equation}
\begin{equation}
\norm{\qm(\E^{-\I \heff\frac{t}{\epsi}} - \E^{-\I h_{0}\frac{t}{\epsi}})\tilde{\chi}(\heff)}_{\mathcal{L}(\hilbert{H})}=\Or\big(\sqrt{M\hspace{-3pt}+\hspace{-3pt}1}\abs{t}\epsi\sqrt{\log(\sigma(\epsi)^{-1})}\big),
\end{equation}
for every $\tilde{\chi}\in\coinf(\field{R}{})$ such that $\tilde{\chi}\chi=\tilde{\chi}$.
\end{theorem}

\begin{corollary}\label{corzeroordertime}
The subspaces associated to the $Q_{M}$s are almost invariant with respect to the dynamics generated by $\heff$ $\forall M\in\field{N}{}$, i.e.,
\begin{equation}
\norm{[\E^{-\I t\heff/\epsi}, \qm]\tilde{\chi}(\heff)}_{\mathcal{L}(\hilbert{H})}=\Or\big(\sqrt{M+1}\abs{t}\epsi\sqrt{\log(\sigma(\epsi)^{-1})}\big).
\end{equation}
\end{corollary}

\begin{remark}\label{adiabaticinvm}\emph{(Adiabatic invariance of $M$-photons dressed particles subspaces).}

Using the unitary $\uone$ we can translate all the previous results from the representation space to the original dynamics. This means that if we define the perturbed dressed projectors
\begin{equation}
P_{M}^{\epsi}:=\hilbert{U}^{*}\qm\uone,
\end{equation}
which satisfy by construction
\begin{equation*}
\norm{P_{M}^{\epsi} - \pizn}_{\mathcal{L}(\hilbert{H})}=\Or(\epsi\sqrt{\log(\sigma(\epsi)^{-1})}),
\end{equation*}
we get that
\begin{equation*}
\norm{[\E^{-\I t\hepsisigma/\epsi}, P_{M}^{\epsi}]\tilde{\chi}(\hepsisigma)}_{\mathcal{L}(\hilbert{H})}=\Or\big(\sqrt{M+1}\abs{t}\epsi\sqrt{\log(\sigma(\epsi))^{-1}}\big),
\end{equation*}
and
\begin{equation*}
\norm{[\E^{-\I t\hepsisigma/\epsi}, \pizn]\tilde{\chi}(\hepsisigma)}_{\mathcal{L}(\hilbert{H})}=\Or\big((1+\abs{t})\epsi\sqrt{\log(\sigma(\epsi)^{-1})}\big), \quad \forall\tilde{\chi}\in\coinf(\field{R}{}) \, .
\end{equation*}
For the original dynamics we have then
\begin{equation*}\begin{split}
& \hspace{-5mm}[\E^{-\I t\hepsi/\epsi}, P_{M}^{\epsi}]\tilde{\chi}(\hepsi) = [(\E^{-\I t\hepsi/\epsi}- \E^{-\I t\hepsisigma/\epsi}), P_{M}^{\epsi}]\chi(\hepsi) +\\
&+ [\E^{-\I t\hepsisigma/\epsi}, P_{M}^{\epsi}](\chi(\hepsi) - \chi(\hepsisigma)) + [\E^{-\I t\hepsisigma/\epsi}, P_{M}^{\epsi}]\tilde{\chi}(\hepsisigma)\, .
\end{split}
\end{equation*}
The first term can be bounded by $\Or(\sigma(\epsi)^{1/2}\epsi^{-1})=\Or(\epsi^{2})$ using proposition \ref{approxsigma} and equation \eqref{unormbounded}. The second one by $\Or(\sigma(\epsi)^{1/2})$ using lemma \ref{cutoffenergysigma}, and the third one has been just estimated above.

Putting together these facts, equation \eqref{almostinvsub} is proved.

For the particular case $M=0$, this result, together with the expression of the zero order Hamiltonian $h_{0}$, was already shown in \cite{Te2}, assuming an infrared regularized interaction and a relativistic dispersion relation for the particles, which automatically implies that they have a bounded maximal velocity, and avoids therefore the introduction of cutoff functions.
\end{remark}

\begin{proof}
First of all we remark that, employing lemma \ref{approxchi}, we can replace $\tilde{\chi}(\heff)$ with $\tilde{\chi}(h_{0})$ up to terms of order $\Or(\epsi)_{\mathcal{L}(\hilbert{H})}$, which are smaller than the error we want to prove.

Since $\heff$ and $h_{0}$ are both self-adjoint on $\hilbert{H}_{0}$, we can apply the Duhamel formula, and obtain
\begin{eqnarray*} 
\lefteqn{  (\E^{-\I t\heff/{\epsi}}-\E^{-\I t h_{0}/{\epsi}})\qm\tilde{\chi}(h_{0}) =}\\&=& -\frac{\I}{\epsi} \int_{0}^{t} ds\, 
\E^{\I (s-t)\heff/\epsi}
(\heff-h_{0})\E^{-\I s h_{0}/\epsi}\qm\tilde{\chi}(h_{0})\\&=& -\I\int_{0}^{t}ds\, \E^{\I (s-t)\heff/\epsi}h_{1,\chi}\E^{-\I s h_{0}/\epsi}\qm\tilde{\chi}(h_{0}) 
+ \Or(\epsi\sqrt{M+1}\sqrt{\log(\sigma^{-1})})\\
&=&-\I\int_{0}^{t}ds\, \E^{\I (s-t)\heff/\epsi}h_{1,\chi}\qm\tilde{\chi}(h_{0})\E^{-\I s h_{0}/\epsi} + \Or(\epsi\sqrt{M+1}\sqrt{\log(\sigma^{-1})})\,,
\end{eqnarray*}
which implies
\begin{eqnarray*}
\lefteqn{\hspace{-1cm} \norm{(\E^{-\I t\heff/\epsi}-\E^{-\I th_{0}/\epsi})\qm\tilde{\chi}(h_{0})}_{\mathcal{L}(\hilbert{H})}\leq}\\&\leq& \abs{t}\cdot\norm{h_{1,\chi}\qm\tilde{\chi}(h_{0})}_{\mathcal{L}(\hilbert{H})}+\Or(\epsi\abs{t}\sqrt{M+1}\sqrt{\log(\sigma^{-1})}).
\end{eqnarray*}
We proceed now as in the proof of lemma \ref{approxchi}. From equation \eqref{honechi} it follows (omitting the terms of the form ``$\pop\cdot\Phi$'', which can be treated in the same way)
\begin{eqnarray*}
\lefteqn{ h_{1,\chi}\qm\tilde{\chi}(h_{0})= \bigg[\frac{1}{2}\Phi(\I\nabla_{x}v_{\sigma})\cdot\pop -\frac{1}{2} \chi(\vsigma^{*}\hepsisigma\vsigma)\phim(\I\nabla_{x}v_{\sigma})\cdot\pop }\\
&&-\,\frac{1}{2}\big(\mathbf{1}-\chi(\vsigma^{*}\hepsisigma\vsigma)\big)\phim(\I\nabla_{x}v_{\sigma})\cdot\pop\,\chi(\vsigma^{*}\hepsisigma\vsigma)\bigg]\qm\tilde{\chi}(h_{0})\\
&=&\frac{1}{2}\Phi(\I\nabla_{x}v_{\sigma})\cdot\pop\,\qm\tilde{\chi}(h_{0})- \frac{1}{2} \chi(\vsigma^{*}\hepsisigma \vsigma)\phim(\I\nabla_{x}v_{\sigma})\cdot\pop\,\qm\tilde{\chi}(h_{0})\\
&&-\,\frac{1}{2}\big(\mathbf{1}-\chi(\vsigma^{*}\hepsisigma\vsigma)\big)\phim(\I\nabla_{x}v_{\sigma})\cdot\pop\,\qm\tilde{\chi}(h_{0})\\
&&-\,\frac{1}{2}\big(\mathbf{1}-\chi(\vsigma^{*}\hepsisigma\vsigma)\big)\phim(\I\nabla_{x}v_{\sigma})\cdot\pop[\chi(\vsigma^{*}\hepsisigma\vsigma)-\chi(h_{0})]\qm\tilde{\chi}(h_{0})\\
&=&-\frac{1}{2}[\mathbf{1}-\chi(h_{0})]\phim(\I\nabla_{x}v_{\sigma})\cdot\pop[\chi(\vsigma^{*}\hepsisigma \vsigma)-\chi(h_{0})]\qm\tilde{\chi}(h_{0})\\
&&+\,\frac{1}{2}[\chi(\vsigma^{*}\hepsisigma\vsigma)-\chi(h_{0})]\phim(\I\nabla_{x}v_{\sigma})\cdot\pop[\chi(\vsigma^{*}\hepsisigma\vsigma)-\chi(h_{0})]\qm\tilde{\chi}(h_{0})\,.
               \end{eqnarray*}
Applying lemma \ref{approxchi}, we have immediately that the last term is of order $\Or(\epsi^{2})$. We examine therefore more closely the second one.

Using equation \eqref{expansionresolvent} and the corollary to lemma \ref{lemmapseudo}, we get that
\begin{eqnarray*}
\lefteqn{ [\chi(\vsigma^{*}\hepsisigma\vsigma)-\chi(h_{0})]\qm\tilde{\chi}(h_{0})=}\\&=&(\mathrm{Q}_{M+1} + \mathrm{Q}_{M-1})\xi(h_{0})[\chi(\vsigma^{*}\hepsisigma\vsigma)
-\chi(h_{0})]\cdot\qm\tilde{\chi}(h_{0}) + \Or(\epsi^{2})_{\mathcal{L}(\hilbert{H})}\,,
\end{eqnarray*}
where the function $\xi$ has a support slightly larger than that of the function $\chi$. Applying again the corollary, we get that, if we choose the support of the function $\tilde{\chi}$ sufficiently smaller than the support of the function $\chi$, then
\begin{equation}\label{honeqnchi}
h_{1,\chi}\qm\tilde{\chi}(h_{0})= \Or(\epsi^{2}),
\end{equation}		 
so we get the first estimate.

For the second one, we apply again the Duhamel formula, but inverting the position of the two unitaries:
\begin{eqnarray*}
\lefteqn{ \qm(\E^{-\I t\heff/\epsi}-\E^{-\I th_{0}/\epsi})\tilde{\chi}(\heff) =}\\&=& -\frac{\I}{\epsi}\int_{0}^{t}ds\, \qm\E^{\I (s-t)h_{0}/\epsi}(\heff-h_{0}) \E^{-\I s \heff/\epsi}\tilde{\chi}(\heff)\\&=&-\I\int_{0}^{t}ds\, \E^{\I (s-t)h_{0}/\epsi}\qm h_{1,\chi}\tilde{\chi}(\heff)\E^{-\I s \heff/\epsi} 
+\Or(\epsi\sqrt{M+1}\sqrt{\log(\sigma^{-1})})\,.
\end{eqnarray*}
It follows from the proof of lemma \ref{approxchi} (in particular equation \eqref{qnhonechichi} and what follows) that
\begin{equation*}
\qm h_{1,\chi}\tilde{\chi}(\heff)=\Or(\epsi\sqrt{M+1}\sqrt{\log(\sigma^{-1})})_{\mathcal{L}(\hilbert{H})},
\end{equation*}
so also the second estimate is proved.\qed
\end{proof}

\begin{lemma}\label{lemtruncheff}
The truncated dressed Hamiltonian
\begin{equation}
\hefftwochi:= h_{0}+\epsi h_{1,\chi}+\epsi^{2}h_{2,\chi}
\end{equation}
is self-adjoint on $\hilbert{H}_{0}$ for $\epsi$ small enough.
\end{lemma}

\begin{proof}
By construction, the coefficients $h_{i,\chi}$ belong to $\mathcal{L}(\hilbert{H}_{0}, \hilbert{H})$, and define symmetric operators on $\hilbert{H}_{0}$. Moreover $h_{0}$ is self-adjoint on $\hilbert{H}_{0}$, therefore
\begin{equation*}
\norm{(\hefftwochi -h_{0})\Psi}_{\hilbert{H}}\leq C\epsi (\norm{h_{0}\Psi}_{\hilbert{H}} + \norm{\psi}_{\hilbert{H}}), \quad\forall\Psi\in\hilbert{H}_{0}. 
\end{equation*}
By a symmetric version of the Kato theorem (\cite{ReSi2}, theorem X.13) the claim follows.\qed
\end{proof}

\begin{theorem}\label{firstorderapprox}\emph{(First order approximation to the time evolution)} Given a function $\tilde{\chi}\in\coinf(\field{R}{})$,
\begin{eqnarray}\label{firstordertime} 
\lefteqn{\E^{-\I t\heff/\epsi}\qm\tilde{\chi}(\heff)=}\nonumber\\
&= &\E^{-\I t\hdiag/\epsi}\qm\tilde{\chi}(\heff)-\I\epsi\int_{0}^{t}ds\ \E^{\I(s-t)h_{0}/\epsi}h_{2, \mathrm{OD}}\E^{-\I sh_{0}/\epsi}\qm\tilde{\chi}(\heff) \nonumber\\
&&+\, \Or(\epsi^{3/2}\abs{t})_{\mathcal{L}(\hilbert{H})}(1-\delta_{M0}) + \Or\big(\epsi^{2}(\abs{t} + \abs{t}^{2})\sqrt{\log(\sigma^{-1})}\big)_{\mathcal{L}(\hilbert{H})}\,,
\end{eqnarray}
where $\delta_{M0}=1$ when $M=0$, $0$ otherwise, $[\hdiag, \qm]=0\,\,\, \forall\, M$,
\begin{eqnarray}\label{hdiag} 
\hdiag&:=& \sum_{l=1}^{N}\frac{1}{2m_{l}^{\epsi}}\popb_{l}^{2}+E(x)+ \hfield \\
&&-\,\frac{\epsi^{2}}{4}\underset{(l_{1}\neq j_{1})}{\sum_{l_{1}, j_{1}=1}^{N}}\int_{\field{R}{3}}dk\, \frac{\hat{\varrho}_{l_{1}}(k)^{*}\hat{\varrho}_{j_{1}}(k)}{\abs{k}^{2}}\big[\E^{\I k\cdot(x_{j_{1}}-x_{l_{1}})}(\kappa\cdot\popb_{l_{1}})(\kappa\cdot\popb_{j_{1}}\nonumber)\\& &\hspace{3cm}+\,(\kappa\cdot\popb_{l_{1}})(\kappa\cdot\popb_{j_{1}})\E^{\I k\cdot(x_{j_{1}}-x_{l_{1}})}\big]\,,\nonumber
\end{eqnarray}
with $m_{l}^{\epsi}:=1/(1+\frac{\epsi^{2}}{2}e_{l})$, $\kappa:= k/\abs{k}$ and 
\begin{equation}
e_{l}:= \frac{1}{4\pi}\int_{\field{R}{3}\times\field{R}{3}}dx\ dy\ \frac{\varrho_{l}(x)\varrho_{l}(y)}{\abs{x-y}} \, .
\end{equation}
The off-diagonal Hamiltonian is defined by
\begin{equation}\begin{split}
& h_{2, \mathrm{OD}}:= \phisigma\cdot\nabla E(x) \, .
\end{split}
\end{equation}
\end{theorem}

\begin{proof}
We split the proof into three parts. In the first one, we show that equation \eqref{firstordertime} is true with a diagonal Hamiltonian $\hdiagtilde$ given by
\begin{eqnarray}\label{hdiagtilde} 
\hdiagtilde&:=& \sum_{l=1}^{N}\frac{1}{2m_{l}^{\epsi}}\popb_{l}^{2}+E(x)+ \hfield + \frac{\epsi^{2}}{4}\sum_{j=1}^{3N}a(\I\deriv{j}{}v_{\sigma})^{*}a(\I\deriv{j}{}v_{\sigma})\\
&&-\,\frac{\epsi^{2}}{4}\underset{(l_{1}\neq j_{1})}{\sum_{l_{1}, j_{1}=1}^{N}}\int_{\field{R}{3}}dk\, \frac{\hat{\varrho}_{l_{1}}(k)^{*}\hat{\varrho}_{j_{1}}(k)}{\abs{k}^{2}}\big[\E^{\I k\cdot(x_{j_{1}}-x_{l_{1}})}(\kappa\cdot\popb_{l_{1}})(\kappa\cdot\popb_{j_{1}}\nonumber)\\& &\hspace{3cm}+\,(\kappa\cdot\popb_{l_{1}})(\kappa\cdot\popb_{j_{1}})\E^{\I k\cdot(x_{j_{1}}-x_{l_{1}})}\big]\,,\nonumber
\end{eqnarray}
and an off-diagonal one $\tilde{h}_{2, \mathrm{OD}}$ defined by
\begin{equation}\label{htwotildeod}\begin{split} 
\tilde{h}_{2, \mathrm{OD}} := \phisigma\cdot\nabla E(x) &- \frac{1}{2}\sum_{j, l =1}^{3N} (\deriv{j}{}\Phi_{0, l}\pop_{l}\pop_{j} + \pop_{l}\pop_{j}\deriv{j}{}\Phi_{0, l})\\ &+ \frac{1}{4} \sum_{j=1}^{3N}\big[a(\I\deriv{j}{}v_{\sigma})^{2} + a(\I\deriv{j}{}v_{\sigma})^{*\,2}] \, .
\end{split}
\end{equation}
In the second part we prove that if one neglects the term
\begin{equation*}
\frac{\epsi^{2}}{4}\sum_{j=1}^{3N}a(\I\deriv{j}{}v_{\sigma})^{*}a(\I\deriv{j}{}v_{\sigma})
\end{equation*}
in $\hdiagtilde$, one gets an error of order $\Or(\epsi^{3/2}\abs{t})$ in the time evolution. Note that this term is exactly zero if the initial state for the field is the Fock vacuum. In the third part, we prove analogously that we can replace $\tilde{h}_{2, \mathrm{OD}}$ with $h_{2, \mathrm{OD}}$. 

More specifically, the terms which we neglect in $\hdiagtilde$ and $\tilde{h}_{2, \mathrm{OD}}$ give rise to higher order contributions to the time evolution, although their norm in $\mathcal{L}(\hilbert{H}_{0}, \hilbert{H})$ is not small. This is caused by the fact that they are strongly oscillating in $\abs{k}$, so that their behavior is determined by the value of the density of states in a neighborhood of $k=0$. For all these terms, the density however vanishes for $k=0$, uniformly in $\sigma$, and this implies that they are of lower order with respect to the leading piece $\phisigma\cdot\nabla E(x)$, whose density instead diverges logarithmically in $\sigma$ (we elaborate on this last observation in a corollary to this theorem). 

We start showing that we can, up to the desired error, replace $\heff$ by $\hefftwochi$. 
By lemma \ref{lemtruncheff}, $\hefftwochi$ is self-adjoint on $\hilbert{H}_{0}$ like $\heff$, therefore we can apply the Duhamel formula and use theorem \ref{thmheff} to get
\begin{eqnarray*}
\E^{-\I t\heff/\epsi}-\E^{-\I t\hefftwochi/\epsi}& = &-\frac{\I}{\epsi}\int_{0}^{t}ds\, \E^{\I (s-t)\heff/\epsi}(\heff-\hefftwochi)\E^{-\I s \hefftwochi/\epsi}\\
&=&\Or(\epsi^{2}\big(\log(\sigma^{-1})\big)^{3/2})\,.
\end{eqnarray*}
Moreover, using lemma \ref{approxchi}, we can replace $\qm\tilde{\chi}(\heff)$ by $\qm\tilde{\chi}(h_{0})\tilde{\tilde{\chi}}(\heff)$.

Since the diagonal Hamiltonian $\hdiagtilde$ is also self-adjoint on $\hilbert{H}_{0}$ for $\epsi$ sufficiently small (the proof can be given along the same lines of lemma \ref{lemtruncheff}), we apply again the Duhamel formula, 
\begin{eqnarray*}
\lefteqn{ (\E^{-\I t\hefftwochi/\epsi}-\E^{-\I t\hdiagtilde/\epsi})\qm\tilde{\chi}(h_{0})\tilde{\tilde{\chi}}(\heff) =}\\ &=&-\,\frac{\I}{\epsi}\int_{0}^{t}ds\ \E^{\I(s-t)\hefftwochi/\epsi}(\hefftwochi-\hdiagtilde)\E^{-\I s\hdiagtilde/\epsi}\qm\tilde{\chi}(h_{0})\tilde{\tilde{\chi}}(\heff)\\
&=&-\,\I\int_{0}^{t}ds\ \E^{\I(s-t)\hefftwochi/\epsi}h_{1,\chi}\E^{-\I s\hdiagtilde/\epsi}\qm\tilde{\chi}(h_{0})\tilde{\tilde{\chi}}(\heff)\\
&&-\,\I\epsi\int_{0}^{t}ds\ \E^{\I(s-t)\hefftwochi/\epsi}(h_{2,\chi}-\tilde{h}_{2, \mathrm{D}})\E^{-\I s\hdiagtilde/\epsi}\qm\tilde{\chi}(h_{0})\tilde{\tilde{\chi}}(\heff)\,,
\end{eqnarray*}
where 
\begin{equation*}\begin{split}
\tilde{h}_{2, \mathrm{D}} :=&  \frac{1}{4}\sum_{j=1}^{3N}a(\I\deriv{j}{}v_{\sigma})^{*}a(\I\deriv{j}{}v_{\sigma})\\
&-\,\frac{\epsi^{2}}{4}\underset{(l_{1}\neq j_{1})}{\sum_{l_{1}, j_{1}=1}^{N}}\int_{\field{R}{3}}dk\, \frac{\hat{\varrho}_{l_{1}}(k)^{*}\hat{\varrho}_{j_{1}}(k)}{\abs{k}^{2}}\big[\E^{\I k\cdot(x_{j_{1}}-x_{l_{1}})}(\kappa\cdot\popb_{l_{1}})(\kappa\cdot\popb_{j_{1}}\nonumber)\\ &\hspace{3cm}+\,(\kappa\cdot\popb_{l_{1}})(\kappa\cdot\popb_{j_{1}})\E^{\I k\cdot(x_{j_{1}}-x_{l_{1}})}\big]\, .
\end{split}
\end{equation*}
To analyze the first term, we remark that, proceeding as in lemma \ref{approxchi}, one can prove that
\begin{equation*}
\tilde{\chi}(h_{0})-\tilde{\chi}(\hdiagtilde)= \Or(\epsi^{2}\sqrt{\log(\sigma^{-1})})_{\mathcal{L}(\hilbert{H})},
\end{equation*}
so 
\begin{equation*}
[\E^{-\I t\hdiagtilde/\epsi}, \tilde{\chi}(h_{0})] = \Or(\epsi^{2}\sqrt{\log(\sigma^{-1})})_{\mathcal{L}(\hilbert{H})},
\end{equation*}
therefore, with  equation \eqref{honeqnchi},
\begin{eqnarray*}
\lefteqn{\hspace{-1cm}-\I\int_{0}^{t}ds\ \E^{\I(s-t)\hefftwochi/\epsi}h_{1,\chi}\E^{-\I s\hdiagtilde/\epsi}\qm\tilde{\chi}(h_{0})\tilde{\tilde{\chi}}(\heff)=}\\
&=&-\,\I\int_{0}^{t}ds\ \E^{\I(s-t)\hefftwochi/\epsi}h_{1,\chi}\qm\tilde{\chi}(h_{0})\E^{-\I s\hdiagtilde/\epsi}\tilde{\tilde{\chi}}(\heff) \\
&&+\, \Or(\epsi^{2}\abs{t}\sqrt{\log(\sigma^{-1})})_{\mathcal{L}(\hilbert{H})}=\Or(\epsi^{2}\abs{t}\sqrt{\log(\sigma^{-1})})_{\mathcal{L}(\hilbert{H})}\,.
\end{eqnarray*}
Concerning the second one, applying once again the Duhamel formula, we have
\begin{eqnarray*}
\lefteqn{ -\I\epsi\int_{0}^{t}ds\ \E^{\I(s-t)\hefftwochi/\epsi}(h_{2,\chi}-\tilde{h}_{2, \mathrm{D}})\E^{-\I s\hdiagtilde/\epsi}\qm\tilde{\chi}(h_{0})\tilde{\tilde{\chi}}(\heff)=}\\
&=& -\,\I\epsi\int_{0}^{t}ds\ \E^{\I(s-t)\hefftwochi/\epsi}(h_{2,\chi}-\tilde{h}_{2, \mathrm{D}})\E^{-\I sh_{0}/\epsi}\qm\tilde{\chi}(h_{0})\tilde{\tilde{\chi}}(\heff) \\
&&+\, \Or(\epsi^{2}\abs{t}^{2}\sqrt{\log(\sigma^{-1})})_{\mathcal{L}(\hilbert{H})}\,,
\end{eqnarray*}
so we have to look at $\E^{\I(s-t)\hefftwochi/\epsi}(h_{2,\chi}-\tilde{h}_{2, \mathrm{D}})\qm\tilde{\chi}(h_{0})$. 
Following a procedure already employed several times, we first observe that, in the expression for $h_{2, \chi}$, equation \eqref{htwochi}, we can replace, making an error of order $\Or(\epsi)$, $\chi(\vsigma^{*}\hepsisigma \vsigma)$ with $\chi(h_{0})$.

Using the corollary to lemma \ref{lemmapseudo}, we can then eliminate from $h_{2, \chi}\qm\tilde{\chi}(h_{0})$ all the terms containing $(\mathbf{1}-\chi(h_{0}))$.
What remains is then
\begin{eqnarray*}
\lefteqn{\E^{\I(s-t)\hefftwochi/\epsi}(h_{2,\chi}-\tilde{h}_{2, \mathrm{D}})\qm\tilde{\chi}(h_{0})=}\\
&=&\E^{\I(s-t)\hefftwochi/\epsi}\sum_{i=-2}^{2}\mathrm{Q}_{M+i}\xi(h_{0})\bigg\{\frac{1}{4}\sum_{j=1}^{3N}\big[a(\I\deriv{j}{}v_{\sigma})^{2} + a(\I\deriv{j}{}v_{\sigma})^{*\,2}]+\phisigmaj\cdot\nabla E(x)\\ &&-\,\deriv{j}{}\phizerojl\pop_{l}\pop_{j}-\frac{1}{2}\sum_{l_{1}=1}^{N}e_{l_{1}}\popb_{l_{1}}^{2}-\frac{1}{4}\underset{(l_{1}\neq j_{1})}{\sum_{l_{1}, j_{1}=1}^{N}}\int_{\field{R}{3}}dk\, \frac{\hat{\varrho}_{l_{1}}(k)^{*}\hat{\varrho}_{j_{1}}(k)}{\abs{k}^{2}}\cdot\\
&&\cdot\big[\E^{\I k\cdot(x_{j_{1}}-x_{l_{1}})}(\kappa\cdot\popb_{l_{1}})(\kappa\cdot\popb_{j_{1}})+ (\kappa\cdot\popb_{l_{1}})(\kappa\cdot\popb_{j_{1}})\E^{\I k\cdot(x_{j_{1}}-x_{l_{1}})}\big]\bigg\}\qm\tilde{\chi}(h_{0})\\
&&+\,\frac{\I}{2}\E^{\I(s-t)\hefftwochi/\epsi}\hspace{-2pt}\sum_{i=-2}^{2}\hspace{-1pt}\mathrm{Q}_{M+i}\xi(h_{0})\big[\chi(h_{0})\phim(\I\nabla_{x}v_{\sigma})\cdot\pop,\ \chi(h_{0})\phisigmaj\cdot\pop \big]\qm\tilde{\chi}(h_{0})   \\
&&+\,\Or(\epsi\sqrt{\log(\sigma^{-1})})_{\mathcal{L}(\hilbert{H})}.
\end{eqnarray*}
Applying theorem \ref{zeroordertime}, we can now replace $\E^{\I(s-t)\hefftwochi/\epsi}$ with $\E^{\I(s-t)h_{0}/\epsi}$ making an error of order $\Or(\abs{t}\epsi\sqrt{\log(\sigma^{-1})})$. 

Concerning the last term, we get, expanding the commutator and applying again the corollary to lemma \ref{lemmapseudo},
\begin{eqnarray*}
\lefteqn{\big[\chi(h_{0})\phim(\I\nabla_{x}v_{\sigma})\cdot\pop,\ \chi(h_{0})\phisigmaj\cdot\pop \big]\qm\tilde{\chi}(h_{0}) =}\\&=&  \sum_{i=-1}^{1}\mathrm{Q}_{M+i}\xi(h_{0}) \big[\phim(\I\nabla_{x}v_{\sigma})\cdot\pop,\ \phisigmaj\cdot\pop \big]\qm\tilde{\chi}(h_{0})\\
&=&\sum_{i=-1}^{1}\mathrm{Q}_{M+i}\xi(h_{0})\big[\phim(\I\deriv{l}{}v_{\sigma}),\ \phisigmajl\big]\pop_{l}\pop_{j}\qm\tilde{\chi}(h_{0})+\Or(\epsi\sqrt{\log(\sigma^{-1})})_{\mathcal{L}(\hilbert{H})}\,.
\end{eqnarray*}

Using equation \eqref{commaphi}, we can check that this cancels exactly the diagonal terms in $h_{2, \chi}-\tilde{h}_{2, \mathrm{D}}$, so that
\begin{eqnarray*}
\lefteqn{ \E^{\I(s-t)\hefftwochi/\epsi}(h_{2,\chi}-\tilde{h}_{2, \mathrm{D}})\qm\tilde{\chi}(h_{0})=}\\&=&\E^{\I(s-t)h_{0}/\epsi}\tilde{h}_{2, \mathrm{OD}}\qm\tilde{\chi}(h_{0}) +
\Or(\epsi\abs{t}\sqrt{\log(\sigma^{-1})})\,.
\end{eqnarray*}

We proceed now to show that we can replace $\hdiagtilde$ with $\hdiag$, up to an error of order $\Or(\epsi^{3/2}\abs{t})_{\mathcal{L}(\hilbert{H})}$.

Applying repeatedly Duhamel formula, and putting $\Psi:= \qm\tilde{\chi}(\heff)\Psi_{0}$, we get
\begin{equation*}\begin{split}
(\E^{-\I t\hdiagtilde/\epsi} - \E^{-\I t\hdiag/\epsi})\Psi =& -\frac{\I\epsi}{4}\sum_{j=1}^{3N}\int_{0}^{t}ds\, \E^{\I(s-t)\hiepsi{0}/\epsi}a(\I\deriv{j}{}v_{\sigma})^{*}a(\I\deriv{j}{}v_{\sigma})\E^{-\I s\hiepsi{0}/\epsi}\Psi \\
&+\Or(\epsi^{2}\abs{t}^{2})_{\mathcal{L}(\hilbert{H})}.
\end{split}
\end{equation*}
To streamline the presentation, we assume that $M=1$, the calculations for $M>1$ are basically the same, but more cumbersome.

The time integral gives 
\begin{equation}\label{estimatetermm1}\begin{split}
&\E^{-\I t\hiepsi{0}}\frac{\hat{\varphi}_{\sigma}(k_{1})}{\abs{k_{1}}^{1/2}}\sum_{j=1}^{N}e_{j}^{2}\int_{\field{R}{3}}dk\, (\kappa_{1}\cdot\kappa) \frac{\hat{\varphi}_{\sigma}(k)}{\abs{k}^{1/2}}\cdot\\
&\cdot\int_{0}^{t}ds\, \E^{\I s(\abs{k_{1}}-\abs{k})/\epsi}\E^{\I s\hiepsi{\mathrm{p}}/\epsi}\E^{\I x_{j}\cdot(k_{1}-k)}\E^{-\I s\hiepsi{\mathrm{p}}/\epsi}\Psi(x, k) =\\
& = \E^{-\I t\hiepsi{0}}\frac{\hat{\varphi}_{\sigma}(k_{1})}{\abs{k_{1}}^{1/2}}\sum_{j=1}^{N}e_{j}^{2}\int_{\field{R}{3}}dk\,  \frac{(\kappa_{1}\cdot\kappa)\hat{\varphi}_{\sigma}(k)}{\abs{k}^{1/2}[1 + \I(\abs{k_{1}}-\abs{k})\epsi^{-1}]}[1 + \I(\abs{k_{1}}-\abs{k})\epsi^{-1}]\cdot\\
&\cdot\int_{0}^{t}ds\, \E^{\I s(\abs{k_{1}}-\abs{k})/\epsi}\E^{\I s\hiepsi{\mathrm{p}}/\epsi}\E^{\I x_{j}\cdot(k_{1}-k)}\E^{-\I s\hiepsi{\mathrm{p}}/\epsi}\Psi(x, k).
\end{split}
\end{equation} 
Integrating by parts we get
\begin{equation*}\begin{split}
& \I(\abs{k_{1}}-\abs{k})\epsi^{-1}\int_{0}^{t}ds\, \E^{\I s(\abs{k_{1}}-\abs{k})/\epsi}\E^{\I s\hiepsi{\mathrm{p}}/\epsi}\E^{\I x_{j}\cdot(k-k_{1})}\E^{-\I s\hiepsi{\mathrm{p}}/\epsi}\Psi(x, k)=\\
&=\E^{\I t(\abs{k_{1}}-\abs{k})/\epsi}\E^{\I t\hiepsi{\mathrm{p}}/\epsi}\E^{\I x_{j}\cdot(k-k_{1})}\E^{-\I t\hiepsi{\mathrm{p}}/\epsi}\Psi - \E^{\I x_{j}\cdot(k-k_{1})}\Psi +\\
&- \frac{\I}{\epsi}\int_{0}^{t}ds\, \E^{\I s(\abs{k_{1}}-\abs{k})/\epsi}\E^{\I s\hiepsi{\mathrm{p}}}[\hiepsi{\mathrm{p}}, \E^{\I x_{j}\cdot(k-k_{1})}]\E^{-\I s\hiepsi{\mathrm{p}}}\Psi,
\end{split}
\end{equation*}
where the commutator is of order $\Or(\epsi)$ when applied to functions of bounded kinetic energy, so that the right-hand side is uniformly bounded in $\epsi$.

We have now to put this expression back in \eqref{estimatetermm1} and estimate the single terms. We show how to do this for the first one, the others being entirely analogous. We ignore the unitary on the left, which does not change the norm, so we have to consider
\begin{equation*}\begin{split}
&\frac{\hat{\varphi}_{\sigma}(k_{1})}{\abs{k_{1}}^{1/2}}e_{j}^{2}\int_{\field{R}{3}}dk\,  \frac{(\kappa_{1}\cdot\kappa)\hat{\varphi}_{\sigma}(k)}{\abs{k}^{1/2}[1 + \I(\abs{k_{1}}-\abs{k})\epsi^{-1}]}\cdot\\
&\cdot\int_{0}^{t}ds\, \E^{\I s(\abs{k_{1}}-\abs{k})/\epsi}\E^{\I s\hiepsi{\mathrm{p}}/\epsi}\E^{\I x_{j}\cdot(k_{1}-k)}\E^{-\I s\hiepsi{\mathrm{p}}/\epsi}\Psi(x, k).
\end{split}
\end{equation*}
for one fixed $j$.
Using twice the Cauchy-Schwarz inequality we get (we put for brevity $f(k) := e_{j}\varphi(k)\abs{k}^{-1/2}$)
\begin{equation*}\begin{split}
& \norm{\cdots}_{\hilbert{H}}^{2} = \int_{\field{R}{3N}}dx\int_{\field{R}{3}}dk_{1}\abs{\cdots}^{2}\leq\abs{t}\int dx\int dk_{1}dk\,\abs{f(k_{1})}^{2}\frac{\abs{f(k)}^{2}}{1 + (\abs{k_{1}} - \abs{k})^{2}\epsi^{-2}}\cdot\\
&\cdot \int dk\int_{0}^{t}ds\, \bigg\lvert\E^{\I s\hiepsi{\mathrm{p}}}\E^{\I x_{j}\cdot(k-k_{1})}\E^{-\I s\hiepsi{\mathrm{p}}}\Psi(x, k)\bigg\rvert^{2}=\\
&=\abs{t}^{2}\norm{\Psi}_{\hilbert{H}}^{2}\int dk_{1}dk\, \frac{\abs{f(k, \lambda)f(k_{1}, \lambda_{1})}^{2}}{1 + (\abs{k_{1}} - \abs{k})^{2}\epsi^{-2}}\leq\\ &\leq C\epsi^{4}\abs{t}^{2}\norm{\Psi}_{\hilbert{H}}^{2}\int_{0}^{\Lambda/\epsi}dk_{1}\int_{0}^{\Lambda/\epsi}dk\, \frac{k_{1}k}{1 + (k_{1} - k)^{2}} = \Or(\epsi\abs{t}^{2}\norm{\Psi}^{2}_{\hilbert{H}}).
\end{split}
\end{equation*}

We examine now separately the last two terms in $\tilde{h}_{2, \mathrm{OD}}$. For the first we get
\begin{eqnarray*}
\lefteqn{ \I\epsi\int_{0}^{t}ds\ \E^{\I(s-t)h_{0}/\epsi}\deriv{j}{}\Phi_{0, l}\pop_{l}\pop_{j}\E^{-\I sh_{0}/\epsi}\qm\tilde{\chi}(\heff)\Psi_{0}=}\\
& =&\I\E^{-\I th_{0}/\epsi}\frac{\epsi}{\sqrt{2}}\int_{0}^{t}ds\ \E^{\I sh_{0}/\epsi}\bigg[a\bigg(\frac{\deriv{l}{}\deriv{j}{}v(x, k)}{\abs{k}}\bigg)^{*} + a\bigg(\frac{\deriv{l}{}\deriv{j}{}v(x, k)}{\abs{k}}\bigg)\bigg]\pop_{j}\pop_{l}\E^{-\I sh_{0}/\epsi}\Psi\\
&=&-\,\I\E^{-\I th_{0}/\epsi}\frac{\epsi}{\sqrt{2}}\int_{0}^{t}ds\, \E^{\I sh_{0}/\epsi}\bigg[a\bigg(\kappa_{j_{2}}\kappa_{l_{2}}\E^{\I k\cdot x_{j_{1}}}\frac{\hat{\varrho}_{j_{1}}(k)}{\abs{k}^{1/2}}\bigg)^{*} \\
&&\hspace{2cm}+ \,a\bigg(\kappa_{j_{2}}\kappa_{l_{2}}\E^{\I k\cdot x_{j_{1}}}\frac{\hat{\varrho}_{j_{1}}(k)}{\abs{k}^{1/2}}\bigg)\bigg]\pop_{(j_{1}, j_{2})}\pop_{(j_{1}, l_{2})}\E^{-\I sh_{0}/\epsi}\Psi \, .
\end{eqnarray*}
The part with the annihilation operator gives
\begin{eqnarray}\label{estimateapp} 
\lefteqn{\E^{-\I th_{0}/\epsi}\frac{\epsi}{\sqrt{2}}\int_{0}^{t}ds\, \E^{\I sh_{0}/\epsi}a\bigg(\kappa_{j_{2}}\kappa_{l_{2}}\E^{\I k\cdot x_{j_{1}}}\frac{\hat{\varrho}_{j_{1}}(k)}{\abs{k}^{1/2}}\bigg)\pop_{(j_{1}, j_{2})}\pop_{(j_{1}, l_{2})}\E^{-\I sh_{0}/\epsi}\Psi=}\nonumber\\
& =& \E^{-\I th_{0}/\epsi}\frac{\epsi}{\sqrt{2}}\sqrt{M}\int_{0}^{t}ds\,\E^{\I s\sum_{\mu=1}^{M-1}\abs{k_{\mu}}/\epsi}\E^{\I s h_{\mathrm{p}}/\epsi}\int_{\field{R}{3}}dk\, \E^{-\I k\cdot x_{j_{1}}}\frac{\hat{\varrho}_{j_{1}}(k)^{*}}{\abs{k}^{1/2}}(\kappa\cdot\popb_{j_{1}})^{2}\cdot\nonumber\\
&&\cdot\E^{-\I s h_{\mathrm{p}}/\epsi}\E^{-\I s\abs{k}/\epsi}\E^{-\I s\sum_{\mu=1}^{M-1}\abs{k_{\mu}}/\epsi}\Psi(x;k, k_{1}, \ldots, k_{M-1})\nonumber\\
&=&\E^{-\I th_{0}/\epsi}\frac{\epsi}{\sqrt{2}}\sqrt{M}\int_{\field{R}{3}}dk\, \frac{\hat{\varrho}_{j_{1}}(k)^{*}}{\abs{k}^{1/2}}\int_{0}^{t}ds\, \E^{\I s h_{\mathrm{p}}/\epsi}\E^{-\I k\cdot x_{j_{1}}}(\kappa\cdot\popb_{j_{1}})^{2}\E^{-\I s h_{\mathrm{p}}/\epsi}\cdot\nonumber\\
&&\cdot\E^{-\I s\abs{k}/\epsi}\Psi(x;k, k_{1}, \ldots, k_{M-1})\nonumber\\
&=&\E^{-\I th_{0}/\epsi}\frac{\epsi}{\sqrt{2}}\sqrt{M}\int_{\field{R}{3}}dk\, \frac{\hat{\varrho}_{j_{1}}(k)^{*}}{\abs{k}^{1/2}(1-\I \abs{k}\epsi^{-1})}(1-\I \abs{k}\epsi^{-1})\int_{0}^{t}ds\, \E^{\I s h_{\mathrm{p}}/\epsi}\cdot\nonumber\\
&&\cdot\E^{-\I k\cdot x_{j_{1}}}(\kappa\cdot\popb_{j_{1}})^{2}\E^{-\I s h_{\mathrm{p}}/\epsi}\E^{-\I s\abs{k}/\epsi}\Psi(x;k, k_{1}, \ldots, k_{M-1})\,.
\end{eqnarray}
Integrating by parts we have
\begin{eqnarray}\label{intparts} 
\lefteqn{ -\I \abs{k}\epsi^{-1}\int_{0}^{t}ds\, \E^{\I s h_{\mathrm{p}}/\epsi}\E^{-\I k\cdot x_{j_{1}}}(\kappa\cdot\popb_{j_{1}})^{2}\E^{-\I s h_{\mathrm{p}}/\epsi}\E^{-\I s\abs{k}/\epsi}\Psi(x;k, k_{1}, \ldots, k_{M-1})=}\nonumber\\
&=&\int_{0}^{t}ds\, \E^{\I s h_{\mathrm{p}}/\epsi}\E^{-\I k\cdot x_{j_{1}}}(\kappa\cdot\popb_{j_{1}})^{2}\E^{-\I s h_{\mathrm{p}}/\epsi}\deriv{s}{}\E^{-\I s\abs{k}/\epsi}\Psi(x;k, k_{1}, \ldots, k_{M-1})\nonumber\\
&=&\E^{\I t h_{\mathrm{p}}/\epsi}\E^{-\I k\cdot x_{j_{1}}}(\kappa\cdot\popb_{j_{1}})^{2}\E^{-\I t h_{\mathrm{p}}/\epsi}\E^{-\I t\abs{k}/\epsi}\Psi(x;k, k_{1}, \ldots, k_{M-1})\nonumber \\
&& -\, \E^{-\I k\cdot x_{j_{1}}}(\kappa\cdot\popb_{j_{1}})^{2}\Psi(x;k, k_{1}, \ldots, k_{M-1})\nonumber\\
&&-\,\frac{\I}{\epsi}\int_{0}^{t}ds\, \E^{\I s h_{\mathrm{p}}/\epsi}[\hp, \E^{-\I k\cdot x_{j_{1}}}(\kappa\cdot\popb_{j_{1}})^{2}]\E^{-\I s h_{\mathrm{p}}/\epsi}\E^{-\I s\abs{k}/\epsi}\Psi(x;k, k_{1}, \ldots, k_{M-1})\nonumber\\
&=&\E^{\I t h_{\mathrm{p}}/\epsi}\E^{-\I k\cdot x_{j_{1}}}(\kappa\cdot\popb_{j_{1}})^{2}\E^{-\I t h_{\mathrm{p}}/\epsi}\E^{-\I t\abs{k}/\epsi}\Psi(x;k, k_{1}, \ldots, k_{M-1})\nonumber \\
& &-\, \E^{-\I k\cdot x_{j_{1}}}(\kappa\cdot\popb_{j_{1}})^{2}\Psi(x;k, k_{1}, \ldots, k_{M-1})\nonumber\\
&&-\,\I\int_{0}^{t}ds\, \E^{\I s h_{\mathrm{p}}/\epsi}\bigg\{\bigg[-\I\nabla_{x_{j_{1}}}(\E^{-\I k\cdot x_{j_{1}}})\cdot\popb_{j_{1}} - \epsi\Delta_{x}(\E^{-\I k\cdot x_{j_{1}}})\bigg](\kappa\cdot\popb_{j_{1}})^{2}\nonumber \\ 
&&+ \,\E^{-\I k\cdot x_{j_{1}}}[E(x), (\kappa\cdot\popb_{j_{1}})^{2}]\bigg]\bigg\} \E^{-\I s h_{\mathrm{p}}/\epsi}\E^{-\I s\abs{k}/\epsi}\Psi(x;k, k_{1}, \ldots, k_{M-1})\nonumber\\
&=& \E^{\I t h_{\mathrm{p}}/\epsi}\E^{-\I k\cdot x_{j_{1}}}(\kappa\cdot\popb_{j_{1}})^{2}\E^{-\I t h_{\mathrm{p}}/\epsi}\E^{-\I t\abs{k}/\epsi}\Psi(x;k, k_{1}, \ldots, k_{M-1})\nonumber \\
&& -\, \E^{-\I k\cdot x_{j_{1}}}(\kappa\cdot\popb_{j_{1}})^{2}\Psi(x;k, k_{1}, \ldots, k_{M-1})\nonumber\\
&&+\,\I\int_{0}^{t}ds\, \E^{\I s h_{\mathrm{p}}/\epsi}\bigg\{\bigg[\E^{-\I k\cdot x_{j_{1}}}\kappa\cdot\popb_{j_{1}} - \epsi\abs{k}\E^{-\I k\cdot x_{j_{1}}}\bigg](\kappa\cdot\popb_{j_{1}})^{2} \nonumber\\ 
&&+\, \E^{-\I k\cdot x_{j_{1}}}[E(x), (\kappa\cdot\popb_{j_{1}})^{2}]\bigg]\bigg\} \E^{-\I s h_{\mathrm{p}}/\epsi}\E^{-\I s\abs{k}/\epsi}\abs{k}\Psi(x;k, k_{1}, \ldots, k_{M-1})\, .
\end{eqnarray}
We have to put now this expression back into \eqref{estimateapp} and estimate the result. We show how to proceed for the most singular term, i.\ e. the one containing
\begin{equation*}
\I\int_{0}^{t}ds\, \E^{\I s h_{\mathrm{p}}/\epsi}\E^{-\I k\cdot x_{j_{1}}}(\kappa\cdot\popb_{j_{1}})^{3}\E^{-\I s h_{\mathrm{p}}/\epsi}\E^{-\I s\abs{k}/\epsi}\abs{k}\Psi(x;k, k_{1}, \ldots, k_{M-1}),
\end{equation*}  
the others can be treated in the same way.
Putting this term back in \eqref{estimateapp}, and ignoring the unitary $\E^{-\I th_{0}/\epsi}$, which does not change the norm, we have to estimate in the end
\begin{eqnarray*}
\lefteqn{ \hspace{-3cm}\frac{\I\epsi}{\sqrt{2}}\sqrt{M}\int_{\field{R}{3}}dk\, \frac{\hat{\varrho}_{j_{1}}(k)^{*}}{\abs{k}^{1/2}(1-\I \abs{k}\epsi^{-1})}\int_{0}^{t}ds\, \E^{\I s h_{\mathrm{p}}/\epsi}\E^{-\I k\cdot x_{j_{1}}}(\kappa\cdot\popb_{j_{1}})^{3}\E^{-\I s h_{\mathrm{p}}/\epsi}\cdot}\\&
&\cdot\E^{-\I s\abs{k}/\epsi}\abs{k}\Psi(x;k, k_{1}, \ldots, k_{M-1})\,.
\end{eqnarray*}
Applying Cauchy-Schwarz inequality we get
\begin{eqnarray*}
 \abs{\cdots}^{2}&\leq& \frac{\epsi^{2}\abs{t}^{2}M}{2}\int_{\field{R}{3}}dk\, \frac{\abs{\hat{\varrho}_{j_{1}}(k)}^{2}}{\abs{k}(1+\abs{k}^{2}\epsi^{-2})}\cdot\\
&&\cdot\int_{\field{R}{3}}dk\, \int_{0}^{t}ds\, \big\lvert\E^{\I s h_{\mathrm{p}}/\epsi}\E^{-\I k\cdot x_{j_{1}}}(\kappa\cdot\popb_{j_{1}})^{3}\E^{-\I s h_{\mathrm{p}}/\epsi}\E^{-\I s\abs{k}/\epsi}\abs{k}\cdot\\
&&\cdot\Psi(x;k, k_{1}, \ldots, k_{M-1})\big\rvert^{2}.
\end{eqnarray*}
The first integral gives
\begin{eqnarray*}
 \int_{\field{R}{3}}dk\, \frac{\abs{\hat{\varrho}_{j_{1}}(k)}^{2}}{\abs{k}(1+\abs{k}^{2}\epsi^{-2})}&=& C\int_{0}^{\Lambda}dk\, \frac{k}{1 + k^{2}\epsi^{-2}}=C\epsi^{2}\int_{0}^{\Lambda/\epsi}\frac{k}{1 + k^{2}}\\[2mm]
&=&\Or(\epsi^{2}\log(1/\epsi))\,,
\end{eqnarray*}
so, calculating the norm we get
\begin{eqnarray*}
\lefteqn{ \hspace{-3mm}\norm{\cdots}_{\hilbert{H}}^{2}\leq C\epsi^{4}\log(1/\epsi)\abs{t}^{2}M\int_{0}^{t}ds\, \big\lVert \E^{\I s h_{\mathrm{p}}/\epsi}\E^{-\I k\cdot x_{j_{1}}}\cdot}\\
&&\cdot(\kappa\cdot\popb_{j_{1}})^{3}\E^{-\I s h_{\mathrm{p}}/\epsi}\E^{-\I s\abs{k}/\epsi}\abs{k}\Psi(x;k, k_{1}, \ldots, k_{M-1})\big\rVert^{2}_{\hilbert{H}}\\
&=&C\epsi^{4}\log(1/\epsi)\abs{t}^{2}M\int_{0}^{t}ds\, \big\lVert(\kappa\cdot\popb_{j_{1}})^{3}\E^{-\I s h_{\mathrm{p}}/\epsi}\abs{k}\Psi(x;k, k_{1}, \ldots, k_{M-1})\big\rVert^{2}_{\hilbert{H}}\\
&\leq& C\epsi^{4}\log(1/\epsi)\abs{t}^{2}M\int_{0}^{t}ds\, \big\lVert(\kappa\cdot\popb_{j_{1}})^{3}\E^{-\I s h_{\mathrm{p}}/\epsi}\abs{k}\qm\tilde{\chi}(\heff)\big\rVert^{2}_{\mathcal{L}(\hilbert{H})}\norm{\Psi_{0}}^{2}_{\hilbert{H}}\,,
\end{eqnarray*}
which shows that the norm of this term is of order $\Or(\epsi^{2}\abs{t}\sqrt{\log(1/\epsi)})$ in $\mathcal{L}(\hilbert{H})$.
The part with the creation operator gives
\begin{eqnarray*}
\lefteqn{ \E^{-\I th_{0}/\epsi}\frac{\epsi}{\sqrt{2}}\int_{0}^{t}ds\, \E^{\I sh_{0}/\epsi}a\bigg(\kappa_{j_{2}}\kappa_{l_{2}}\E^{\I k\cdot x_{j_{1}}}\frac{\hat{\varrho}_{j_{1}}(k)}{\abs{k}^{1/2}}\bigg)^{*}\pop_{(j_{1}, j_{2})}\pop_{(j_{1}, l_{2})}\E^{-\I sh_{0}/\epsi}\Psi=}\\
& =& \E^{-\I th_{0}/\epsi}\frac{\epsi}{\sqrt{2}}\int_{0}^{t}ds\, \E^{\I s\sum_{\mu=1}^{M+1}\abs{k_{\mu}}/\epsi} \E^{\I s\hp/\epsi}\frac{1}{\sqrt{M+1}}\sum_{\mu=1}^{M+1}\kappa_{\mu, j_{2}}\kappa_{\mu, l_{2}}\E^{\I\kappa_{\mu}\cdot x_{j_{1}}}\frac{\hat{\varrho}_{j_{1}}(k_{\mu})}{\abs{k_{\mu}}^{1/2}}\\
&&\cdot \pop_{(j_{1}, j_{2})}\pop_{(j_{1}, l_{2})}\E^{-\I s\sum_{\nu=1(\nu\neq\mu)}^{M+1}\abs{k_{\nu}}/\epsi}\E^{-\I s\hp/\epsi}\Psi(x; k_{1}, \ldots, \hat{k}_{\mu}, \ldots, k_{M+1})\\
&= &\E^{-\I th_{0}/\epsi}\frac{\epsi}{\sqrt{2(M+1)}}\frac{\hat{\varrho}_{j_{1}}(k_{\mu})}{\abs{k_{\mu}}^{1/2}}\int_{0}^{t}ds\, \E^{\I s\hp/\epsi}\E^{\I\kappa_{\mu}\cdot x_{j_{1}}}(\kappa_{\mu}\cdot\popb_{j_{1}})^{2}\E^{-\I s\hp/\epsi}\E^{-\I s\abs{k_{\mu}}/\epsi}\cdot\\
&&\cdot\Psi(x; k_{1}, \ldots, \hat{k}_{\mu}, \ldots, k_{M+1}) \\
&=&\E^{-\I th_{0}/\epsi}\frac{\epsi}{\sqrt{2(M+1)}}\frac{\hat{\varrho}_{j_{1}}(k_{\mu})}{\abs{k_{\mu}}^{1/2}(1-\I\abs{k_{\mu}}\epsi^{-1})}(1-\I\abs{k_{\mu}}\epsi^{-1})\int_{0}^{t}ds\, \E^{\I s\hp/\epsi}\E^{\I\kappa_{\mu}\cdot x_{j_{1}}}\cdot\\
&&\cdot(\kappa_{\mu}\cdot\popb_{j_{1}})^{2}\E^{-\I s\hp/\epsi}\E^{-\I s\abs{k_{\mu}}/\epsi}\Psi(x; k_{1}, \ldots, \hat{k}_{\mu}, \ldots, k_{M+1})\, .
\end{eqnarray*}
We can now integrate by parts using equation \eqref{intparts}. As for the annihilation part, we examine just one term,
\begin{eqnarray*}
\lefteqn{ \hspace{-2cm}\frac{\epsi}{\sqrt{2(M+1)}}\frac{\hat{\varrho}_{j_{1}}(k_{\mu})\abs{k_{\mu}}^{1/2}}{(1-\I\abs{k_{\mu}}\epsi^{-1})}\int_{0}^{t}ds\, \E^{\I s h_{\mathrm{p}}/\epsi}\E^{\I k_{\mu}\cdot x_{j_{1}}}(\kappa_{\mu}\cdot\popb_{j_{1}})^{3}\E^{-\I s h_{\mathrm{p}}/\epsi}\cdot}\\
&&\cdot\E^{-\I s\abs{k_{\mu}}/\epsi}\Psi(x;k_{1}, \ldots, \hat{k}_{\mu}, \ldots, k_{M+1})\,.
\end{eqnarray*}
The norm squared is given by
\begin{eqnarray*}
\lefteqn{ \frac{\epsi^{2}}{2(M+1)}\int dx\int dk_{1}\ldots dk_{M+1} \frac{\abs{\hat{\varrho}_{j_{1}}(k_{\mu})}^{2}\abs{k_{\mu}}}{1+\abs{k_{\mu}}^{2}\epsi^{-2}}\bigg\lvert\int_{0}^{t}ds\,\E^{-\I s\abs{k_{\mu}}/\epsi}\E^{\I s\hp/\epsi}\E^{\I k_{\mu}\cdot x_{j_{1}}}\cdot}\\
&&\cdot (\kappa_{\mu}\cdot\popb_{j_{1}})^{3}\E^{-\I s\hp/\epsi}\Psi(x;k_{1}, \ldots, \hat{k}_{\mu}, \ldots, k_{M+1})\bigg\rvert^{2}\\
&=&\frac{\epsi^{6}}{2(M+1)}\int \hspace{-1pt}dx\hspace{-1pt}\int\hspace{-1pt} dk_{1}\ldots dk_{M+1} \frac{\abs{\hat{\varrho}_{j_{1}}(\epsi k_{\mu})}^{2}\abs{k_{\mu}}}{1+\abs{k_{\mu}}^{2}}\bigg\lvert\int_{0}^{t}ds\,\E^{-\I s\abs{k_{\mu}}}\E^{\I s\hp/\epsi}\E^{\I \epsi k_{\mu}\cdot x_{j_{1}}}\hspace{-1pt}\cdot\\
&&\cdot (\kappa_{\mu}\cdot\popb_{j_{1}})^{3}\E^{-\I s\hp/\epsi}\Psi(x;k_{1}, \ldots, \hat{k}_{\mu}, \ldots, k_{M+1})\bigg\rvert^{2}\\
&\leq& \frac{\epsi^{6}\abs{t}}{2(M+1)}\int dk_{1}\ldots dk_{M+1} \frac{\abs{\hat{\varrho}_{j_{1}}(\epsi k_{\mu})}^{2}\abs{k_{\mu}}}{1+\abs{k_{\mu}}^{2}}\int_{0}^{t}ds\,\big\lVert\E^{-\I s\abs{k_{\mu}}}\E^{\I s\hp/\epsi}\E^{\I \epsi k_{\mu}\cdot x_{j_{1}}}\cdot\\
&&\cdot (\kappa_{\mu}\cdot\popb_{j_{1}})^{3}\E^{-\I s\hp/\epsi}\Psi(x;k_{1}, \ldots, \hat{k}_{\mu}, \ldots, k_{M+1})\big\rVert^{2}\\
&=&\frac{\epsi^{6}\abs{t}}{2(M+1)}\int dk_{1}\ldots dk_{M+1} \frac{\abs{\hat{\varrho}_{j_{1}}(\epsi k_{\mu})}^{2}\abs{k_{\mu}}}{1+\abs{k_{\mu}}^{2}}\int_{0}^{t}ds\,\big\lVert\E^{-\I s\abs{k_{\mu}}}\E^{\I s\hp/\epsi}\E^{\I \epsi k_{\mu}\cdot x_{j_{1}}}\cdot\\
&&\cdot (\kappa_{\mu}\cdot\popb_{j_{1}})^{3}\E^{-\I s\hp/\epsi}\Psi(x;k_{1}, \ldots, \hat{k}_{\mu}, \ldots, k_{M+1})\big\rVert^{2}_{L^{2}(\field{R}{3N}, dx)}\\
&=&\frac{\epsi^{6}\abs{t}}{2(M+1)}\int dk_{1}\ldots dk_{M+1} \frac{\abs{\hat{\varrho}_{j_{1}}(\epsi k_{\mu})}^{2}\abs{k_{\mu}}}{1+\abs{k_{\mu}}^{2}}\int_{0}^{t}ds\,\big\lVert(\kappa_{\mu}\cdot\popb_{j_{1}})^{3}\E^{-\I s\hp/\epsi}\cdot\\
&&\cdot\Psi(x;k_{1}, \ldots, \hat{k}_{\mu}, \ldots, k_{M+1})\big\rVert^{2}_{L^{2}(\field{R}{3N}, dx)}\\
&\leq &\frac{C\epsi^{6}\abs{t}}{2(M+1)}\int_{0}^{\varrho_{0}/\epsi} d\abs{k_{\mu}} \frac{\abs{k_{\mu}}^{3}}{1+\abs{k_{\mu}}^{2}}\int d\Omega_{\mu}\int_{0}^{t}ds\,\big\lVert(\kappa_{\mu}\cdot\popb_{j_{1}})^{3}\E^{-\I s\hp/\epsi}\cdot\\
&&\cdot\Psi(x;k_{1}, \ldots, \hat{k}_{\mu}, \ldots, k_{M+1})\big\rVert^{2}_{\hilbert{H}}\\
&=&\frac{C\epsi^{6}\abs{t}\varrho_{0}^{2}}{4(M+1)}[\epsi^{-2}-\log(\epsi^{-2}+1)]\int d\Omega_{\mu}\int_{0}^{t}ds\,\big\lVert(\kappa_{\mu}\cdot\popb_{j_{1}})^{3}\E^{-\I s\hp/\epsi}\cdot\\
&&\cdot\qm\tilde{\chi}(\heff)\Psi_{0}\big\rVert^{2}_{\hilbert{H}},
\end{eqnarray*}
so we get that this term gives a contribution of order $\Or(\epsi^{2}\abs{t})$ in the norm of $\mathcal{L}(\hilbert{H})$.

We separate also in the second term of $\tilde{h}_{2, \mathrm{OD}}$ the annihilation and the creation part. For the annihilation part we get
\begin{eqnarray*}
\lefteqn{ -\frac{\I\epsi}{4}\int_{0}^{t}ds\ \E^{\I(s-t)h_{0}/\epsi}\sum_{j=1}^{3N}a(\I\deriv{j}{}v_{\sigma})^{2}\E^{-\I sh_{0}/\epsi}\Psi=}\\
&=&-\frac{\I\epsi}{4}\sqrt{M(M-1)}\E^{-\I th_{0}/\epsi}\sum_{j_{1}, j_{2}}\int d\xi\int d\zeta\, \hat{\xi}_{j_{2}}\hat{\zeta}_{j_{2}} \frac{\hat{\varrho}_{j_{1}}^{\sigma}(\xi)^{*}\hat{\varrho}_{j_{1}}^{\sigma}(\zeta)^{*}}{\abs{\xi}^{1/2}\abs{\zeta}^{1/2}}\cdot\\ &&\cdot\int_{0}^{t}ds\ \E^{-\I s(\abs{\xi}+\abs{\zeta})/\epsi} \E^{\I s\hp/\epsi}\E^{-\I(\xi+\zeta)\cdot x_{j_{1}}}\E^{-\I s\hp/\epsi}\Psi \, .
\end{eqnarray*}
We proceed now in the same way as we did for the first term. Integrating by parts we get
\begin{eqnarray*}\lefteqn{
\frac{-\I(\abs{\xi}+\abs{\zeta})}{\epsi}\int_{0}^{t}ds\, \E^{-\I s(\abs{\xi}+\abs{\zeta})/\epsi} \E^{\I s\hp/\epsi}\E^{-\I(\xi+\zeta)\cdot x_{j_{1}}}\E^{-\I s\hp/\epsi}\Psi=}\\
&=&\E^{-\I t(\abs{\xi}+\abs{\zeta})/\epsi} \E^{\I t\hp/\epsi}\E^{-\I(\xi+\zeta)\cdot x_{j_{1}}}\E^{-\I t\hp/\epsi}\Psi - \E^{-\I(\xi+\zeta)\cdot x_{j_{1}}}\Psi \\
&&-\,\int_{0}^{t}ds\, \E^{-\I s(\abs{\xi}+\abs{\zeta})/\epsi} \E^{\I s\hp/\epsi}\E^{-\I(\xi+\zeta)\cdot x_{j_{1}}}[(\xi+\zeta)\cdot\popb_{j_{1}} - \epsi\abs{\xi+\zeta}^{2}]\E^{-\I s\hp/\epsi}\Psi\,.
\end{eqnarray*}
As in the previous case, we examine just one term, 
\begin{eqnarray*}\lefteqn{\hspace{-1cm}
 \frac{\I\epsi}{4}\sqrt{M(M-1)}\int d\xi\int d\zeta\, \hat{\xi}_{j_{2}}\hat{\zeta}_{j_{2}} \frac{\hat{\varrho}_{j_{1}}^{\sigma}(\xi)^{*}\hat{\varrho}_{j_{1}}^{\sigma}(\zeta)^{*}}{\abs{\xi}^{1/2}\abs{\zeta}^{1/2}[1-\I(\abs{\xi}+\abs{\zeta})\epsi^{-1}]}\cdot}\\ &&\cdot\int_{0}^{t}ds\, \E^{-\I s(\abs{\xi}+\abs{\zeta})/\epsi} \E^{\I s\hp/\epsi}\E^{-\I(\xi+\zeta)\cdot x_{j_{1}}}(\xi+\zeta)\cdot\popb_{j_{1}}\E^{-\I s\hp/\epsi}\cdot\\
&&\cdot\Psi(x;\xi, \zeta, k_{1}, \ldots, k_{M-2})\,.
\end{eqnarray*}
the others can be treated in the same way.
Using Cauchy-Schwarz inequality we get that, independently of $\sigma$,
\begin{eqnarray*}\lefteqn{
 \norm{\cdots}_{\hilbert{H}}^{2}\leq C\epsi^{2}\abs{t}M(M-1)\int_{\sigma}^{\Lambda} d\abs{\xi}\int_{\sigma}^{\Lambda} d\abs{\zeta} \frac{\abs{\xi}\abs{\zeta}(\abs{\xi}^{2}+\abs{\zeta}^{2})}{1+(\abs{\xi}+\abs{\zeta})^{2}\epsi^{-2}}\cdot}\\
&&\cdot \int_{0}^{t}ds\, \bigg\lVert \frac{\xi+\zeta}{\abs{\xi+\zeta}}\cdot\popb_{j_{1}}\E^{-\I s\hp/\epsi}\qn\tilde{\chi}(\heff)\bigg\rVert_{\mathcal{L}(\hilbert{H})}^{2}\norm{\Psi_{0}}_{\hilbert{H}}^{2}=\Or(\epsi^{4}\abs{t}^{2})_{\hilbert{H}}\,.
\end{eqnarray*}

The creation part can be estimated in a way entirely analogous to the one already employed for the first term of $\tilde{h}_{2, \mathrm{OD}}$.\qed
\end{proof}

\begin{corollary}\label{radiatedpiece} The radiated piece (i.\ e. the piece of the wave function which makes a transition between the almost invariant subspaces) for a system starting in the Fock vacuum is given by
\begin{equation}\begin{split}
&\tilde{\Psi}_{\rm rad}(t) :=   (\id - Q_{0})e^{-\I \frac{t}{\epsi}\heff}\psi(x)\Omega_{\mathrm{F}} =\\
&-\frac{\epsi}{\sqrt{2}}\E^{-\I th_{0}/\epsi}\sum_{j=1}^{N}\frac{\hat{\varrho}_{j}^{\sigma(\epsi)}(k)}{\abs{k}^{3/2}}\kappa\cdot\int_{0}^{t}ds\, \E^{\I s\abs{k}/\epsi}\opw(\ddot{\vec{x}}_{j}^{c}(s; x, p))\psi(x) +\\
&+ \tilde{R}(t, \epsi)\, ,
\end{split}
\end{equation}
where 
\begin{equation*}
\norm{\tilde{R}(t, \epsi)}_{\hilbert{H}}\leq C\epsi^{2}\log(\epsi^{-1})(\abs{t}+\abs{t}^{2})(\norm{\psi}_{\hilbert{H}} + \norm{\abs{x}\psi}_{\hilbert{H}} + \norm{\abs{\pop}\psi}_{\hilbert{H}})\, ,
\end{equation*}
and $\vec{x}_{j}^{c}$ is the solution to the classical equations of motion
\begin{equation}\begin{split}
& \ddot{\vec{x}}_{j}^{c}(s; x, p) = -\nabla_{\vec{x}_{j}}E(x^{c}(s; x, p)),\\
& \vec{x}_{j}^{c}(0; x, p)=\vec{x}_{j},\qquad \dot{\vec{x}}_{j}^{c}(0; x, p)=\vec{p}_{j},\quad j=1, \ldots, n\, .
\end{split}
\end{equation}

We get the leading order of the radiated piece corresponding to the original Hamiltonian $\hepsi$, for a system starting in the approximate dressed vacuum $\Omega_{\sigma(\epsi)}(x)$, applying to this wave function the dressing operator $V_{\sigma(\epsi)}(x)$.
\end{corollary}

\begin{proof}
Applying equation \eqref{firstordertime} for the case $M=0$ we get at the leading order
\begin{eqnarray*}
\lefteqn{Q_{0}^{\perp}\E^{-\I t\heff/\epsi}\psi(x)\Omega_{\mathrm{F}}=-\I\epsi\int_{0}^{t}ds\ \E^{\I(s-t)h_{0}/\epsi}h_{2, \mathrm{OD}}\E^{-\I sh_{0}/\epsi}\psi(x)\Omega_{\mathrm{F}}=}\\
&=&-\frac{\epsi}{\sqrt{2}}\int_{0}^{t}ds\, \E^{\I(s-t)\hp/\epsi}\E^{\I(s-t)\abs{k}/\epsi}\nabla E(x)\cdot \frac{\I\nabla_{x}v_{\sigma}(x, k)}{\abs{k}}\E^{-\I s\hp/\epsi}\psi(x)\\
&=&\frac{\epsi}{\sqrt{2}}\sum_{j=1}^{N}\frac{\hat{\varrho}_{j}^{\sigma}(k)}{\abs{k}^{3/2}}\kappa\cdot\int_{0}^{t}ds\, \E^{\I(s-t)\abs{k}/\epsi}\E^{\I (s-t)\hp/\epsi}\E^{\I k\cdot \vec{x}_{j}}\nabla_{\vec{x}_{j}} E(x) \E^{-\I s\hp/\epsi}\psi(x)\\
&=&\frac{\epsi}{\sqrt{2}}\sum_{j=1}^{N}\frac{\hat{\varrho}_{j}^{\sigma}(k)}{\abs{k}^{3/2}}\kappa\cdot\int_{0}^{t}ds\, \E^{\I(s-t)\abs{k}/\epsi}\E^{\I (s-t)\hp/\epsi}(\E^{\I k\cdot \vec{x}_{j}}-1)\nabla_{\vec{x}_{j}} E(x)\E^{-\I s\hp/\epsi}\psi(x)\\
&&+\,\frac{\epsi}{\sqrt{2}}\sum_{j=1}^{N}\frac{\hat{\varrho}_{j}^{\sigma}(k)}{\abs{k}^{3/2}}\kappa\cdot\int_{0}^{t}ds\, \E^{\I(s-t)\abs{k}/\epsi}\E^{\I (s-t)\hp/\epsi}\nabla_{\vec{x}_{j}} E(x)\E^{-\I s\hp/\epsi}\psi(x)\\
&=&\frac{\epsi}{\sqrt{2}}\sum_{j=1}^{N}\frac{\hat{\varrho}_{j}^{\sigma}(k)}{\abs{k}^{3/2}}\kappa\cdot\int_{0}^{t}ds\, \E^{\I(s-t)\abs{k}/\epsi}\E^{\I (s-t)\hp/\epsi}(\E^{\I k\cdot \vec{x}_{j}}-1)\nabla_{\vec{x}_{j}} E(x)\E^{-\I s\hp/\epsi}\psi(x)\\
&&-\,\E^{-\I t h_{0}} \frac{\epsi}{\sqrt{2}}\sum_{j=1}^{N}\frac{\hat{\varrho}_{j}^{\sigma}(k)}{\abs{k}^{3/2}}\kappa\cdot\int_{0}^{t}ds\, \E^{\I s\abs{k}/\epsi}\opw(\ddot{\vec{x}}_{j}(s; x, p))\psi(x)\\ 
&&+\,\Or(\epsi^{2}\abs{t})_{\mathcal{L}(\hilbert{H})}\norm{\psi}_{L^{2}(\field{R}{3n})}\,,
\end{eqnarray*}
where we have used Egorov's theorem to approximate $\E^{\I s\hp/\epsi}\nabla_{x_{j}} E(x)\E^{-\I s\hp/\epsi}$ (see, e. g., \cite{Ro}). To end the proof we have to show that the norm of the first term is small. The procedure to employ is identical to the one applied several times in the proof of theorem \ref{firstorderapprox}: First integrate by parts with respect to $s$ and then estimate the resulting terms. For the sake of completeness we show how to estimate one of these terms
\begin{equation*}
\frac{\epsi\hat{\varrho}_{j}^{\sigma}(k)\kappa}{\sqrt{2}\abs{k}^{3/2}(1 + \I\abs{k}\epsi^{-1})}\cdot\int_{0}^{t}ds\, \E^{\I(s-t)\abs{k}/\epsi}\E^{\I (s-t)\hp/\epsi}(\E^{\I k\cdot \vec{x}_{j}}-1)\nabla_{\vec{x}_{j}} E(x)\E^{-\I s\hp/\epsi}\psi(x)
\end{equation*}
because the others have an analogous structure. Its norm satisfies 
\begin{equation*}\begin{split}
&\norm{\cdots}^{2}_{\hilbert{H}}\\
&\leq\frac{\epsi^{2}\abs{t}}{2}\int dk\,\frac{\abs{\hat{\varrho}_{j}^{\sigma}(\epsi k)}^{2}}{\abs{k}^{3}(1 + \abs{k}^{2})}\int_{0}^{t}ds\,\bigg\lVert(\E^{\I \epsi k\cdot \vec{x}_{j}}-1)\nabla_{\vec{x}_{j}}E\,\E^{-\I s\hp/\epsi}\psi\bigg\rVert^{2}_{L^{2}(\field{R}{3N})}\\
&\leq \tilde{C}\epsi^{4}\abs{t} \int_{\sigma}^{\Lambda/\epsi} d\abs{k}\, \frac{\abs{k}}{1 + \abs{k}^{2}}\int_{0}^{t}ds\, \bigg\lVert \abs{\vec{x}_{j}}\nabla_{\vec{x}_{j}}E\,\E^{-\I s\hp/\epsi}\psi\bigg\rVert^{2}_{L^{2}(\field{R}{3N})}\\
&= \tilde{C}\epsi^{4}\abs{t}\sup_{x\in\field{R}{3N}}\abs{\nabla_{\vec{x}_{j}}E(x)}\log\bigg(\frac{1 +\Lambda\epsi^{-2}}{1 +\sigma(\epsi)^{2}}\bigg)\int_{0}^{t}ds\, \bigg\lVert \abs{\vec{x}_{j}}\E^{-\I s\hp/\epsi}\psi\bigg\rVert^{2}_{L^{2}(\field{R}{3N})}\\
&\leq  C\epsi^{4}\abs{t}(\abs{t} + \abs{t}^{2})\log\bigg(\frac{1 +\Lambda\epsi^{-2}}{1 +\sigma(\epsi)^{2}}\bigg)(\norm{\abs{{\popb}_{j}}\psi}^{2} + \norm{\abs{\vec{x}_{j}}\psi}^{2} +\norm{\psi}^{2}) \, ,\quad
\end{split}
\end{equation*}
where in the last inequality we have applied theorem $2.1$ from \cite{RaSi}. \qed
\end{proof}

\begin{remark}\label{remarkradiatedpiece}
The norm squared of the leading part of the radiated piece is 
\begin{equation*}\begin{split}
& \frac{2\epsi^{2}}{3\pi}\int dx \int_{0}^{\infty} dk\, \frac{\abs{\hat{\varphi}_{\sigma}(k)}^{2}}{k}\bigg\lVert \opw\bigg(\int_{0}^{t}ds\, \E^{-\I sk/\epsi}\sum_{j=1}^{N}\ddot{\vec{x}}^{c}_{j}(s; x,p)\bigg)\psi(x)\bigg\rVert_{\field{C}{3}}^{2}\geq\\
&\geq \frac{\epsi^{2}}{12\pi^{4}}\int dx \int_{\sigma}^{\Lambda} dk\, \frac{1}{k}\bigg\lVert \opw\bigg(\int_{0}^{t}ds\, \E^{-\I sk/\epsi}\sum_{j=1}^{N}\ddot{\vec{x}}^{c}_{j}(s; x,p)\bigg)\psi(x)\bigg\rVert_{\field{C}{3}}^{2}\geq\\
& \geq \frac{\epsi^{2}}{12\pi^{4}}\int dx \int_{\sigma(\epsi)\epsi^{-1}}^{\Lambda} dk\, \frac{1}{k}\bigg\lVert \opw\bigg(\int_{0}^{t}ds\, \E^{-\I sk}\sum_{j=1}^{N}\ddot{\vec{x}}^{c}_{j}(s; x,p)\bigg)\psi(x)\bigg\rVert_{\field{C}{3}}^{2}\, .
\end{split}
\end{equation*}
The symbol which appears in the Weyl quantization
\begin{equation*}
\int_{0}^{t}ds\, \E^{-\I sk}\sum_{j=1}^{N}\ddot{\vec{x}}^{c}_{j}(s; x,p)
\end{equation*}
is independent of $\epsi$ and for $k=0$ is a non null function, 
\begin{equation*}
\sum_{j=1}^{N}[\dot{\vec{x}}_{j}^{c}(t; x, p) - \dot{\vec{x}}_{j}^{c}(0, x, p)]=\sum_{j=1}^{N}[\dot{\vec{x}}_{j}^{c}(t; x, p) - \vec{p}_{j}],
\end{equation*}
therefore the corresponding operator will be also non null.
Therefore, we expect for a generic state $\psi$ that
\begin{equation}\label{conditionpsi}
\inf_{0\leq k \leq \Lambda}\bigg\lVert \opw\bigg(\int_{0}^{t}ds\, \E^{-\I sk}\sum_{j=1}^{n}\ddot{\vec{x}}^{c}_{j}(s; x,p)\bigg)\psi(x)\bigg\rVert_{L^{2}(\field{R}{3N})\otimes\field{C}{3}}>0 \, .
\end{equation}
Then the norm of the radiated piece will be bigger than
\begin{equation*}\begin{split}
&\hspace{-5mm} \inf_{0\leq k \leq \Lambda}\bigg\lVert \opw\bigg(\int_{0}^{t}ds\, \E^{-\I sk}\sum_{j=1}^{N}\ddot{\vec{x}}^{c}_{j}(s; x,p)\bigg)\psi(x)\bigg\rVert_{L^{2}(\field{R}{3n})\otimes\field{C}{3}}\cdot\\
&\hspace{5mm}\cdot\bigg(\frac{\epsi^{2}}{12\pi^{4}} \int_{\sigma(\epsi)\epsi^{-1}}^{\Lambda} dk\, \frac{1}{k}\bigg)^{1/2}=\Or\big(\epsi\log(\epsi\sigma(\epsi)^{-1})\big),
\end{split}
\end{equation*}
which gives a lower bound on the transition almost of the same order of the upper bound.
\end{remark}

\begin{remark}\label{remradiatedpower}
The radiated energy, defined in equation \eqref{radiatedenergy}, can be written at the leading order as
\begin{equation*}
E_{\textrm{rad}}(t) = \langle \E^{-\I t\heff/\epsi}\psi_{\textrm{d}}(x)\fockvac, (\id\otimes\hfield)\E^{-\I t\heff/\epsi}\psi_{\textrm{d}}(x)\fockvac\rangle,
\end{equation*}
where $\psi_{\textrm{d}}$ is defined in \eqref{psidressed}.
Using the expression for the radiated piece we get  
\begin{eqnarray*}
\lefteqn{\hspace{-9mm} E_{\textrm{rad}}(t) = \langle Q_{0}^{\perp}\E^{-\I t\heff/\epsi}\psi_{\textrm{d}}\fockvac, (\id\otimes\hfield)Q_{0}^{\perp}\E^{-\I t\heff/\epsi}\psi_{\textrm{d}}\fockvac\rangle\cong}\\
&\cong&\frac{\epsi^{2}}{2}\sum_{i, j=1}^{N}\int_{\field{R}{3}}dk\, \frac{\hat{\varrho}_{i}^{\sigma(\epsi)*}\hat{\varrho}_{j}^{\sigma(\epsi)}}{\abs{k}^{2}}\int_{0}^{t}ds\,\int_{0}^{t}ds'\, \E^{\I(s-s')\abs{k}/\epsi}\cdot\\
&&\cdot\langle \kappa\cdot\opw(\vec{\ddot{x}}_{i}^{c}(s'))\psi_{\textrm{d}}, \kappa\cdot\opw(\vec{\ddot{x}}_{j}^{c}(s))\psi_{\textrm{d}}\rangle_{L^{2}(\field{R}{3N})}\\
&=&\frac{\epsi^{2}}{12\pi^{2}}\sum_{i, j=1}^{N}e_{i}e_{j}\int_{0}^{t}ds\,\int_{0}^{t}ds'\, \frac{\epsi}{\I(s-s')}(\E^{\I(s-s')\Lambda/\epsi}-1)\cdot\\
&&\cdot\langle \psi_{\textrm{d}}, \opw(\vec{\ddot{x}}_{i}^{c}(s')\cdot\vec{\ddot{x}}_{j}^{c}(s))\psi_{\textrm{d}}\rangle_{L^{2}(\field{R}{3N})}\,,
\end{eqnarray*}
where we have inserted the explicit expression of the form factor given in \eqref{formfactor} and used the product formula \eqref{productformula} at leading order.
The radiated power is then
\begin{equation*}\begin{split}
&\hspace{-2cm} P_{\textrm{rad}}(t) = \frac{d}{dt}E_{\textrm{rad}}(t) = \frac{\epsi^{3}}{6\pi^{2}}\sum_{i, j=1}^{N}\int_{0}^{t}ds\, \frac{\sin[(t-s)\Lambda/\epsi]}{t-s}\cdot\\
&\hspace{1cm}\cdot\langle \psi_{\textrm{d}}, \opw(\vec{\ddot{x}}_{i}^{c}(t)\cdot\vec{\ddot{x}}_{j}^{c}(s))\psi_{\textrm{d}}\rangle_{L^{2}(\field{R}{3N})}
\end{split}
\end{equation*}
which converges formally to the expression given in \eqref{radiatedpower} when $\epsi\to 0^{+}$.
\end{remark}

\begin{corollary} Let 
\begin{equation*}
\omega(t):= \E^{-\I t\heff/\epsi}\omega_{0}\E^{\I t\heff/\epsi},
\end{equation*}
where $\omega_{0}\in\hilbert{I}_{1}(\qm\tilde{\chi}(\heff)\hilbert{H})$, the Banach space of trace class operators on $\qm\tilde{\chi}(\heff)\hilbert{H}$, and let $\omegap$ be the partial trace over the field states
\begin{equation*}
\omegap(t):= \mathrm{tr}_{\fock}\,\omega(t),
\end{equation*}
then
\begin{eqnarray*} 
 \omegap(t) &=& \E^{-\I t\hdiagp/\epsi}\omegap(0)\E^{\I t\hdiagp/\epsi}+ \Or(\epsi^{3/2}\abs{t})_{\hilbert{I}_{1}(L^{2}(\field{R}{3N}))}(1-\delta_{M0})+ \\&&+\,\Or(\epsi^{2}\abs{t}\sqrt{\log(\sigma(\epsi)^{-1})})_{\hilbert{I}_{1}(L^{2}(\field{R}{3N}))}  \hspace{-1.9pt}+\Or(\epsi^{2}\abs{t}^{2}\sqrt{\log(\sigma(\epsi)^{-1})})_{\hilbert{I}_{1}(L^{2}(\field{R}{3N}))},\nonumber
\end{eqnarray*}
where
\begin{eqnarray*}
\hdiagp&:=& \sum_{l=1}^{N}\frac{1}{2m_{l}^{\epsi}}\popb_{l}^{2}+E(x) \\
&&-\,\frac{\epsi^{2}}{4}\underset{(l_{1}\neq j_{1})}{\sum_{l_{1}, j_{1}=1}^{N}}\int_{\field{R}{3}}dk\, \frac{\hat{\varrho}_{l_{1}}(k)^{*}\hat{\varrho}_{j_{1}}(k)}{\abs{k}^{2}}\big[\E^{\I k\cdot(\vec{x}_{j_{1}}-\vec{x}_{l_{1}})}(\kappa\cdot\popb_{l_{1}})(\kappa\cdot\popb_{j_{1}})\\ &&\hspace{2cm}+\,(\kappa\cdot\popb_{l_{1}})(\kappa\cdot\popb_{j_{1}})\E^{\I k\cdot(\vec{x}_{j_{1}}-\vec{x}_{l_{1}})}\big]\,,
\end{eqnarray*}
and $\hilbert{I}_{1}(L^{2}(\field{R}{3N}))$ denotes the space of trace class operators on $L^{2}(\field{R}{3N})$. 
\end{corollary}

\begin{proof}
The proof follows from the following three facts:
\begin{enumerate}
\item the term of order $\epsi$ in equation \eqref{firstordertime} is off-diagonal with respect to the $\qm$;

\item the diagonal Hamiltonian $\hdiag$, defined in \eqref{hdiag}, is equal to $\hdiagp\otimes\mathbf{1}+\mathbf{1}\otimes\hfield$, so we have that
\begin{equation*}
\mathrm{tr}_{\fock}\big(\E^{-\I t\hdiag/\epsi}\omega_{0}\E^{\I t\hdiag/\epsi}\big)=\E^{-\I t\hdiagp/\epsi}\mathrm{tr}_{\fock}(\omega_{0})\E^{\I t\hdiagp/\epsi}\, ;
\end{equation*}

\item the following well know inequality,which holds for any Hilbert space $\hilbert{H}$:
\begin{equation*}
\norm{AB}_{\hilbert{I}_{1}(\hilbert{H})}\leq \norm{A}_{\hilbert{I}_{1}(\hilbert{H})}\cdot\norm{B}_{\mathcal{L}(\hilbert{H})}\,.\quad\qed
\end{equation*}

\end{enumerate}
\end{proof}

\begin{theorem}\label{densitymatrix}
Given a macroscopic observable for the particles, $\opw(a)$, where $a$ is a smooth function bounded with all its derivatives, and a density matrix $\omega\in \hilbert{I}_{1}(P_{M}^{\epsi}\chi(\hepsi)\hilbert{H})$
whose time evolution is defined by
\begin{equation}
\omega(t):= e^{-\I t\hepsi/\epsi}\omega e^{\I t\hepsi/\epsi},
\end{equation}
then
\begin{eqnarray}\label{trace} 
\lefteqn{\hspace{-2.3cm} \tr_{\hilbert{H}}\bigg(\big(\opw(a)\otimes\id_{\fock}\big)\omega(t)\bigg)=\tr_{L^{2}(\field{R}{3N})}\bigg(\opw(a)e^{-\I t H^{(2)}_{\mathrm{D, p}}}\tr_{\fock}\tilde{\omega}(0)e^{\I t H^{(2)}_{\mathrm{D, p}}}\bigg) +}\nonumber\\ && +\Or(\epsi^{3/2}\abs{t})(1-\delta_{M0})+\,\Or\big(\epsi^{2}(\abs{t} + \abs{t}^{2})\log(\sigma(\epsi)^{-1})\big),
\end{eqnarray}
where
\begin{equation}\label{omegatilde}
\tilde{\omega}(0):= V_{\sigma(\epsi)}^{*}\omega V_{\sigma(\epsi)}\, .
\end{equation}
\end{theorem}

\begin{proof}
First of all we observe that, using proposition \ref{approxsigma} and lemma \ref{cutoffenergysigma}, we have
\begin{equation*}\begin{split}
& \tr_{\hilbert{H}}\bigg(\big(\opw(a)\otimes\id_{\fock}\big)\omega(t)\bigg) = \tr_{\hilbert{H}}\bigg(\big(\opw(a)\otimes\id_{\fock}\big)e^{-\I t\hepsisigmaepsi/\epsi}\omega_{\sigma(\epsi)}\cdot\\
&\cdot e^{\I t\hepsisigmaepsi/\epsi}\bigg) + \Or(\sigma(\epsi)^{1/2}) + \Or(\sigma(\epsi)\epsi^{-1})\,,
\end{split}
\end{equation*}
where $\omega_{\sigma(\epsi)}\in \hilbert{I}_{1}(P_{M}^{\epsi}\chi(\hepsisigmaepsi)\hilbert{H})$.
By the definition of the dressed Hamiltonian and the cyclicity of the trace we have then at the leading order
\begin{eqnarray*} \lefteqn{\hspace{-5mm}
 \tr_{\hilbert{H}}\bigg(\big(\opw(a)\otimes\id_{\fock}\big)\omega(t)\bigg)=\Or(\sigma(\epsi)^{1/2}) + \Or(\sigma(\epsi)\epsi^{-1})+}\\&+& \tr_{\hilbert{H}}\bigg(\hilbert{U}\big(\opw(a)\otimes\id_{\fock}\big)\hilbert{U}^{*}e^{-\I t\heff/\epsi}\hilbert{U}\omega_{\sigma(\epsi)}\hilbert{U}^{*}e^{\I t\heff/\epsi}\bigg) \, .
\end{eqnarray*}
The transformed observable, using the definition of $\hilbert{U}$ and lemma \ref{approxchi}, is given by
\begin{eqnarray*} 
\lefteqn{ \hspace{-5mm}\hilbert{U}\big(\opw(a)\otimes\id_{\fock}\big)\hilbert{U}^{*} = \vsigma^{*}\big(\opw(a)\otimes\id_{\fock}\big)\vsigma + \I\epsi\chi(h_{0})\phisigmaj\cdot\pop\vsigma^{*}\cdot}\\&&
\cdot\big(\opw(a)\otimes\id_{\fock}\big)\vsigma +\I\epsi\big(\id - \chi(h_{0})\big)\phisigmaj\cdot\pop\chi(h_{0})\vsigma^{*}\big(\opw(a)\otimes\id_{\fock}\big)\vsigma \\
&&-\I\epsi\vsigma^{*}\big(\opw(a)\otimes\id_{\fock}\big)\vsigma\chi(h_{0})\pop\cdot\phisigmaj\\&& -\,\I\epsi\vsigma^{*}\big(\opw(a)\otimes\id_{\fock}\big)\vsigma\big(\id - \chi(h_{0})\big)\pop\cdot\phisigmaj\chi(h_{0}) +\Or(\epsi^{2}\log(\sigma^{-1})) \, .
\end{eqnarray*}
The operator $\vsigma$ can be considered as the Weyl quantization of the operator valued symbol
\begin{equation*}
(x, p) \to \E^{-\I\Phi(\I v_{\sigma}(x, \cdot))}\in\mathcal{L}(\fock) \,.
\end{equation*}
It is not in the standard symbol classes because the derivative is an unbounded operator, but if we multiply it by $\qm$ we get a smooth bounded symbol. In the calculation of the trace $\vsigma$ is always multiplied by $\qm$ or $Q_{M\pm 1}$, therefore we can proceed as if it were in a standard symbol class.

A simple application of the product rule for pseudodifferential operators (equation (A.11) \cite{Te3}) gives then
\begin{equation*}\begin{split}
& \hspace{-5mm}\vsigma^{*}\big(\opw(a)\otimes\id_{\fock}\big)\vsigma = \opw(a)\otimes\id_{\fock} + \vsigma^{*}[\big(\opw(a)\otimes\id_{\fock}\big), \vsigma] =\\
&= \opw(a)\otimes\id_{\fock} - \I\epsi\vsigma^{*}\opw\big(\{a, \vsigma\}\big) + \Or(\epsi^{3})_{\mathcal{L}(\hilbert{H})}\,,
\end{split}
\end{equation*}
where $\{\cdot, \cdot\}$ denotes the Poisson bracket. We get therefore
\begin{equation*}
\{a, \vsigma\} = (\nabla_{p}a)\cdot\I\Phi(\I\nabla_{x}v_{\sigma})\vsigma
\end{equation*}
and
\begin{eqnarray*} 
 \vsigma^{*}\big(\opw(a)\otimes\id_{\fock}\big)\vsigma &=& \opw(a)\otimes\id_{\fock} +\epsi\Phi(\I\nabla_{x}v_{\sigma})\big(\opw(\nabla_{p}a)\otimes\id_{\fock}\big) \\
&&+\, \Or(\epsi^{2}\log(\sigma^{-1}))\, .
\end{eqnarray*}
Using this expression we have 
\begin{eqnarray*}
\lefteqn{ \hilbert{U}\big(\opw(a)\otimes\id_{\fock}\big)\hilbert{U}^{*} = \opw(a)\otimes\id_{\fock} +\epsi\Phi(\I\nabla_{x}v_{\sigma})\big(\opw(\nabla_{p}a)\otimes\id_{\fock}\big) +}\\& &+\,\I\epsi\chi(h_{0})\phisigmaj\cdot\pop\big(\opw(a)\otimes\id_{\fock}\big)+\I\epsi\big(\id - \chi(h_{0})\big)\phisigmaj\cdot\pop\chi(h_{0})\big(\opw(a)\otimes\id_{\fock}\big) \\
&&-\,\I\epsi\big(\opw(a)\otimes\id_{\fock}\big)\chi(h_{0})\pop\cdot\phisigmaj\\&& -\,\I\epsi\big(\opw(a)\otimes\id_{\fock}\big)\big(\id - \chi(h_{0})\big)\pop\cdot\phisigmaj\chi(h_{0}) +\Or(\epsi^{2}\log(\sigma^{-1})) \, .
\end{eqnarray*}
All the terms of order $\epsi$ in the previous expression are off-diagonal with respect to the $\qm$s, and the same holds for the term of order $\epsi$ in \eqref{firstordertime}. Therefore, they all vanish when we calculate the trace. Using point $2$ and $3$ of last corollary we get then \eqref{trace} with
\begin{equation*}
\tilde{\omega}(0) = \hilbert{U}\omega_{\sigma(\epsi)}\hilbert{U}^{*}.
\end{equation*}
Using again lemma \ref{cutoffenergysigma} and the fact that the terms of order $\epsi$ in the expansion of $\hilbert{U}$ are off-diagonal we get in the end also \eqref{omegatilde}.\qed
\end{proof}

\section*{Acknowledgment}
We thank Alessandro Pizzo for showing us the strategy to prove proposition \ref{approxsigma} and the German Research Foundation DFG for financial support.

\end{document}